\documentclass[twocolumn,tighten]{aastex63}

\usepackage{natbib}
\usepackage{multirow}
\usepackage{amsmath}
\usepackage{hyperref}
\hypersetup{citecolor=blue,urlcolor=blue}

\shorttitle{ALMA observations of the $z = 7.07$ LLQSO J1243$+$0100}
\shortauthors{T. Izumi et al.}
\begin{document}
\title{Subaru High-$z$ Exploration of Low-Luminosity Quasars (SHELLQs). XIII. \\ 
Large-scale Feedback and Star Formation in a Low-Luminosity Quasar at $z = 7.07$ \\ 
on the Local Black Hole to Host Mass Relation}

\correspondingauthor{Takuma Izumi}
\email{takuma.izumi@nao.ac.jp}

\author[0000-0001-9452-0813]{Takuma Izumi}
\altaffiliation{NAOJ Fellow}
\affil{National Astronomical Observatory of Japan, 2-21-1 Osawa, Mitaka, Tokyo 181-8588, Japan}
\affil{Department of Astronomical Science, The Graduate University for Advanced Studies, SOKENDAI, 2-21-1 Osawa, Mitaka, Tokyo 181-8588, Japan}
\author{Yoshiki Matsuoka} 
\affil{Research Center for Space and Cosmic Evolution, Ehime University, 2-5 Bunkyo-cho, Matsuyama, Ehime 790-8577, Japan}
\author[0000-0001-7201-5066]{Seiji Fujimoto}
\affil{Cosmic Dawn Center (DAWN), Copenhagen, Denmark}
\affil{Niels Bohr Institute, University of Copenhagen, Jagtvej 128, 2200 Copenhagen N}
\author[0000-0003-2984-6803]{Masafusa Onoue} 
\affil{Max-Planck-Institut f\"{u}r Astronomie, K\"{o}nigstuhl 17, D-69117 Heidelberg, Germany}
\author[0000-0002-0106-7755]{Michael A. Strauss}
\affil{Princeton University Observatory, Peyton Hall, Princeton, NJ 08544, USA}
\author[0000-0003-1937-0573]{Hideki Umehata}
\affil{RIKEN Cluster for Pioneering Research, 2-1 Hirosawa, Wako, Saitama 351-0198, Japan}
\author[0000-0001-6186-8792]{Masatoshi Imanishi}
\affil{National Astronomical Observatory of Japan, 2-21-1 Osawa, Mitaka, Tokyo 181-8588, Japan}
\affil{Department of Astronomical Science, The Graduate University for Advanced Studies, SOKENDAI, 2-21-1 Osawa, Mitaka, Tokyo 181-8588, Japan}
\author[0000-0002-4052-2394]{Kotaro Kohno}
\affil{Institute of Astronomy, Graduate School of Science, The University of Tokyo, 2-21-1 Osawa, Mitaka, Tokyo 181-0015, Japan}
\affil{Research Center for the Early Universe, Graduate School of Science, The University of Tokyo, 7-3-1 Hongo, Bunkyo-ku, Tokyo 113-0033, Japan}
\author[0000-0002-3866-9645]{Toshihiro Kawaguchi}
\affil{Department of Economics, Management and Information Science, Onomichi City University, Hisayamada 1600-2, Onomichi, Hiroshima 722-8506, Japan}
\author[0000-0002-6808-2052]{Taiki Kawamuro}
\affil{Nu\'{c}leo de Astronom\'{i}a de la Facultad de Ingenier\'{i}a, Universidad Diego Portales, Av. Ej\'{e}ercito Libertador 441, Santiago, Chile}
\author[0000-0002-9850-6290]{Shunsuke Baba}
\affil{National Astronomical Observatory of Japan, 2-21-1 Osawa, Mitaka, Tokyo 181-8588, Japan}
\author[0000-0002-7402-5441]{Tohru Nagao}
\affil{Research Center for Space and Cosmic Evolution, Ehime University, 2-5 Bunkyo-cho, Matsuyama, Ehime 790-8577, Japan}
\author[0000-0002-3531-7863]{Yoshiki Toba}
\affiliation{Department of Astronomy, Kyoto University, Kitashirakawa-Oiwake-cho, Sakyo-ku, Kyoto 606-8502, Japan}
\affiliation{Academia Sinica Institute of Astronomy and Astrophysics, 11F of Astronomy-Mathematics Building, AS/NTU, No.1, Section 4, Roosevelt Road, Taipei 10617, Taiwan}
\affiliation{Research Center for Space and Cosmic Evolution, Ehime University, 2-5 Bunkyo-cho, Matsuyama, Ehime 790-8577, Japan}
\author[0000-0001-9840-4959]{Kohei Inayoshi}
\affil{Kavli Institute for Astronomy and Astrophysics, Peking University, Beijing 100871, People’s Republic of China}
\author[0000-0002-0000-6977]{John D. Silverman}
\affil{Kavli Institute for the Physics and Mathematics of the Universe (Kavli-IPMU, WPI), The University of Tokyo Institutes for Advanced Study, The University of Tokyo, 5-1-5 Kashiwanoha, Kashiwa, Chiba 277-8583, Japan}
\affil{Department of Astronomy, School of Science, The University of Tokyo, 7-3-1 Hongo, Bunkyo-ku, Tokyo 113-0033, Japan}
\author[0000-0002-7779-8677]{Akio K. Inoue}
\affil{Department of Physics, School of Advanced Science and Engineering, Faculty of Science and Engineering, Waseda University, 3-4-1, Okubo, Shinjuku, Tokyo 169-8555, Japan} 
\affil{Waseda Research Institute for Science and Engineering, Faculty of Science and Engineering, Waseda University, 3-4-1, Okubo, Shinjuku, Tokyo 169-8555, Japan}
\author{Soh Ikarashi}
\affil{Centre for Extragalactic Astronomy, Department of Physics, Durham University, South Road, Durham DH1 3LE, UK}
\author[0000-0002-4923-3281]{Kazushi Iwasawa}
\affil{ICREA and Institut de Ci\`{e}ncies del Cosmos, Universitat de Barcelona, IEEC-UB, Mart\'{i} i Franqu\`{e}s, 1,E-08028 Barcelona, Spain}
\author[0000-0003-3954-4219]{Nobunari Kashikawa}
\affil{Department of Astronomy, School of Science, The University of Tokyo, 7-3-1 Hongo, Bunkyo-ku, Tokyo 113-0033, Japan}
\author[0000-0002-0898-4038]{Takuya Hashimoto}
\affil{Tomonaga Center for the History of the Universe (TCHoU), Faculty of Pure and Applied Sciences, University of Tsukuba, Tsukuba, Ibaraki 305-8571, Japan}
\author[0000-0002-6939-0372]{Kouichiro Nakanishi}
\affil{National Astronomical Observatory of Japan, 2-21-1 Osawa, Mitaka, Tokyo 181-8588, Japan}
\affil{Department of Astronomical Science, The Graduate University for Advanced Studies, SOKENDAI, 2-21-1 Osawa, Mitaka, Tokyo 181-8588, Japan}
\author[0000-0001-7821-6715]{Yoshihiro Ueda}
\affil{Department of Astronomy, Kyoto University, Kitashirakawa-Oiwake-cho, Sakyo-ku, Kyoto 606-8502, Japan}
\author[0000-0001-7825-0075]{Malte Schramm}
\affil{Graduate school of Science and Engineering, Saitama University, 255 Shimo-Okubo, Sakura-ku, Saitama City, Saitama 338-8570, Japan} 
\author[0000-0003-1700-5740]{Chien-Hsiu Lee} 
\affil{NSF's National Optical-Infrared Astronomy Research Laboratory, 950 North Cherry Avenue, Tucson, AZ 85719, USA}
\author[0000-0002-2536-1633]{Hyewon Suh}
\affil{Subaru Telescope, National Astronomical Observatory of Japan (NAOJ), 650 North A'ohoku Place, Hilo, HI 96720, USA}
\affil{Gemini Observatory/NSF’s NOIRLab, 670 N. A’ohoku Place, Hilo, HI 96720, USA}

\begin{abstract}
We present ALMA [\ion{C}{2}] 158 $\micron$ line and underlying far-infrared (FIR) continuum emission observations 
($0\arcsec.70 \times 0\arcsec.56$ resolution) toward HSC J124353.93$+$010038.5 (J1243$+$0100) at $z = 7.07$, 
the only low-luminosity ($M_{\rm 1450} > -25$ mag) quasar currently known at $z > 7$. 
The FIR continuum is bright (1.52 mJy) and resolved with a total luminosity of $L_{\rm FIR} = 3.5 \times 10^{12}~L_\odot$. 
The spatially extended component is responsible for $\sim 40\%$ of the emission. 
The area-integrated [\ion{C}{2}] spectrum shows a broad wing 
(${\rm FWHM} = 997$ km s$^{-1}$, $L_{\rm [CII]} = 1.2 \times 10^9~L_\odot$) 
as well as a bright core (${\rm FWHM} = 235$ km s$^{-1}$, $L_{\rm [CII]} = 1.9 \times 10^9~L_\odot$). 
This wing is the first detection of a galactic-scale quasar-driven outflow 
(atomic outflow rate $> 447~M_\odot$ yr$^{-1}$) at $z > 7$. 
The estimated large mass loading factor of the total outflow 
(e.g., $\gtrsim 9$ relative to the [\ion{C}{2}]-based SFR) suggests that this outflow will soon quench the star-formation of the host. 
The core gas dynamics are governed by rotation, 
with a rotation curve suggestive of a compact bulge ($\sim 3.3 \times 10^{10}~M_\odot$), although it is not yet spatially resolved. 
Finally, we found that J1243$+$0100 has a black hole mass-to-dynamical mass ratio 
(and -to-bulge mass ratio) of $\sim 0.4\%$ ($\sim 1\%$), consistent with the local value within uncertainties. 
Our results therefore suggest that the black hole-host co-evolution relation is already in place at $z \sim 7$ for this object. 
\end{abstract}

\section{Introduction}\label{sec1}
The mass accretion onto a supermassive black hole (SMBH, with a mass of $M_{\rm BH} \gtrsim 10^{5-6}~M_\odot$) 
is invoked to explain the enormous luminosity observed as an active galactic nucleus (AGN). 
In the local universe, SMBHs have been identified at the centers of massive galaxies, 
and there is a tight correlation between $M_{\rm BH}$ and the properties of the host galaxy 
such as bulge mass and stellar velocity dispersion 
\citep[e.g.,][]{1998AJ....115.2285M,2000ApJ...539L...9F,2003ApJ...589L..21M,2013ARA&A..51..511K,2015ApJ...813...82R}. 
These relations strongly suggest that the formation and growth of SMBHs and host galaxies are intimately linked, a {\em co-evolution} of these two components of galaxies. 
Although the detailed mechanisms by which the correlation arises remain unclear, 
some theoretical models suggest that strong negative AGN feedback on star formation, 
which is connected to the merger histories of galaxies, plays a key role in driving the co-evolution  
\citep{1988ApJ...325...74S,1998A&A...331L...1S,2005Natur.433..604D,2006ApJS..163....1H}. 
Detections of galaxy-scale AGN-driven outflows in multiphase gas 
\citep[e.g.,][]{2008A&A...491..407N,2012A&A...537A..44A,2012ApJ...746...86G,2012MNRAS.425L..66M,2013MNRAS.436.2576L,2014A&A...562A..21C,2016A&A...591A..28C,2017ApJ...850..140T}, 
a higher AGN fraction in interacting/merging systems \citep[e.g., ][]{2011MNRAS.418.2043E,2011ApJ...743....2S,2018PASJ...70S..37G}, 
as well as the global similarity in star-formation and SMBH accretion histories over cosmic time 
\citep[][for a review]{2014ARA&A..52..415M}, support this view. 

As theoretical models in principle make specific predictions 
for the time evolution of galaxy properties, 
observations of high redshift SMBHs and their host galaxies 
play a crucial role in testing and refining our understanding of the co-evolution process
\citep{2017PASA...34...22G,2017PASA...34...31V,2020ARA&A..58...27I}. 
Massive quiescent galaxies already exist in significant numbers at $z \sim 2-3$ 
\citep[e.g.,][]{2014ApJ...783L..14S,2020ApJ...898..171E}, 
suggesting that AGN feedback is important at even higher redshifts. 
Thus $z > 6$ quasars, seen when the universe was less than a billion years old, 
are a unique beacon to study SMBH and galaxy formation. 
To date, more than 200 quasars with rest-frame ultraviolet (UV) magnitude $M_{\rm 1450} \lesssim -22$ mag 
are known at $z > 5.7$ \citep{2020ARA&A..58...27I}, most of which were discovered by wide-field optical and near-infrared surveys 
\citep[e.g.,][]{2001AJ....122.2833F,2003AJ....125.1649F,2016ApJ...833..222J,2016ApJS..227...11B,2007AJ....134.2435W,
2010AJ....139..906W,2016ApJ...828...26M,2018PASJ...70S..35M,2018ApJS..237....5M,2019ApJ...883..183M}. 
The sample includes eight quasars at $z > 7$ 
\citep{2011Natur.474..616M,2018Natur.553..473B,2018ApJ...869L...9W,
2021arXiv210103179W,2019AJ....157..236Y,2020ApJ...897L..14Y,2019ApJ...883..183M,2019ApJ...872L...2M}. 
They have SMBH masses of $M_{\rm BH} \sim 10^9~M_\odot$, 
challenging models for the formation and initial growth of SMBHs at high redshift. 
It is noteworthy that already at $z > 7$, some quasars show fast {\it nuclear winds} 
as evidenced by broad absorption line (BAL) features and blueshifted ionized line emission 
\citep{2018ApJ...869L...9W,2021arXiv210103179W,2020ApJ...898..105O,2020ApJ...905...51S}. 

Sub/mm observations of the rest-frame far-infrared (FIR) continuum 
and C$^+$ $^2P_{3/2}$ $\rightarrow$ $^2P_{1/2}$ 157.74 $\micron$ ([\ion{C}{2}] 158 $\micron$; 
one of the prime coolants of the cold interstellar medium/ISM) line emission 
have revealed that high-redshift quasar host galaxies possess copious amount of 
cold gas ($\sim 10^{10}~M_\odot$) and dust ($\sim 10^8~M_\odot$), 
with high star-formation rates (SFR) of $\gtrsim 100-1000~M_\odot$ yr$^{-1}$ 
\citep[e.g.,][]{2010ApJ...714..699W,2013ApJ...773...44W,2016ApJ...816...37V,2017ApJ...845..154V,2020ApJ...904..130V}. 
\citet{2012MNRAS.425L..66M} discovered a massive AGN-driven [\ion{C}{2}] outflow 
(with an estimated neutral outflow rate $>1400\,M_\odot$ yr$^{-1}$) in the $z = 6.42$ quasar J1148$+$5251 
that extends over $r > 10$ kpc \citep{2015A&A...574A..14C}. However, this remains the only 
 individual $z > 6$ quasar in which [\ion{C}{2}] outflow has been seen. 
\citet{2019A&A...630A..59B} stacked [\ion{C}{2}] data cubes 
of 48 quasars at $z > 4.5$, and claimed to detect a broad (FWHM $\sim 1700$ km s$^{-1}$) component, 
which they interpreted as a modest AGN-driven outflow ($\sim 100~M_\odot$ yr$^{-1}$) in the average object. 
However, \citet{2020ApJ...904..131N} did a similar stacking analysis of 27 $z \gtrsim 6$ quasars, but they found no evidence for outflows. 

High resolution interferometric observations predominantly 
performed by the Atacama Large Millimeter/submillimeter Array (ALMA) 
have allowed studies of cold gas dynamics in quasar host galaxies 
\citep{2013ApJ...773...44W,2016ApJ...816...37V,2018ApJ...854...97D,2020A&A...637A..84P}. 
These studies revealed that $z \gtrsim 6$ optically luminous ($M_{\rm 1450} \lesssim -26$ mag) quasars have, 
on average, $\sim 10\times$ more massive SMBHs than the local co-evolution relations for a given 
velocity dispersion and/or dynamical mass of  the host, suggesting that SMBHs 
were formed significantly earlier than their host galaxies. This result is in tension 
with hydrodynamic simulations of quasars 
\citep[e.g.,][]{2006ApJS..163....1H,2019MNRAS.488.4004L,2020MNRAS.499.3819M}. 
However, our understanding of $z \gtrsim 6$ quasars has been biased 
to the most luminous (and presumably most massive) SMBH population  \citep{2007ApJ...670..249L,2011MNRAS.417.2085V,2014MNRAS.438.3422S}. 
Indeed, ALMA observations of low-luminosity quasars ($M_{\rm 1450} \gtrsim -25$ mag) find that their star formation rates are lower ($\lesssim 100~M_\odot$ yr$^{-1}$) and their 
SMBH-to-host mass ratios are roughly consistent with the local value 
\citep{2013ApJ...770...13W,2015ApJ...801..123W,2017ApJ...850..108W,2018PASJ...70...36I,2019PASJ...71..111I}. 
Therefore, sensitive observations of lower luminosity objects, even at $z>7$, 
are necessary to gain a less-biased picture of early SMBH/galaxy evolution.

\subsection{Our target: J1243$+$0100}\label{sec1.1}
The wide-field optical deep imaging survey data (the HSC-Subaru Strategic Program \citep[HSC-SSP,][]{2018PASJ...70S...8A})
obtained with the Hyper Suprime-Cam \citep[HSC,][]{2012SPIE.8446E..0ZM,2018PASJ...70S...1M} 
mounted on the 8.2 m Subaru telescope have yielded a large number of low-luminosity quasars at redshifts above 6. 
We have established a multi-wavelength follow-up consortium for $z \gtrsim 6$ HSC quasars, 
{\it the Subaru High-z Exploration of Low-Luminosity Quasars (SHELLQs)}. 
SHELLQs \citep[e.g.,][]{2016ApJ...828...26M,2018PASJ...70S..35M,2018ApJS..237....5M} 
has so far discovered $> 90$ low-luminosity quasars down to $M_{\rm 1450} \sim -22$ mag at $z \gtrsim 6$. 

Our target of this work, HSC J124353.93$+$010038.5 (hereafter J1243$+$0100), 
is the only low-luminosity quasar known at $z > 7$, discovered by \citet{2019ApJ...872L...2M}. 
Optical-to-near infrared spectroscopy allowed determination of the redshift ($z_{\rm MgII} = 7.07 \pm 0.01$), 
UV absolute magnitude ($M_{\rm 1450} = -24.13 \pm 0.08$ mag) 
and bolometric luminosity ($L_{\rm Bol} = (1.4 \pm 0.1) \times 10^{46}$ erg s$^{-1}$), 
the \ion{Mg}{2}-based single epoch black hole mass ($M_{\rm BH} = (3.3 \pm 2.0) \times 10^8~M_\odot$), 
and the corresponding Eddington ratio ($\lambda_{\rm Edd} = 0.34 \pm 0.20$). 
The luminosity of J1243$+$0100 is an order of magnitude lower than the other $z > 7$ quasars known to date \citep{2011Natur.474..616M,2018Natur.553..473B,2018ApJ...869L...9W,2021arXiv210103179W,2019AJ....157..236Y,2020ApJ...897L..14Y}. 
Its $M_{\rm 1450}$ is close to the knee/characteristic magnitude ($M^*_{\rm 1450}$) 
of the quasar luminosity function (QLF) at $z \sim 6$ \citep{2018ApJ...869..150M}. 
Thus if the quasar luminosity function doesn't evolve significantly from $z \sim 7$ to $z \sim 6$, 
we can regard J1243$+$0100 as the first example of an {\it representative} quasar at $z > 7$. 
In addition, J1243$+$0100 has a \ion{C}{4} $\lambda$1549 emission line blueshifted by $-2400$ km s$^{-1}$ relative to \ion{Mg}{2} $\lambda$2800, 
as well as BAL features indicative of fast nuclear outflows. 

In this paper, we present ALMA observations of the [\ion{C}{2}] 158 $\micron$ line 
and the underlying rest-frame FIR continuum emission of J1243$+$0100. 
This is the thirteenth in a series of publications presenting the results of SHELLQs. 
Throughout this work, we adopt the concordant Lambda cold dark matter ($\Lambda$CDM) 
cosmology with $H_0$ = 70 km s$^{-1}$ Mpc$^{-1}$, 
$\Omega_{\rm M} = 0.3$, and $\Omega_{\rm \Lambda} = 0.7$. 
At the redshift of the source ($z = 7.07$), the age of the universe is 0.74 Gyr 
and an angular size of $1\arcsec$ corresponds to 5.2 kpc.

\section{ALMA Observations}\label{sec2}
We observed the redshifted [\ion{C}{2}] line and FIR continuum emission of J1243$+$0100 
in ALMA Band 6 ($\lambda_{\rm obs} = 1.3$ mm) on 2019 October 16 and 22 (ID = 2019.1.00074.S, PI = T. Izumi), as a Cycle 7 program. 
Our observations were conducted in a single pointing 
with a $\sim 24\arcsec$ diameter field of view (FoV), with 41--43 antennas. 
Three spectral windows (each $\sim 1.875$ GHz wide) were placed 
on one side-band to maximize the contiguous frequency coverage. 
We set the phase-tracking center of this pointing to 
($\alpha_{\rm ICRS}$, $\delta_{\rm ICRS}$) = (12$^{\rm h}$43$^{\rm m}$53$^{\rm s}$.930, $+$01\arcdeg00\arcmin38\arcsec.50), 
which corresponds to the optical quasar position tied to the {\it Gaia} astrometry. 
The baseline length ranged from 15.1 m to 740.4 or 783.5 m, 
resulting in a maximum recoverable scale of $\sim 6\arcsec$. 
J1058$+$0133 and J1427$-$4206 were observed as flux and bandpass calibrators, 
and J1232$-$0224 was monitored to calibrate the complex gain variation. 
The total on-source time was 115 minutes. 
Table \ref{tbl1} summarizes these observations. 

The data were processed using \verb|CASA| \citep{2007ASPC..376..127M} version 5.6. 
All images were reconstructed with the \verb|tclean| task using natural weighting to maximize the sensitivity. 
For the [\ion{C}{2}] cube, we averaged several channels to obtain a velocity resolution of 75 km s$^{-1}$, 
which resulted in a 1$\sigma$ channel sensitivity of 0.10 mJy beam$^{-1}$ (beam size = $0\arcsec.70 \times 0\arcsec.56$, P.A. = $-58\arcdeg.4$). 
Note that we first deconvolved the line cube including the continuum emission down to the $3\sigma$ level to determine the line position and to identify the channels free of line emission. 
These line-free channels were integrated to generate a continuum map 
($0\arcsec.70 \times 0\arcsec.56$, P.A. = $-68\arcdeg.1$, 1$\sigma$ = 13.6 $\mu$Jy beam$^{-1}$), 
which we subtracted in the $uv$ plane using the task \verb|uvcontsub| (with a first-order polynomial function), before making the line cube. 
In this paper, we show only statistical errors unless mentioned otherwise. 
The absolute flux uncertainty is $\sim 10\%$ (ALMA Cycle 7 Proposer's Guide). 
We also used the \verb|MIRIAD| software \citep{1995ASPC...77..433S} for some of the analyses in this paper. 

\begin{deluxetable*}{ccccccc}
\tabletypesize{\small}
\tablecaption{Journal of ALMA Observations\label{tbl1}}
\tablewidth{0pt}
\tablehead{
\colhead{Date} & \colhead{Antenna} & \colhead{Baseline} & \colhead{Integration} & \multicolumn{3}{c}{Calibrator} \\ \cline{5-7}
\colhead{(UT)} & \colhead{Number} & \colhead{(m)} & \colhead{(min)} & Bandpass & Flux & Phase 
}
\decimalcolnumbers 
\startdata
2019 Oct 16 & 41 & 15.1--740.4 & 38.5 & J1058$+$0133 & J1058$+$0133 & J1232$-$0224 \\ 
2019 Oct 22 & 43 & 15.1--783.5 & 38.5 & J1058$+$0133 & J1058$+$0133 & J1232$-$0224 \\ 
2019 Oct 22 & 43 & 15.1--783.5 & 38.5 & J1427$-$4206 & J1427$-$4206 & J1232$-$0224 \\ 
\enddata
\tablecomments{(1) Our observations were taken in three sessions, on the UT dates listed.
(2) Number of antennas used in the observation. 
(3) Minimum and maximum baseline lengths in meters. 
(4) Net on-source integration time in minutes. 
(5)-(7) Calibrators used in the observation.}
\end{deluxetable*}

\section{Results}\label{sec3}
Both the FIR continuum and the [\ion{C}{2}] line emission are clearly detected. 
We detail their properties in the following, 
and summarize the results in Table \ref{tbl2}. 
The derived properties are compared with those of other $z > 6$ quasars in $\S~4$. 

\begin{figure*}
\begin{center}
\includegraphics[width=\linewidth]{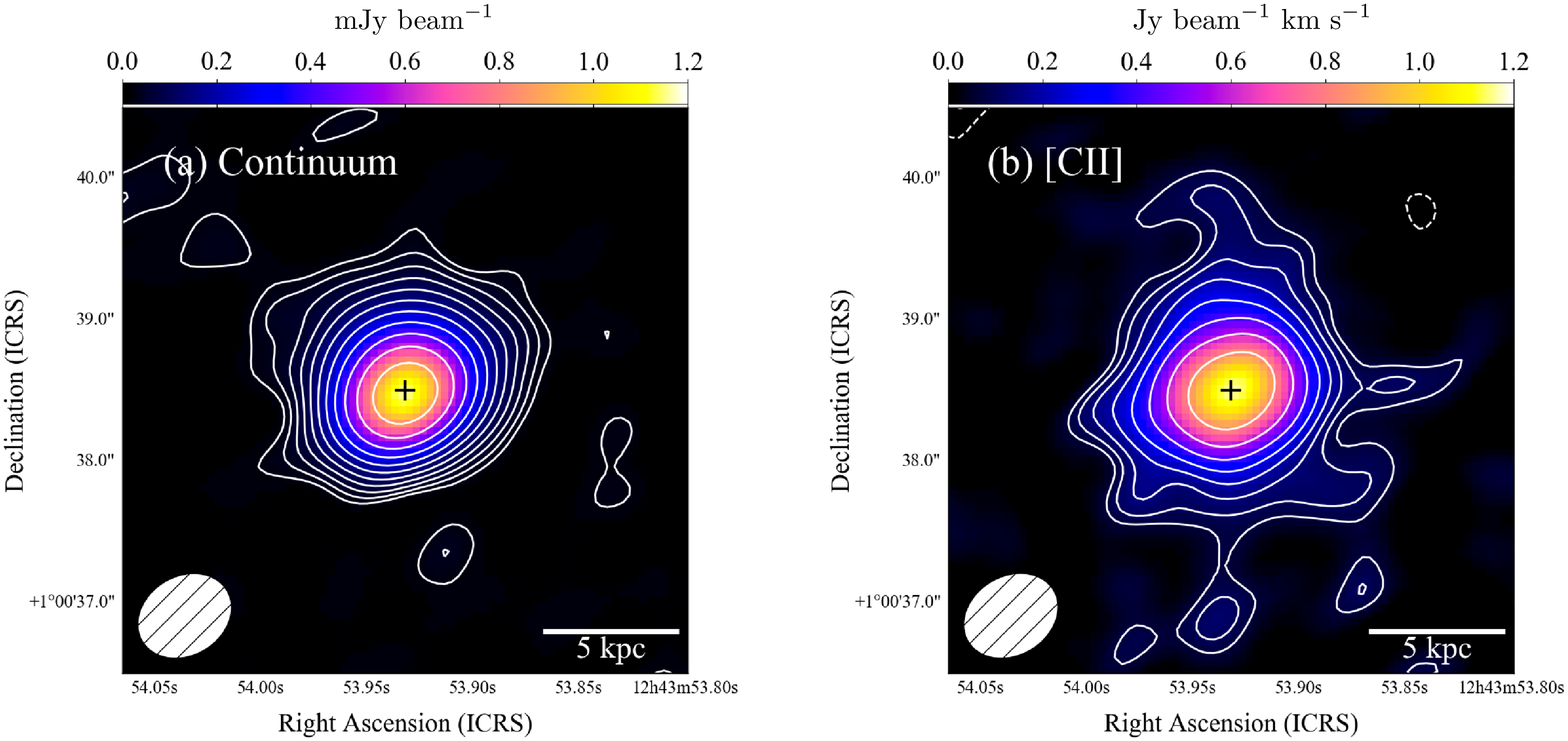}
\caption{
(a) Spatial distribution of the rest-FIR continuum emission of J1243$+$0100. The image is 4$\arcsec$ on a side.  
Contours start at $\pm 2\sigma$ ($1\sigma = 13.6$ $\mu$Jy beam$^{-1}$) and increase by factors $\sqrt{2}$. 
(b) Spatial distribution of the velocity-integrated intensity of [\ion{C}{2}] 158 $\micron$. 
Contours start at $\pm 2\sigma$ ($1\sigma = 0.037$ Jy beam$^{-1}$ km s$^{-1}$) and increase by factors of $\sqrt{2}$. 
In each panel, the synthesized beam is shown in the bottom-left corner, 
and no significant negative emission is found. 
The central black plus denotes the FIR continuum peak position. 
}
\label{fig1}
\end{center}
\end{figure*}

\begin{deluxetable*}{c|ccc}
\tabletypesize{\small}
\tablecaption{Properties of the Host Galaxy of J1243$+$0100 \label{tbl2}}
\tablewidth{0pt}
\tablehead{
 & \multicolumn{3}{c}{Area-integrated [\ion{C}{2}] 158 $\mu$m Line Emission} \\ \cline{2-4}
 & \colhead{\multirow{2}{*}{Single Gaussian}} & \multicolumn{2}{c}{Double Gaussian (fit to the spectrum)} \\ \cline{3-4}
 & & Core & Wing
}
\startdata
$z_{\rm [CII]}$ & 7.0749 $\pm$ 0.0001 & 7.0749 $\pm$ 0.0001 & Fixed to the Core \\ 
FWHM$_{\rm [CII]}$ (km s$^{-1}$) & 280 $\pm$ 12 & 235 $\pm$ 17 & 997 $\pm$ 227 \\ 
$S_{\rm [CII]}\Delta V$ (Jy km s$^{-1}$) & 2.11 $\pm$ 0.08 & 1.59 $\pm$ 0.17 & 1.03 $\pm$ 0.21 \\ 
$L_{\rm [CII]}$ (10$^9$ $L_\odot$) & 2.52 $\pm$ 0.10 & 1.90 $\pm$ 0.20 & 1.23 $\pm$ 0.25 \\ 
SFR$_{\rm [CII]}$ ($M_\odot$ yr$^{-1}$)$^\dag$ & 220 $\pm$ 8 & 165 $\pm$ 17 & 107 $\pm$ 22$^\ddag$ \\ \cline{1-4}
 & \multicolumn{3}{c}{Continuum Emission ($T_{\rm dust}$ = 47 K, $\beta$ = 1.6, $\kappa_\lambda = 0.77 ({\rm 850 \micron}/\lambda)^\beta$ cm$^2$ g$^{-1}$)} \\ 
 & Total (imfit) & Extended ($uv$-plot)$^\#$ & Point Source ($uv$-plot)$^\#$ \\ \cline{1-4}
$f_{\rm 1.3mm}$ (mJy) & 1.52 $\pm$ 0.03 & 0.63 $\pm$ 0.04 & 0.85 $\pm$ 0.04 \\ 
$L_{\rm FIR}$ (10$^{12}$ $L_\odot$) & 3.5 $\pm$ 0.1 & 1.5 $\pm$ 0.1 & 2.0 $\pm$ 0.1  \\ 
$L_{\rm TIR}$ (10$^{12}$ $L_\odot$) & 5.0 $\pm$ 0.1 & 2.1 $\pm$ 0.1 & 2.8 $\pm$ 0.1 \\ 
$M_{\rm dust}$ (10$^8$ $M_\odot$) & 2.5 $\pm$ 0.1 & 1.0 $\pm$ 0.1 & 1.4 $\pm$ 0.1 \\ 
SFR$_{\rm TIR}$ ($M_\odot$ yr$^{-1}$) & 742 $\pm$ 16 & 307 $\pm$ 20$^\clubsuit$ & 414 $\pm$ 20$^\spadesuit$ \\ 
\enddata
\tablecomments{$^\dag$ Based on the calibration for local \ion{H}{2}/starburst galaxies (De Looze et al. 2014). 
$^\ddag$This SFR$_{\rm [CII]}$ is valid only if the broad wing is due to emission from companion galaxies. 
$^\#$We decomposed the continuum emission to a spatially extended component 
and a point source based on the $uv$-plane analysis. 
$^\spadesuit$This SFR$_{\rm TIR}$ is appropriate if this emission is due to star formation, but the dust may be heated by the quasar itself. 
$^\clubsuit$The extended component must be powered by star formation, so the inferred rate for this component represents our conservative estimate for ${\rm SFR_{TIR}}$.}
\end{deluxetable*}

\subsection{Continuum properties}\label{sec3.1}
Figure \ref{fig1}a shows the spatial distribution of the rest-FIR continuum emission. 
It is very bright with a peak flux density of 1.15 mJy beam$^{-1}$, detected at $\sim 85\sigma$ ($1\sigma$ = 13.6 $\mu$Jy beam$^{-1}$).  This high signal to noise ratio (S/N) is well above the threshold of 10 required to make a robust size measurement \citep{2018ApJ...854...97D,2018ApJ...866..159V}. 
We measured the properties of the source with CASA task \verb|imfit|, which fits a 2D Gaussian to the observed map in the image plane. 
The emission peaks at ($\alpha_{\rm ICRS}$, $\delta_{\rm ICRS}$) = (12$^{\rm h}$43$^{\rm m}$53$^{\rm s}$.932, $+$01\arcdeg00\arcmin38\arcsec.49), 
consistent with the optical quasar position (\S~2). 
Thus, we adopt this FIR continuum peak position as the quasar position. 
The observed size of the emitting region is 
$(0\arcsec.79 \pm 0\arcsec.01) \times (0\arcsec.67 \pm 0\arcsec.01)$. 
After deconvolving by the beam, we obtain an intrinsic size of  
$(0\arcsec.38 \pm 0\arcsec.03) \times (0\arcsec.36 \pm 0\arcsec.04)$ or 
$(2.0 \pm 0.2)$ kpc $\times$ $(1.8 \pm 0.2)$ kpc at $z = 7.07$ (Table \ref{tbl_add1}). 
This lies with the range of FIR sizes ($\sim 1-6$ kpc) of previously observed $z \gtrsim 6$ quasars 
 \citep{2019PASJ...71..111I,2020ApJ...904..130V}. 

The area-integrated (= total) flux density of this component is $1.52 \pm 0.03$ mJy. 
With this we first determine the area-integrated FIR luminosity ($L_{\rm FIR}$; 42.5--122.5 $\micron$) 
and the total IR luminosity ($L_{\rm TIR}$; 8--1000 $\micron$) 
assuming an optically thin modified black body spectrum. 
Following previous studies of $z > 6$ quasars, 
we assume an intrinsic dust temperature ($T_{\rm dust}$) of 47 K 
and a dust spectral emissivity index of $\beta = 1.6$, 
values which are characteristic of high redshift optically luminous quasars \citep{2006ApJ...642..694B,2013ApJ...772..103L}. 
We also correct for the contrast ($\times 1/f_{\rm CMB}$) and the additional heating effects 
of the cosmic microwave background (CMB) radiation \citep{2013ApJ...766...13D}, 
\begin{equation}
f_{\rm CMB} = 1 - B_{\rm \nu_{rest}}(T_{\rm CMB,z})/B_{\rm \nu_{rest}}(T_{\rm dust,z})
\end{equation}
\begin{equation}
T_{\rm dust,z} = (T^{4+\beta}_{\rm dust} + T^{4+\beta}_{\rm CMB,z=0} [(1+z)^{4+\beta} - 1])^{\frac{1}{4+\beta}},
\end{equation}
where $T_{\rm CMB,z} = 2.73(1+z) = 22.0$ K at $z = 7.07$ 
and $\nu_{\rm rest}$ is the [\ion{C}{2}] rest frequency (1900.54 GHz). 
With these corrections, we find $L_{\rm FIR} = (3.5 \pm 0.1) \times 10^{12}~L_\odot$ 
and $L_{\rm TIR} = (5.0 \pm 0.1) \times 10^{12}~L_\odot$. 
We also find a dust mass of $M_{\rm dust} = (2.5 \pm 0.1) \times 10^8~M_\odot$ 
by adopting a rest-frame mass absorption coefficient of 
$\kappa_\lambda = 0.77 ({\rm 850 \micron}/\lambda)^\beta$ cm$^2$ g$^{-1}$ \citep{2000MNRAS.315..115D}. 
Note that these results are sensitive to the assumed values of $T_{\rm dust}$ and $\beta$, which are known to be different in different sources
\citep{2013ApJ...772..103L,2018ApJ...866..159V,2019MNRAS.489.1397L}. 
For example, varying $T_{\rm dust}$ over the range $35-60$ K \citep[e.g.,][]{2013ApJ...772..103L,2010MNRAS.409...75H,2020MNRAS.494.3828D} with $\beta = 1.6$ 
results in $L_{\rm FIR} = (1.7-6.1) \times 10^{12}~L_\odot$. 
Continuum measurements over a range of wavelengths are needed to constrain $T_{\rm dust}$. 
In what follows, we do not include the systematic uncertainty due to the assumed dust temperature. 

If we further assume that this IR continuum emission is entirely due to 
star formation \citep[e.g.,][]{2006ApJ...649...79S,2014ApJ...785..154L}, we can derive its SFR. 
We use the conversion ${\rm SFR_{TIR}} = 1.49 \times 10^{-10} L_{\rm TIR}/L_\odot$ \citep{2011ApJ...737...67M}, 
and obtain ${\rm SFR_{TIR}} = (742 \pm 16)~M_\odot$ yr$^{-1}$. 
This conversion is based on the Kroupa initial mass function \citep[IMF,][]{2001MNRAS.322..231K}, 
but is in accord with other studies \citep[e.g.,][]{1998ARA&A..36..189K} after accounting for the differing IMFs assumed therein. 
If the dust is partially heated by the quasar itself, the true SFR would be lower, so this value is an upper limit \citep[e.g.,][]{2016MNRAS.459..257S}. 

\begin{figure}
\begin{center}
\includegraphics[width=\linewidth]{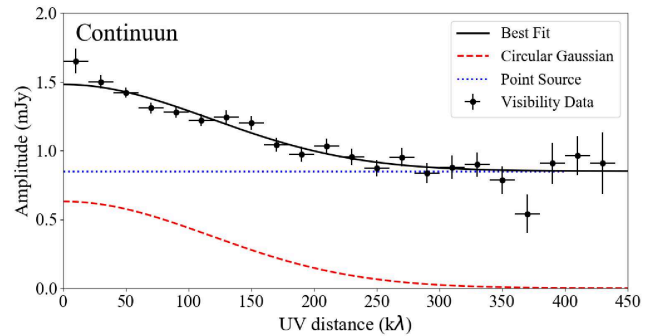}
\caption{
Real part of the continuum visibilities as a function of the $uv$ distance. 
We modeled this distribution with a combination of a circular Gaussian 
(${\rm FWHM} = 0\arcsec.66 \pm 0.\arcsec06$) and a point source function. 
The visibilities are binned in steps of 20 k$\lambda$. 
}
\label{fig_add1}
\end{center}
\end{figure}

To further explore this possibility, we fit the observed visibilities 
to a model of an unresolved point source and an extended circular Gaussian (Figure \ref{fig_add1}). 
Doing this fit in the $uv$ plane avoids uncertainties in the deconvolution process 
\citep[e.g.,][]{2015ApJ...810..133I,2017ApJ...835..286I,2019ApJ...887..107F,2020ApJ...900....1F}. 
Annular averages of the $uv$ dataset were created in 20 k$\lambda$ bins by using the MIRIAD task \verb|uvamp|, 
after shifting the phase center to the exact FIR continuum peak position. 
The results are summarized in Table \ref{tbl_add1}. 
The Gaussian component has a FWHM of $0\arcsec.66 \pm 0\arcsec.06$ ($3.4 \pm 0.3$ kpc), 
likely tracing the star-forming region of this galaxy. 
The flux density of this component is $0.63 \pm 0.04$ mJy, 
which is equivalent to ${\rm SFR_{TIR}} = 307 \pm 20~M_\odot$ yr$^{-1}$ ($T_{\rm dust} = 47$ K, $\beta = 1.6$). 
The point source has a higher flux density ($0.85 \pm 0.04$ mJy) 
than the Gaussian component, resulting in ${\rm SFR_{TIR}}$ of $414 \pm 20~M_\odot$ yr$^{-1}$. 
However, given that it is unresolved, it may be heated by the quasar itself. 
Note that, \citet{2018ApJ...866..159V} did not find a significant correlation 
between $L_{\rm Bol}$ and $L_{\rm FIR}$ \citep[see also][]{2020ApJ...904..130V} in optically luminous quasars. 
On the other hand, \citet{2021arXiv210101199I} did find that correlation after expanding the range of $L_{\rm Bol}$, 
which may be suggestive of a certain level of quasar contribution to $L_{\rm FIR}$. 
Hence a {\it conservative} estimate of ${\rm SFR_{TIR}}$ ($\equiv {\rm SFR^{cons}_{TIR}}$) is that obtained from the Gaussian component alone.

\begin{deluxetable}{c|c}
\tabletypesize{\small}
\tablecaption{Continuum Spatial Extent (FWHM) \label{tbl_add1}}
\tablewidth{0pt}
\tablehead{
Domain & Size
 }
\startdata 
Image-plane & $(0\arcsec.38 \pm 0\arcsec.03) \times (0\arcsec.36 \pm 0\arcsec.04)$ \\ 
(deconvolved) & or $(2.0 \pm 0.2) \times (1.8 \pm 0.2)$ kpc$^2$ \\ \hline
$uv$-plane & $0\arcsec.66 \pm 0\arcsec.06$ \\ 
 & or $3.4 \pm 0.3$ kpc \\ 
\enddata
\tablecomments{The $uv$-plane fit also includes a point source component (see Figure \ref{fig_add1}).}
\end{deluxetable}

\subsection{[\ion{C}{2}] line properties}\label{sec3.2}
\subsubsection{Global gas distribution}\label{sec3.2.1}
\begin{figure}
\begin{center}
\includegraphics[width=\linewidth]{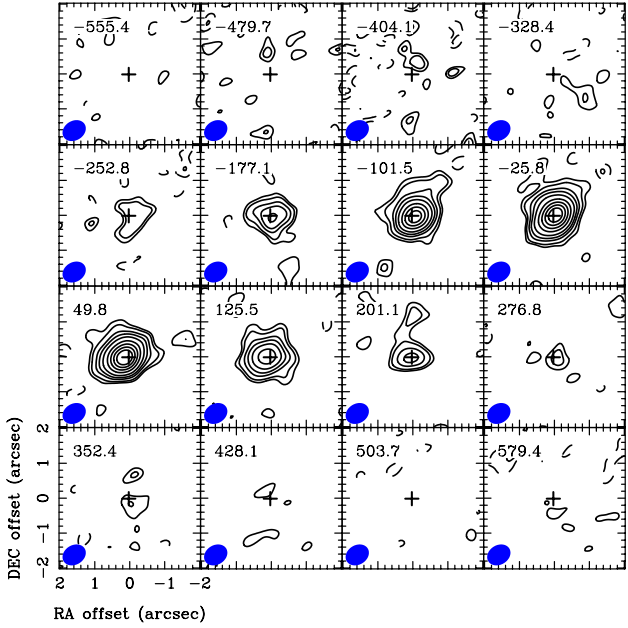}
\caption{
Velocity channel maps of the [\ion{C}{2}] line emission of J1243$+$0100. 
Each channel is labeled with its central velocity in km s$^{-1}$. 
The central plus sign in each panel denotes the FIR continuum peak position (\S~3.1). 
Contours are drawn at $-$3, $-$2, 2, 3, 5, 7, 10, 15, 20, 25, and 30$\sigma$, where 1$\sigma$ = 0.10 mJy beam$^{-1}$. 
The synthesized beam is shown in the bottom left corner. 
}
\label{fig2}
\end{center}
\end{figure}

Figure \ref{fig1}b shows the velocity-integrated [\ion{C}{2}] moment 0 map of J1243$+$0100. We integrate over $\pm 900$ km s$^{-1}$ relative to the systemic redshift 
given the broad wing component in the {\it area-integrated} spectrum (\S~\ref{sec3.2.2}).  
Hence this choice of the velocity range is the result of an iterative process. 
Note that, however, the high-velocity component is weak, 
below $3\sigma$ in the (native resolution) velocity channel maps (Figure \ref{fig2}). 
The [\ion{C}{2}] spatial distribution is clearly extended and complex. 
We applied the CASA task \verb|imfit| to this moment-0 map, 
which gave a beam-deconvolved size of $(0\arcsec.69 \pm 0\arcsec.09) \times (0\arcsec.67 \pm 0\arcsec.10)$ or 
$(3.6 \pm 0.5)$ kpc $\times$ $(3.5 \pm 0.5)$ kpc. 
The [\ion{C}{2}] flux peak position and the spatial extent are identical within the uncertainties to 
those of the spatially extended component of the FIR continuum emission  (Table \ref{tbl_add1}). 
We will perform further detailed size measurements in \S~\ref{sec3.2.4}. 

\subsubsection{Line spectrum}\label{sec3.2.2}
\begin{figure}
\begin{center}
\includegraphics[width=\linewidth]{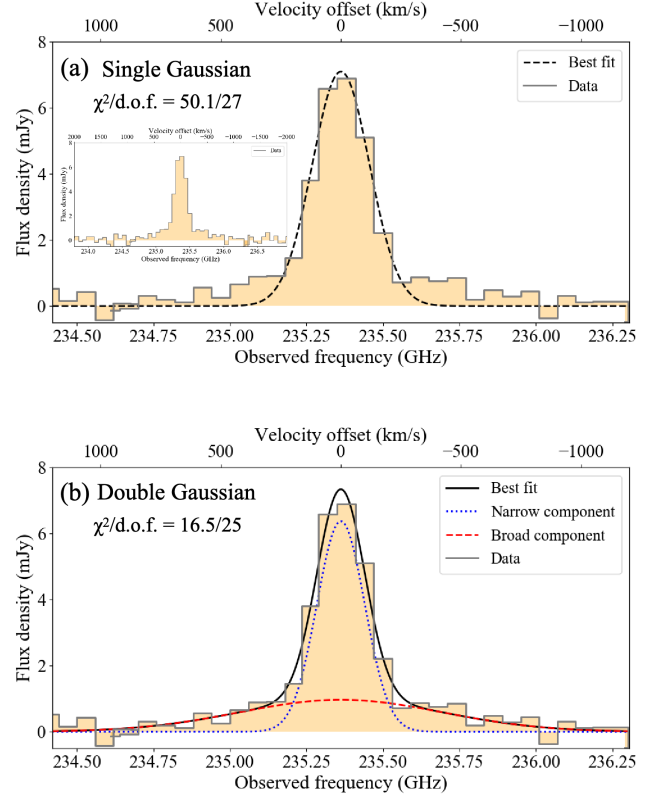}
\caption{
Area-integrated [\ion{C}{2}] line spectrum of J1243$+$0100 measured 
over a region of [\ion{C}{2}] integrated intensity $>2\sigma$ $\cap$ $r < 2\arcsec$. 
The sensitivity is 0.31 mJy per channel. 
(a) Single Gaussian fit and (b) Double Gaussian fit to the observed spectrum. 
The resultant parameters are listed in Table \ref{tbl2}. 
The fit in (b) shows much lower $\chi^2$ than that in (a). 
The inset of (a) shows a wider-frequency view, which illustrates the reliability of our continuum subtraction. 
}
\label{fig3}
\end{center}
\end{figure}

We extract the [\ion{C}{2}] line spectrum (Figure \ref{fig3}) by integrating the signal 
within the 2$\sigma$ contours around the center of the moment-0 map. 
We hereafter refer to this as the {\it area-integrated} spectrum. 
This method gives lower noise than measurement within a circular aperture, 
particularly when the source is resolved and complex \citep[see detailed discussion in][]{2020A&A...643A...2B}. 
The corresponding $1\sigma$ noise level is 0.31 mJy, which was measured 
from spectral windows above and below the window containing the line emission. 

The [\ion{C}{2}] spectrum peaks at a flux density of $\sim 7$ mJy. 
This is much brighter than the other HSC quasars observed by ALMA 
\citep[their peaks are mostly $< 2$ mJy,][]{2018PASJ...70...36I,2019PASJ...71..111I}. 
The line spectrum shows a broad component that extends over $\pm 900$ km s$^{-1}$. 
As we described in \S~\ref{sec3.2.1} above, we integrated over this full velocity range 
to create Figure \ref{fig1}b after finding this broad component. 
Given this broad component, the spectrum is poorly fit with a single Gaussian (Figure \ref{fig3}a, Table \ref{tbl2}) 
with a returned $\chi^2$/d.o.f = 50.1/27 (estimated over $\pm 1000$ km s$^{-1}$). 
A double Gaussian fit (Figure \ref{fig3}b, Table \ref{tbl2}), both centered on the same frequency, 
gives much better results: $\chi^2$/d.o.f. = 16.5/25. 
We will discuss the nature of this broad component in \S~\ref{sec4.2}. 

The narrow core component is likely due to the host galaxy of J1243$+$0100. 
The line center redshift is $z_{\rm [CII]} = 7.0749 \pm 0.0001$.  
We hereafter regard this as the systemic redshift of this quasar; note that it is 
consistent with the \ion{Mg}{2}-based redshift of $7.07 \pm 0.01$ \citep{2019ApJ...872L...2M}. 
With this, we also confirm the large blueshift ($\sim -2400$ km s$^{-1}$) of the \ion{C}{4} emission line \citep{2019ApJ...872L...2M}, 
indicating the presence of a fast quasar wind. 
The line width (${\rm FWHM^{core}} = 235 \pm 17$ km s$^{-1}$) is on the small end of values found for $z > 6$ quasars of all luminosities 
\citep[e.g.,][]{2018ApJ...854...97D,2019PASJ...71..111I}.   As we described in  (\S~\ref{sec3.2.4}), this is likely due to rotation of a disk at a small inclination angle. 
The [\ion{C}{2}] line luminosity of this component, following \citet{2005ARA&A..43..677S}, 
is $L^{\rm core}_{\rm [CII]} = (1.90 \pm 0.20) \times 10^9~L_\odot$. 

By further assuming that the [\ion{C}{2}] line is excited primarily by star formation, 
we can estimate the SFR using the \citet{2014A&A...568A..62D} calibration based on local \ion{H}{2}/starburst galaxies: 
$\log ({\rm SFR_{[CII]}}/M_\odot~{\rm yr^{-1}}) = -7.06 + 1.0 \times \log (L_{\rm [CII]}/L_\odot)$, with a factor of two calibration uncertainty.   
We obtain ${\rm SFR^{core}_{[CII]}} = 165 \pm 17~M_\odot$ yr$^{-1}$, 
which is consistent within the calibration uncertainty with the result we found from the spatially extended continuum component ${\rm SFR^{cons}_{TIR}}$. 
This relation is applicable to high redshift ($z \sim 4-8$) star-forming galaxies 
as recently demonstrated by \citet{2020A&A...643A...3S,2020A&A...643A...1L}. 
If some of the [\ion{C}{2}] excitation is in fact due to the quasar, our derived SFR is again an upper limit. 

With this core line luminosity and the area-integrated $L_{\rm FIR}$ 
(i.e., imfit-based value) derived in \S~\ref{sec3.1}, we obtain $\log (L_{\rm [CII]}/L_{\rm FIR}) = -3.27$. 
This value is comparable to those of optically luminous $z \gtrsim 6$ quasars 
\citep[e.g.,][]{2013ApJ...773...44W,2016ApJ...816...37V}, 
and is $\sim 6\times$ smaller than the canonical Milky Way value \citep[$\sim 3 \times 10^{-3}$,][]{2013ARA&A..51..105C}. 
Thus this quasar follows the so-called {\it [\ion{C}{2}]-deficit} trend found in ULIRG-class objects 
\citep[e.g.,][]{1997ApJ...491L..27M,2010ApJ...724..957S,2013ApJ...774...68D}, 
and in $z \gtrsim 6$ quasars \citep[e.g.,][]{2019PASJ...71..111I,2020ApJ...904..130V}. 
This deficit is likely correlated with a high FIR surface density 
\citep[see discussion in, e.g.,][]{2018ApJ...854...97D,2019PASJ...71..111I,2020ApJ...904..130V}. 
However, if we use the extended component of the FIR continuum emission 
(i.e., excluding the point source, Table \ref{tbl_add1}) alone, 
we obtain $\log (L_{\rm [CII]}/L_{\rm FIR}) = -2.89$, which is now close to the Milky Way value. 
This suggests that the quasar itself contributes to the unresolved component of $L_{\rm FIR}$, 
which causes the $L_{\rm [CII]}/L_{\rm FIR}$ ratio to be lower in quasars than in starbursts. 

We previously emphasized the uncertainty in $L_{\rm FIR}$ due to our lack of knowledge of $T_{\rm dust}$, 
which eventually affects our interpretation of $L_{\rm [CII]}/L_{\rm FIR}$. 
To circumvent this issue, we also measure the [\ion{C}{2}] equivalent widths. 
If we use the imfit-based total continuum flux density, we obtain 
${\rm EW_{[CII]}} = 0.55 \pm 0.06$ $\micron$, a value only a factor of $\sim 2$ smaller 
than the median ${\rm EW_{[CII]}}$ of local starburst galaxies \citep[$\sim 1.0~\micron$,][]{2014ApJ...790...15S}. 
If we instead use the decomposed extended flux density, we find 
${\rm EW_{[CII]}} = 1.33 \pm 0.17$ $\micron$, fully consistent with local starbursts. 
This implies that the ISM physical conditions are not 
very different between J1243$+$0100 and the local starbursts. 
If that is the case, our inferred value of $L_{\rm FIR}$, and the canonical value of $T_{\rm dust} = 47$ K may be overestimates. 
However, without multi-band FIR data, we will continue to use this canonical dust temperature in what follows. 

The broad wing component has ${\rm FWHM^{wing}} = 997 \pm 227$ km s$^{-1}$, 
with a brightness of $65 \pm 15\%$ of the core component. 
The positive and negative velocity wings have identical shapes within the errors, 
as was seen in the [\ion{C}{2}] outflow profile found in J1148$+$5251 \citep{2012MNRAS.425L..66M}. 
We will argue in  (\S~\ref{sec4.2}) that these wings are indeed due to cold outflowing gas.

\subsubsection{Global gas dynamics}\label{sec3.2.3}
\begin{figure*}
\begin{center}
\includegraphics[width=\linewidth]{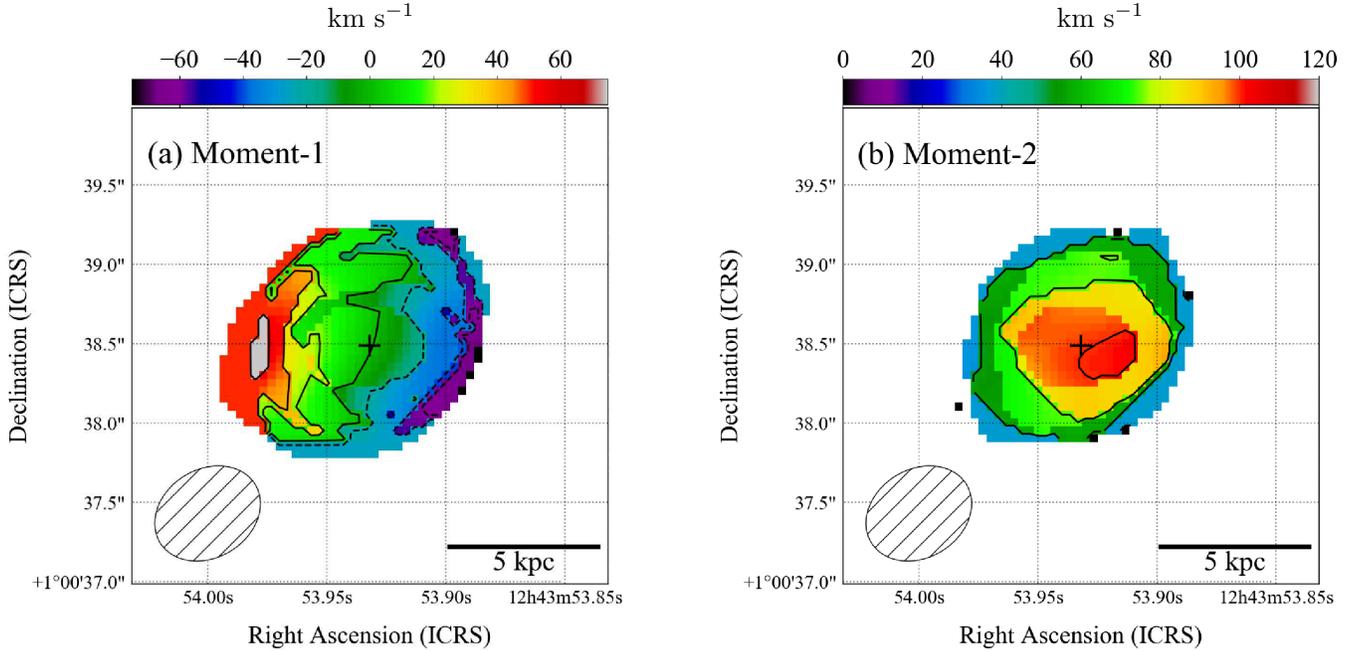}
\caption{
(a) Intensity-weighted [\ion{C}{2}] mean velocity map of the central 3\arcsec region of J1243$+$0100. 
The contours indicate the velocity relative to the systemic redshift, 
ranging from $-$60 to $+$60 km s$^{-1}$ in steps of 20 km s$^{-1}$. 
(b) Intensity-weighted velocity dispersion map of the same region. 
Here the contours indicate 25, 50, 75, and 100 km s$^{-1}$. 
These maps were made with a conservative 5$\sigma$ clipping. 
In each panel, the bottom-left ellipse corresponds to our synthesized beam. 
The central black plus sign denotes the quasar position. 
}
\label{fig4}
\end{center}
\end{figure*}

\begin{figure}
\begin{center}
\includegraphics[width=\linewidth]{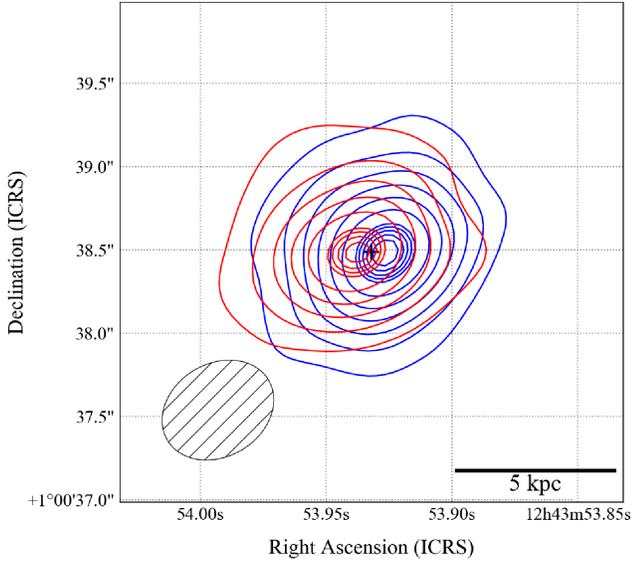}
\caption{
The [\ion{C}{2}] intensity distributions integrated over the blue channels 
($-253$ to $-26$ km s$^{-1}$; blue contours) 
and the red channels ($+50$ to $+277$ km s$^{-1}$, red contours) separately. 
The blue contours indicate 5, 10, ..., 35, 36, ..., and 39$\sigma$, 
whereas the red ones indicate 5, 10, ..., 30, 31, 32, and 33$\sigma$, 
where $1\sigma = 0.015$ Jy beam$^{-1}$ km s$^{-1}$. 
The central plus sign denotes the quasar position. 
}
\label{fig5}
\end{center}
\end{figure}

We made an intensity-weighted mean velocity map and a velocity dispersion map (Figure \ref{fig4}), 
using the CASA task \verb|immoments| with a conservative $5\sigma$ clipping to avoid noisy pixels. 
Thus these maps do not reflect the contribution from the broad wing component. 
Although the data are convolved with the beam, 
Figure \ref{fig4}a shows a large-scale velocity gradient across the galaxy. 
This gradient is also apparent in the channel maps (Figure \ref{fig2}) 
as the peak positions of the [\ion{C}{2}] emission move 
from west to east as a function of channel velocity. 
We highlight this motion by considering the blue and the red sides of the line spectrum: 
Figure \ref{fig5} is made by separately integrating the second row and the third row channels of Figure \ref{fig2}. 
We found a clear spatial offset of $\sim 0\arcsec.2$ (1 kpc) along the east-west direction between the blue and the red peaks. 
Similar velocity gradients over this spatial scale have been observed in some optically luminous quasars 
\citep[e.g.,][]{2013ApJ...773...44W,2016ApJ...816...37V,2017ApJ...845..138S,2018ApJ...854...97D}, but some other quasars 
including J1342$+$0928 at $z = 7.54$ and J1120$+$0641 at $z = 7.09$ are dispersion-dominated systems  
\citep{2017ApJ...837..146V,2019ApJ...880....2W,2019ApJ...881L..23B}. 
The velocity dispersion in Figure \ref{fig4}b peaks at $\sim 100$ km s$^{-1}$ 
at a position slightly offset from the quasar nucleus. 
Note that, beam-smearing and the strong rotation gradient near the nucleus artificially boosts the apparent dispersion.  
Indeed, our dynamical modeling (\S~\ref{sec4.4}) suggests that this galaxy is {\it rotation-dominated}. 

\subsubsection{Decomposed [\ion{C}{2}] spatial extent}\label{sec3.2.4}
\begin{figure}
\begin{center}
\includegraphics[width=\linewidth]{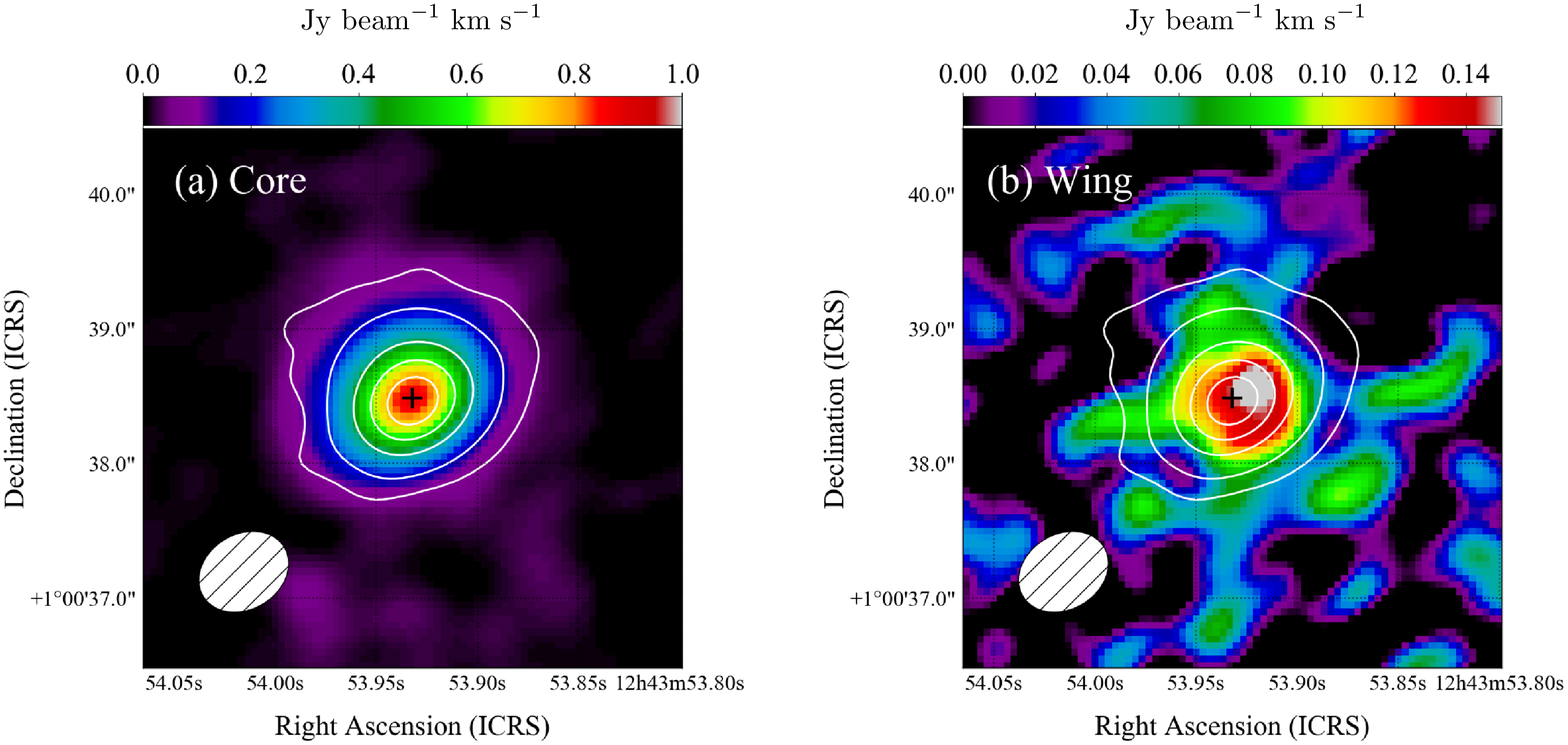}
\caption{
Velocity-integrated [\ion{C}{2}] intensity map of 
(a) the core component and (b) the wing component. 
These are made by integrating the $-102$ to $+126$ km s$^{-1}$ channels for (a), 
and $-934$ to $-253$ km s$^{-1}$ plus $+277$ to $+882$ km s$^{-1}$ channels for (b), 
respectively (see also Figures \ref{fig2} and \ref{fig3}). 
The central plus sign indicates the quasar location. 
The contours indicate 3, 10, 30, 50, and 70$\sigma$ of the FIR continuum emission (see also Figure \ref{fig1}a). 
The 1$\sigma$ sensitivity is 0.015 and 0.033 Jy beam$^{-1}$ km s$^{-1}$ for (a) and (b), respectively. 
}
\label{fig6}
\end{center}
\end{figure}

Table~\ref{tbl3} lists our measurements of the spatial extent of the narrow core component 
and the broad wing component of the [\ion{C}{2}] line emission, respectively.  For this analysis we regard the {\it core} component as the line emission within 
the 235 km s$^{-1}$ FWHM range determined by our double Gaussian fit (Table \ref{tbl2}), 
and the {\it wing} as the line emission outside of that range and within $\pm$(250--900) km s$^{-1}$. 

First, we constructed moment 0 maps of these components 
by separately integrating relevant velocity channels (Figure \ref{fig6}). 
It is evident that the core component is spatially resolved.  While most of the flux lies within an arcsecond of the center, there is also a further larger 
($\sim 2\arcsec$) and fainter structure (Figure \ref{fig6}a). 
We first used {\tt imfit} to perform a 2D elliptical Gaussian fit to this image
which returned its beam-deconvolved size (FWHM) of $(3.4 \pm 0.2) \times (3.0 \pm 0.2)$ kpc$^2$ (Table \ref{tbl3}). 
This is consistent with the measured [\ion{C}{2}]-emitting region sizes of other 
$z \gtrsim 6$ quasars  \citep[e.g.,][]{2018ApJ...854...97D,2019PASJ...71..111I}. 

\begin{figure}
\begin{center}
\includegraphics[width=\linewidth]{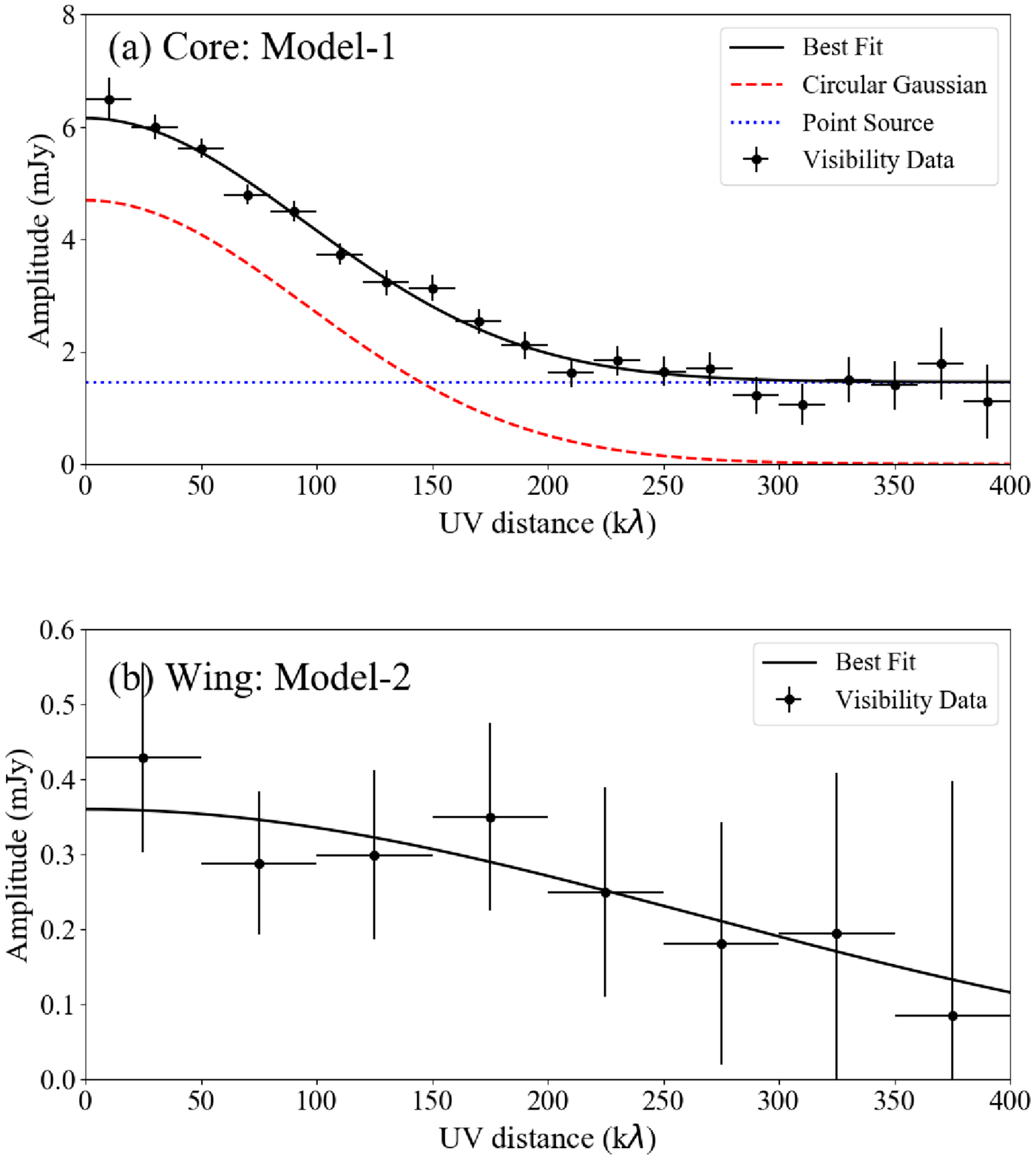}
\caption{
Real part of the [\ion{C}{2}] visibilities as a function of the $uv$ distance. 
(a) Our best-fit model to the core component, consisting of a circular Gaussian 
(${\rm FWHM} = 0\arcsec.81 \pm 0\arcsec.04$) and an unresolved point source. 
(b) Best-fit model for the wing component (${\rm FWHM} = 0\arcsec.29 \pm 0\arcsec.17$). 
The extent of the source is less than $3\sigma$ from zero. 
In (a), the observed visibilities are binned in steps of 20 k$\lambda$, whereas in (b) we binned in steps of 50 k$\lambda$ given the faintness of the wing component. 
}
\label{fig7}
\end{center}
\end{figure}

We also modeled the observed visibilities following the analysis in \S~\ref{sec3.1}.
Figure \ref{fig7}a shows the $uv$-plot of the core component 
(averaged over the range 235.270--235.455 GHz = 235 km s$^{-1}$ FWHM range around the line center). 
A decline of the visibilities from 0 to $\sim 200$ k$\lambda$  
indicates the existence of an extended (resolved) component, while the contribution from a compact (unresolved) source is apparent at $\gtrsim 250$ k$\lambda$. 
The solid line indicates our best-fit model of a point source and a single circular Gaussian distribution (Model-1), as has been used in previous works 
on $z > 6$ quasars \citep{2012MNRAS.425L..66M,2015A&A...574A..14C}. 
The resultant FWHM ($0\arcsec.81 \pm 0\arcsec.04$) is $\sim 20\%$ larger than the \verb|imfit| result (Table \ref{tbl3}), 
as we now explicitly modeled the point source, reducing the central concentration of the  Gaussian component.  

We also measured the spatial extent of the wing component. 
While the moment-0 map (Figure \ref{fig6}b) is noisy, 
it suggests that the bulk of the high velocity flux originates 
from the central $r < 0\arcsec.5$ (i.e., inside the FIR continuum-emitting region). 
Hence, J1243$+$0100 itself, rather than companion objects, 
is likely to be the source of this [\ion{C}{2}] wing (\S~\ref{sec4.2}). 
We modeled the visibilities in the wing in 50 k$\lambda$ bins 
over the range 234.970--235.166 GHz and 235.559--235.756 GHz, 
corresponding to $\pm$(250--500) km s$^{-1}$ (bright part of the wing) 
with a single circular Gaussian fit over $<300$ k$\lambda$. 
We inferred a spatial extent of ${\rm FWHM} = 0\arcsec.29 \pm 0\arcsec.17$ or $1.5 \pm 0.9$  kpc. 
This extent is only $1.7\sigma$ from zero, so Table~\ref{tbl3} also lists the $3\sigma$ limit of the extent (2.7 kpc).

\begin{deluxetable}{c|c}
\tabletypesize{\small}
\tablecaption{[\ion{C}{2}] Spatial Extent (FWHM) \label{tbl3}}
\tablewidth{0pt}
\tablehead{
\multicolumn{2}{c}{Core}
 }
\startdata 
Moment-0 & $(0\arcsec.66 \pm 0\arcsec.04) \times (0\arcsec.58 \pm 0\arcsec.03)$ \\ 
(deconvolved) & or $(3.4 \pm 0.2) \times (3.0 \pm 0.2)$ kpc$^2$ \\ \hline
Model-1 & $0\arcsec.81 \pm 0\arcsec.04$ or $4.2 \pm 0.2$ kpc \\ \hline
 \multicolumn{2}{c}{Wing} \\ \hline 
\multirow{2}{*}{Model-2} & $0\arcsec.29 \pm 0\arcsec.17$ or $1.5 \pm 0.9$ kpc \\ 
 & $3\sigma$ limit $< 0\arcsec.52$ or $< 2.7$ kpc \\ \hline
\enddata
\tablecomments{{\it Model-N} indicates a direct circular Gaussian fit result to the visibilities. 
In Model-1, we fit a single Gaussian function and a point source. 
In Model-2, we only fit a single Gaussian, due to the low S/N of the data in the wing component.} 
\end{deluxetable}

\subsubsection{Dynamical mass}\label{sec3.2.5}
With the size of the core component determined from the moment-0 analysis (\verb|imfit|, Table \ref{tbl3}) 
and the line FWHM (Table \ref{tbl2}), we can estimate the host galaxy dynamical mass ($M_{\rm dyn}$). 
We fit to the data in the image plane, as is standard in $z \gtrsim 6$ quasar studies 
\citep[e.g.,][]{2013ApJ...773...44W,2015ApJ...801..123W,2016ApJ...816...37V,2019PASJ...71..111I}. 
We assume that the line emission originates in a thin rotating disk: 
the rotation-dominated line-of-sight velocity distribution (Figure \ref{fig4} and \S~\ref{sec4.4}) favors this assumption. 
The inclination angle  of the disk ($i=20\arcdeg.5$, where $0\arcdeg$ is face-on) is determined from the axis ratio of the deconvolved size. 
The circular velocity is given by $v_{\rm circ} = 0.75{\rm FWHM}/\sin i$ (i.e., half width at 20\% line maximum). 
The disk diameter is given by $D = 1.5 \times a_{\rm maj}$, where $a_{\rm maj}$ is the deconvolved size 
of the spatial Gaussian major axis, and the factor 1.5 accounts for spatially extended low-level emission \citep{2013ApJ...773...44W}: 
we indeed see such an extended component (Figures \ref{fig1} and \ref{fig6}). 
The $M_{\rm dyn}$ within $D$ is then 
\begin{equation}
\left( \frac{M_{\rm dyn}}{M_\odot} \right) = 1.16 \times 10^5 \left( \frac{v_{\rm circ}}{{\rm km~s^{-1}}} \right)^2 \left( \frac{D}{{\rm kpc}} \right). 
\end{equation}
With the values determined above, we find $M_{\rm dyn} = (7.6 \pm 0.9) \times 10^{10}~M_\odot$. 
The quoted error does not include the uncertainties of 
the inclination angle or the geometry of the line-emitting region. 
The inferred dynamical mass is similar to that found for other 
$z \gtrsim 6$ quasar host galaxies of both high and low nuclear luminosity
\citep[e.g.,][]{2013ApJ...773...44W,2015ApJ...801..123W,2016ApJ...816...37V,2019PASJ...71..111I}. 
Note that the two other $z > 7$ quasars (J1343$+$0928 and J1120$+$0641) 
observed with ALMA are dispersion-dominated systems with 
$M_{\rm dyn} \lesssim (3-4) \times 10^{10}~M_\odot$ \citep{2017ApJ...837..146V,2017ApJ...851L...8V}, 
masses $\gtrsim 2-3\times$ smaller than we have found for the host galaxy of J1243$+$0100.

\subsection{Continuum sources in the FoV?}\label{sec3.3}
\begin{figure}
\begin{center}
\includegraphics[width=\linewidth]{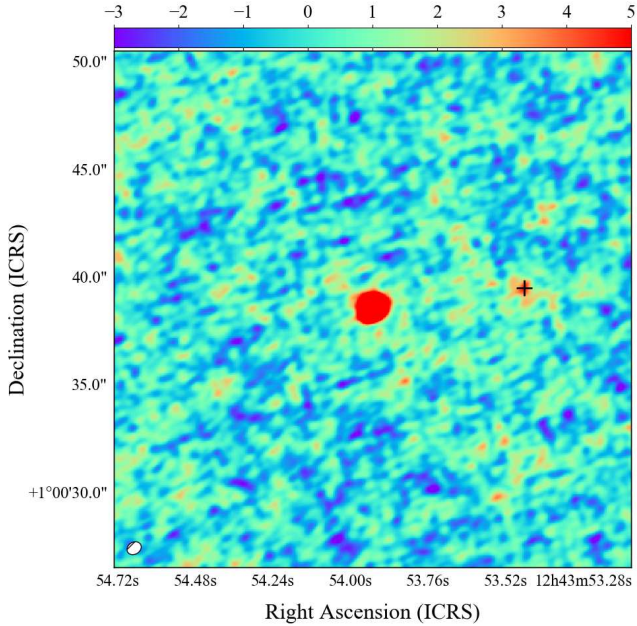}
\caption{
A large-scale S/N map of the rest-FIR continuum emission around J1243$+$0100.  No primary beam-correction has been made.  Other than the quasar host galaxy itself, no source is detected at 
$>5\sigma$ ($1\sigma = 13.6$ $\mu$Jy beam$^{-1}$).  One apparent object at $4.5\sigma$ (i.e., below our detection threshold) is indicated with a plus sign; it lies 7\arcsec.1 ($\sim 37$ kpc) from the quasar in projection.  
The color scale has units of mJy beam$^{-1}$. 
}
\label{fig8}
\end{center}
\end{figure}

ALMA observations have discovered star-forming companion/merging galaxies 
to some $z \gtrsim 5$ quasars \citep[e.g.,][]{2017ApJ...836....8T,2017Natur.545..457D,2017ApJ...850..108W,2019ApJ...882...10N}, 
in accord with the hierarchical galaxy evolution scenario in which 
quasar activity is driven by mergers \citep{1988ApJ...325...74S,2006ApJS..163....1H}. 
We thus searched for companion continuum emitters in our FoV ($\sim 0.13$ arcmin$^2$), 
using a S/N map of the region (Figure \ref{fig8}). 
We found one emission {\it candidate} at $4.5\sigma$ (i.e., below our $5\sigma$ detection threshold), 
at ($\alpha_{\rm ICRS}$, $\delta_{\rm ICRS}$) = (12$^{\rm h}$43$^{\rm m}$53$^{\rm s}$.463, $+$01\arcdeg00\arcmin39\arcsec.47), 
which is 7\arcsec.1 (or $\sim 37$ kpc in projection) from the quasar. 
No significant line emission is found at this location over our spectral coverage, 
and no optical counterpart is identified in our HSC maps ($g, r, i, z, y$ bands). 
No other source was detected in the field. 
Given the field number count of sources at 1.2 mm \citep[e.g.,][]{2016ApJS..222....1F}, 
we would predict  $\sim 1-3$ emitters in our FoV; 
given small number statistics and the cosmic variance; our non-detection is consistent with this result. 

Similarly, within the data cube created in \S~2, we did not find any 
[\ion{C}{2}] line emitter\footnote{We define a line emitter as an object with a peak line flux density of $> 5\sigma$ 
and with at least two contiguous velocity channels with $>3\sigma$ emission.} within our FoV, 
and within a velocity range of $\pm 1000$ km s$^{-1}$ relative to the quasar. 
We will present an analysis using [\ion{C}{2}] cubes with different velocity resolutions in a future paper.

\section{Discussion}\label{sec4}
\subsection{Comparison of the star-forming nature}\label{sec4.1}
Under the assumptions made in \S~3, it is intriguing that the ${\rm SFR}_{\rm TIR}$ of J1243$+$0100 
(including the decomposed conservative value ${\rm SFR}^{cons}_{\rm TIR}$) is as high as those of $z > 6$ 
optically luminous quasars \citep[e.g,][]{2018ApJ...866..159V,2020ApJ...904..130V}, 
despite the fact that its quasar nucleus is $>10\times$ fainter than those of the latter. 
As a reference for future higher redshift observations, 
we summarize the observational properties of the currently known $z > 7$ quasars in Table \ref{tbl4}. 

Seven intrinsically low-luminosity ($M_{\rm 1450} > -25$ mag) 
HSC quasars have observed with ALMA \citep{2018PASJ...70...36I,2019PASJ...71..111I}. 
Six of them have inferred ${\rm SFR_{TIR}}$ a factor of $3-10$ lower than that of J1243$+$0100. 
The seventh, J2239$+$0207 at $z = 6.25$ has a FIR luminosity, $L_{\rm FIR} = 2.2 \times 10^{12}~L_\odot$, 
comparable to that of J1243$+$0100, but it has a close companion galaxy \citep{2019PASJ...71..111I}, which may have triggered its starburst. 
Note that another HSC quasar J1205$-$0000 ($z = 6.72$) also shows a comparably high $L_{\rm FIR}$ \citep{2021arXiv210101199I}, 
but this source is dust-reddened, and is indeed as optically luminous as 
SDSS-class quasars when dust extinction is taken into account \citep{2020PASJ...72...84K}. 
The luminosity of the [\ion{C}{2}] line (spectral) core component of J1243$+$0100 is higher than all other HSC quasars 
\citep{2018PASJ...70...36I,2019PASJ...71..111I}, which all have $L_{\rm [CII]} \leq 1.0 \times 10^9~L_\odot$. 
The quasar VIMOS2911 \citep{2017ApJ...850..108W} is the only other  
optically low-luminosity quasar known at $z>6$ with FIR luminosity comparable to J1243$+$0100. 
Thus optically faint but FIR luminous quasars are a rare population at $z > 6-7$. 

\begin{deluxetable*}{c|cccccccc}
\tabletypesize{\small}
\tablecaption{Properties of $z > 7$ Quasars Known to Date \label{tbl4}}
\tablewidth{0pt}
\tablehead{
\multirow{2}{*}{Object} & \multirow{2}{*}{Redshift} & M$_{\rm 1450}$ & $M_{\rm BH}$ & \multirow{2}{*}{$L_{\rm Bol}/L_{\rm Edd}$} & ${\rm SFR_{TIR}}$ & $M_{\rm dyn}$ & \multirow{2}{*}{Ref.} \\
 & & (mag) & ($M_\odot$) & & ($M_\odot$ yr$^{-1}$) & ($M_\odot$) & 
 }
\startdata 
J0313$-$1806 & $7.6423 \pm 0.0013$ ([\ion{C}{2}]) & $-26.13 \pm 0.05$ & $(1.6 \pm 0.4) \times 10^8$ & $0.67 \pm 0.14$ & $225 \pm 25$ & -- & 1 \\ 
J1342$+$0928 & $7.5413 \pm 0.0007$ ([\ion{C}{2}]) & $-26.76 \pm 0.04$ & $9.1^{+1.4}_{-1.3} \times 10^8$ & $1.1 \pm 0.2$ & $150 \pm 30$ & $<3.2 \times 10^{10}$ & 2, 3, 4, 5 \\ 
J1007$+$2115 & $7.5149 \pm 0.0004$ ([\ion{C}{2}]) & $-26.66 \pm 0.07$ & $(1.5 \pm 0.2) \times 10^9$ & $1.1 \pm 0.2$ & 700 & -- & 6 \\ 
J1120$+$0461 & $7.0851 \pm 0.0005$ ([\ion{C}{2}]) & $-26.6 \pm 0.1$ & $(2.4 \pm 0.2) \times 10^9$ & $0.48 \pm 0.04$ & $315 \pm 25$ & $<4.3 \times 10^{10}$ & 7, 8, 9 \\ 
J1243$+$0100 & $7.0749 \pm 0.0001$ ([\ion{C}{2}]) & $-24.13 \pm 0.08$ & $(3.3 \pm 2.0) \times 10^8$ & $0.34 \pm 0.20$ & 307--742 & $(7.6 \pm 0.9) \times 10^{10}$ & 10, 11 \\ 
J0038$-$1527 & $7.021 \pm 0.005$ (SED) & $-27.10 \pm 0.08$ & $(1.33 \pm 0.25) \times 10^9$ & $1.25 \pm 0.19$ & -- & -- & 12 \\ 
J0252$-$0503 & $7.02$ (SED) & $-25.77 \pm 0.09$ & -- & -- & -- & -- & 13 \\ 
J2356$+$0017 & $7.01$ (Ly$\alpha$) & $-25.31 \pm 0.04$ & -- & -- & -- & -- & 14 \\ 
\enddata
\tablecomments{The literature values of ${\rm SFR_{TIR}}$ (or $L_{\rm TIR}$) are computed in the same manner as described in \S~\ref{sec3.1}. 
Reference. (1) \citet{2021arXiv210103179W}, (2) \citet{2018Natur.553..473B}, (3) \citet{2020ApJ...898..105O}, (4) \citet{2017ApJ...851L...8V}, (5) \citet{2019ApJ...881...63N}, 
(6) \citet{2020ApJ...897L..14Y}, (7) \citet{2011Natur.474..616M}, (8) \citet{2014ApJ...790..145D}, (9) \citet{2017ApJ...837..146V}, (10) \citet{2019ApJ...872L...2M}, (11) this work, 
(12) \citet{2018ApJ...869L...9W}, (13) \citet{2019AJ....157..236Y}, (14) \citet{2019ApJ...883..183M}.}
\end{deluxetable*}

\begin{figure}
\begin{center}
\includegraphics[width=\linewidth]{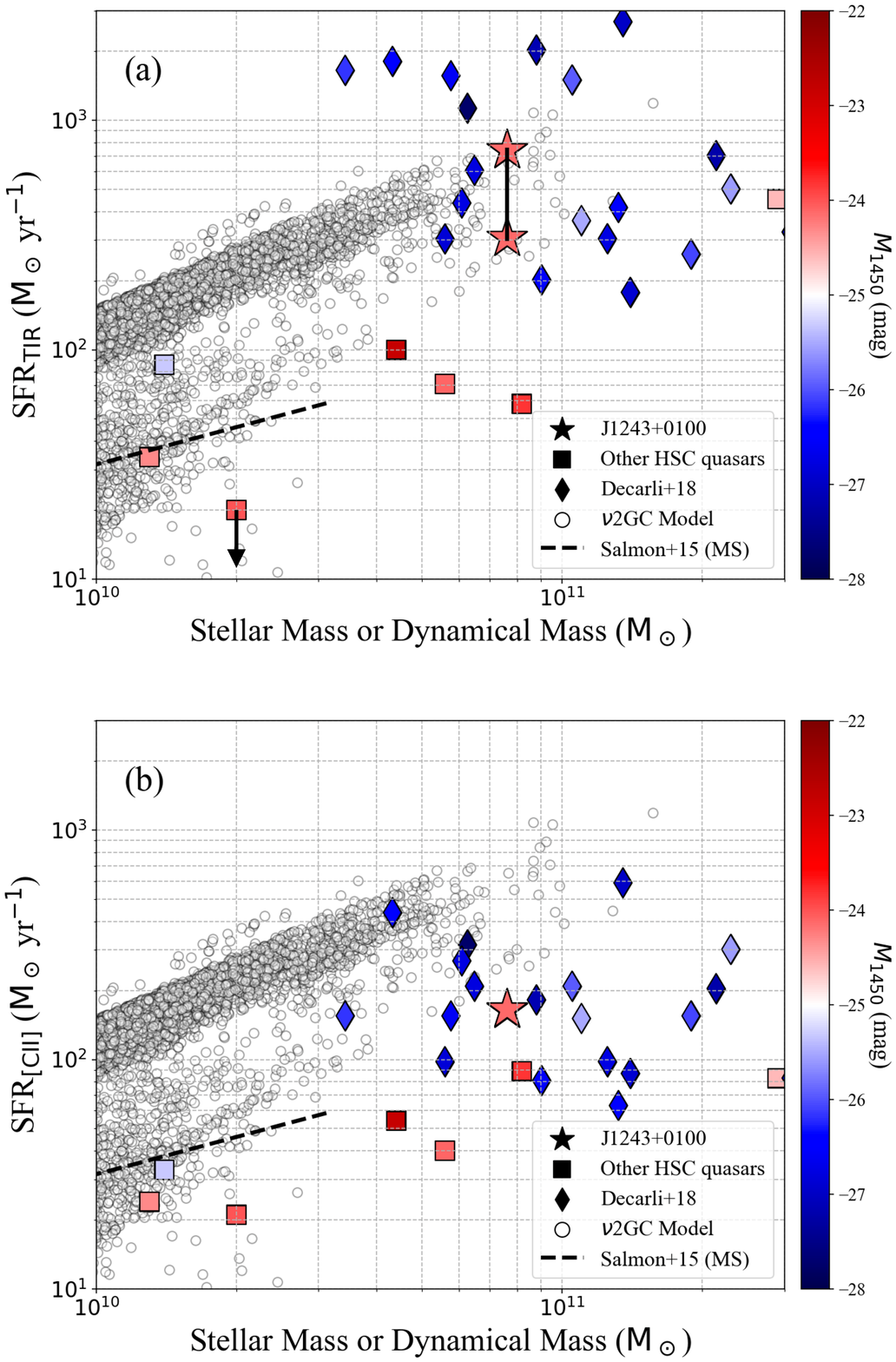}
\caption{
(a) TIR-based and (b) [\ion{C}{2}]-based SFR 
as a function of $M_{\rm dyn}$ for the HSC quasars \citep{2018PASJ...70...36I,2019PASJ...71..111I} 
and optically luminous $z \gtrsim 6$ quasars \citep{2018ApJ...854...97D}. 
For the ${\rm SFR_{TIR}}$ of J1243$+$0100, we plot both the imfit-based total value 
and the $uv$-plot-based decomposed value for the extended component (${\rm SFR^{cons}_{TIR}}$).  
These quasars are color-coded by their $M_{\rm 1450}$. 
Assuming that $M_{\rm dyn} = M_\star$, we also plot total SFR and stellar mass for galaxies 
from the $\nu^2$GC semi-analytic model \citep{2019MNRAS.482.4846S}. 
Two sequences, namely the starburst- and the star-forming main-sequences (MS), are visible in the model. 
The diagonal dashed line indicates the observed MS suggested from 
rest-frame UV-to-NIR photometric observations \citep{2015ApJ...799..183S}. 
}
\label{fig9}
\end{center}
\end{figure}

We summarize these findings in Figure \ref{fig9} in the context of 
the star-forming main sequence (MS): the majority of normal star-forming galaxies 
are found to populate a sequence on the galaxy stellar mass $M_\star$--SFR plane 
\citep[see $z \sim 1-2$ studies in, e.g.,][]{2007ApJ...670..156D,2007ApJ...660L..43N}. 
Galaxies lying  above (below) this MS are considered to be starburst (quiescent) systems. 
The evolution of the MS over cosmic time has been extensively studied up to $z \sim 5-6$ 
\citep[e.g.,][]{2014ApJS..214...15S,2014ApJ...791L..25S,2015ApJ...799..183S}. 
While the MS is not well-constrained at $z \gtrsim 5$, 
we compare the SFR of J1243$+$0100 and other HSC quasars \citep{2018PASJ...70...36I,2019PASJ...71..111I} 
and optically luminous quasars at $z \gtrsim 6$ \citep{2018ApJ...854...97D} 
with the MS at $z \sim 6$ \citep{2015ApJ...799..183S}. 
Here we assume $M_{\rm dyn} = M_\star$, as is frequently done in $z > 6$ quasar studies 
\citep[e.g.,][]{2013ApJ...773...44W,2015ApJ...801..123W,2016ApJ...816...37V,2020A&A...637A..84P}. 
The dynamical masses of course have large uncertainties, and represent an upper limit to the stellar mass. 
Note that the dynamical masses for the quasars from the literature are computed in the same manner as described in \S~\ref{sec3.2.5}. 

As observational constraints on the MS at high redshifts $z > 5-6$ are still limited, we also show the $M_\star$--SFR relation 
predicted by a semi-analytic model, the {\it new numerical galaxy catalog} \citep[$\nu^2$GC,][]{2016PASJ...68...25M}. 
In this model, the merging histories of dark matter haloes are based on large cosmological 
$N$-body simulations \citep{2015PASJ...67...61I}: 
we adopt the results from the subset of the models with the largest volume (1.12 $h^{-1}$ comoving Gpc box) 
and the dark matter mass resolution of $2.20 \times 10^8~h^{-1}~M_\odot$ (8192$^3$ particles). 
The model prescriptions for star formation, gas heating and cooling, supernova feedback, 
SMBH growth, and AGN feedback are described in, e.g., \citet{2016PASJ...68...25M} and \citet{2019MNRAS.482.4846S}. 
Here we selected $\sim 41000$ galaxies at $z \sim 6$ that host $M_{\rm BH} \geq 10^7~M_\odot$ SMBHs. 
The simulation shows both a main sequence and a starburst sequence;
the gap between these two is artificial due to the limited mass and time resolutions of the model \citep{2019MNRAS.482.4846S}. 

While in more quasars, the TIR-based SFR lies above the [\ion{C}{2}]-based value, 
the optically luminous quasars of \citet{2018ApJ...854...97D} are clearly located {\it on or above} the MS, 
whereas the HSC quasars lie {\it on or below} the MS. 
Hence these two samples populate different regimes in this diagram, 
likely representing different evolutionary stages: 
some HSC quasars seem to be ceasing their star formation already at $z > 6$, 
and they may evolve into compact and massive quiescent galaxies 
like those found at $z \sim 3-4$ \citep{2014ApJ...783L..14S,2017Natur.544...71G,2020ApJ...898..171E}. 
In contrast, J1243$+$0100 is clearly on the starburst sequence based either on ${\rm SFR}_{\rm TIR}$  or ${\rm SFR^{cons}_{TIR}}$; 
it is near the MS if we instead adopt ${\rm SFR}_{\rm [CII]}$. 
In either case, J1243$+$0100 is forming stars at a significant rate. 
This argument would be conservative as $M_{\rm dyn}$ is actually the upper limit of $M_\star$. 

One intriguing difference between J1243$+$0100 and optically luminous 
$z \gtrsim 6-7$ quasars would be their Eddington ratios ($\lambda_{\rm Edd}$). 
A large fraction of $z \gtrsim 6-7$ quasars have $\lambda_{\rm Edd}$ close to 1 
\citep[e.g.,][]{2019ApJ...880...77O}, while $\lambda_{\rm Edd} = 0.34 \pm 0.20$ \citep{2019ApJ...872L...2M}, 
a value $\sim 3$ times smaller is found for J1243$+$0100. 
\citet{2019ApJ...879..117K} found similarly high $L_{\rm FIR}$ (or ${\rm SFR}_{\rm TIR}$) 
in some $z \sim 6$ quasars with $\lambda_{\rm Edd}< 0.2$, 
and suggested that feedback was starting to quench the nuclear activity of these objects. 
We suggest that J1243$+$0100 is similar. 
We observe both accretion disk-scale feedback (in the form of BAL winds), and galaxy-scale outflows (\S~\ref{sec4.2}). 
That is, we suggest that J1243$+$0100 has recently started to clear out the (circum-)nuclear gas 
via feedback, which may soon cause the galaxy-scale star-formation to cease ({\it inside-out} feedback).

\subsection{Nature of the broad wing component}\label{sec4.2}
We here discuss the physical origin of the broad [\ion{C}{2}] wing seen in Figure \ref{fig3}. 
It could be due to (A) close companion/merger galaxies or (B) fast [\ion{C}{2}] outflows \citep{2021arXiv210101199I}. However, we do not consider cenario (A) to be likely for a number of reasons: 

\begin{itemize}
\item[i.] Companion: no companions are seen in optical, FIR continuum, or [\ion{C}{2}] within $3\arcsec$ of the quasar. 
In the optical, there are no detected companions 
in our deep Subaru/HSC $y$-band map with $5\sigma_{\rm AB}$ limiting 
magnitude of 24.65 mag \citep{2019ApJ...872L...2M}. 
\item[ii.] Symmetric line profile: even if there is a close [\ion{C}{2}] companion, 
its redshift is still (slightly) offset from J1243$+$0100, which will naturally result in an asymmetric line profile. 
But the observed broad [\ion{C}{2}] wing is almost symmetric, 
like that seen in the fast [\ion{C}{2}] outflow of J1148$+$5251 \citep{2012MNRAS.425L..66M}. 
To make this apparent broad symmetric component 
we do need multiple companions at a range of velocity offsets within this small region, 
but it is surprising if the net effect were a symmetric feature. 
Hence this scenario would be unlikely. 
\item[iii.] Size of the wing component: the estimated spatial extent of the broad wing 
is small $< 0\arcsec.52$ (Table \ref{tbl3}), and is located within the continuum-emitting region (Figure \ref{fig4}). 
Thus, the mechanism responsible for this broad wing must lie {\it inside} this galaxy. 
\item[iv.] Likely rotation-dominated host galaxy: we considered whether 
violent shocks from a galaxy merger has caused this wing. 
However, this is unlikely: Figure \ref{fig4}a shows 
that the gas motions appear to be dominated by rotation, as we quantify in \S~\ref{sec4.4}. 
A merger, on the other hand, would likely show dispersion-dominant gas dynamics 
\citep[e.g.,][]{2011ApJ...730....4B,2016ApJ...816L...6D,2018Sci...362.1034D,2020ApJ...890..149T}. 
\end{itemize}

We thus conclude that the broad [\ion{C}{2}] wing is due to a {\it fast neutral outflow}. 
This is the second individual detection of a [\ion{C}{2}] outflow at $z > 6$ 
after J1148$+$5251 at $z = 6.4$ \citep{2012MNRAS.425L..66M,2015A&A...574A..14C}, 
and J1243$+$0100 is now the highest redshift galaxy yet known with large-scale outflows. 
Both the peak flux density ratio and the velocity-integrated flux ratio of the broad-to-narrow [\ion{C}{2}] components 
of J1243$+$0100 are comparable to those of J1148$+$5251. 
We now estimate the outflow properties in a similar way to previous analyses 
\citep[e.g.,][]{2012MNRAS.425L..66M,2019A&A...630A..59B}.

\begin{deluxetable}{c|c}
\tabletypesize{\small}
\tablecaption{Outflow Properties \label{tbl5}}
\tablewidth{0pt}
\tablehead{
Quantity & Value
 }
\startdata 
$v_{\rm out}$ (km s$^{-1}$) & 499 $\pm$ 113 \\ 
$R_{\rm out}$ (kpc) & $< 1.3$ \\ 
$\tau_{\rm out}$ (10$^6$ yr) & $< 2.6 \pm 0.6$ \\ 
$M_{\rm out}$ ($10^9~M_\odot$) & $> 1.2 \pm 0.2$ \\ 
$\dot{M}_{\rm out}$ ($M_\odot$ yr$^{-1}$) & $> 447 \pm 137$ \\ 
$\dot{E}_{\rm out}$ ($10^{43}$ erg s$^{-1}$) & $> 3.5 \pm 1.6$ \\ 
$\dot{E}_{\rm out}/L_{\rm Bol}$ & $\gtrsim 0.25\%$ \\
$\dot{P}_{\rm out}/(L_{\rm Bol}/c)$ & $\gtrsim 3.0$ \\ \hline 
$\dot{M}^{\rm tot}_{\rm out}$ ($M_\odot$ yr$^{-1}$) & $\gtrsim 1410$ \\ 
\enddata
\tablecomments{The above quantities refer to the neutral atomic outflow except for $\dot{M}^{\rm tot}_{\rm out}$, 
which is estimated using the relation in \citet{2019MNRAS.483.4586F}.}
\end{deluxetable}

First, we define the neutral outflow rate by assuming a constant flow \citep{2020A&A...633A.134L} as 
\begin{equation}
\dot{M}_{\rm out} = M_{\rm out}/\tau_{\rm out} = M_{\rm out} v_{\rm out}/R_{\rm out}
\end{equation}
where $M_{\rm out}$ is the outflowing mass, $\tau_{\rm out}$ is the flow time-scale, 
$v_{\rm out}$ is the outflow velocity, and $R_{\rm out}$ is the spatial extent (radius) of the outflow. 
The wing is symmetric and centered on the [\ion{C}{2}] core component 
(indeed we forced the two components to have the same center in our fits). 
We thus adopt $v_{\rm out} =  {\rm FWHM}/2$. 
From the values in Table~\ref{tbl2}, we find $v_{\rm out} = 499 \pm 113$ km s$^{-1}$. 
$R_{\rm out}$ is defined as half of the spatial FWHM of the wing component. 
We use the $3\sigma$ upper limit of the wing extent (Table \ref{tbl3}), 
i.e., $R_{\rm out} < 1.3$ kpc and do not consider its uncertainty for simplicity. 
This provides a {\it lower limit} of $\dot{M}_{\rm out}$. 
The outflow mass in neutral hydrogen gas \citep{2010ApJ...714L.162H} is computed as 
\begin{equation}
\begin{split}
\frac{M_{\rm out}}{M_\odot} &= 0.77 \left( \frac{0.7 L_{\rm [CII],broad}}{L_\odot} \right) \left( \frac{1.4 \times 10^{-4}}{X_{\rm C^+}} \right) \\
& \times \frac{1 + 2e^{-91/T_{\rm ex}} + n_{\rm crit}/n}{2e^{-91/T_{\rm ex}}}
\end{split}
\end{equation}
where $X_{\rm C^+}$ is the ratio of C$^+$ abundance to H, $T_{\rm ex}$ is the gas excitation temperature in K, 
$n$ is the gas volume density in cm$^{-3}$, and $n_{\rm crit}$ is the critical density of the line, $\sim 3 \times 10^3$ cm$^{-3}$. 
The factor 0.7 in the first parenthesis is the typical fraction of [\ion{C}{2}] arising from photodissociation regions \citep[PDRs,][]{1997ARA&A..35..179H}. 
As we do not know the actual gas density, we compute a {\it lower limit} on $M_{\rm out}$ by assuming the high density-limit ($n \gg n_{\rm crit}$). 
We also adopt a typical abundance of $X_{\rm C^+} = 1.4 \times 10^{-4}$ and $T_{\rm ex} = 200$ K in PDRs \citep{1997ARA&A..35..179H} 
following previous studies of [\ion{C}{2}] outflows \citep[e.g.,][]{2012MNRAS.425L..66M,2015A&A...574A..14C,2020A&A...633A..90G}. 
The inferred value of $M_{\rm out}$  is quite insensitive to the assumed value of $T_{\rm ex}$. 
Using the measurements in Table \ref{tbl2}, we obtain $M_{\rm out} = (1.2 \pm 0.2) \times 10^9~M_\odot$ and 
$\dot{M}_{\rm out} > 447 \pm 137~M_\odot$ yr$^{-1}$, respectively. 
This $\dot{M}_{\rm out}$ is a strict lower limit because of our treatments of $R_{\rm out}$ and $M_{\rm out}$. 

The above $\dot{M}_{\rm out}$  refers to the neutral atomic component only. 
However, \citet{2019MNRAS.483.4586F} observationally studied multi-phase outflows in local star-forming galaxies and AGNs, 
and found that a large fraction of the outflowing mass is in molecular phase 
\citep[see also][]{2018Galax...6..138R,2005ARA&A..43..769V,2020A&ARv..28....2V}.  
Hydrodynamic simulations of a $z = 7.5$ quasar find the same trend, 
particularly at the central kpc-scale regions \citep{2018MNRAS.481.4877N}.  
Based on these results, we estimate that the full outflow rate 
is roughly a factor of three larger than the atomic-only value. 
This suggests that the total outflow rate for J1243$+$0100 
is $\dot{M}^{\rm tot}_{\rm out} \gtrsim 1410~M_\odot$ yr$^{-1}$. 

The estimated $\dot{M}_{\rm out}$ ($\dot{M}^{\rm tot}_{\rm out}$) is 
$\sim 3\times$ (or $\sim 9\times$) larger than the ${\rm SFR}_{\rm [CII]}$ of the spectral core component. 
This $\dot{M}_{\rm out}$ ($\dot{M}^{\rm tot}_{\rm out}$) is also $\sim 180\times$ (or $\sim 570\times$) 
greater than the mass accretion rate onto the SMBH ($\sim 2.5~M_\odot$ yr$^{-1}$)
\footnote{We estimated this by using the bolometric luminosity (\S~\ref{sec1.1}) and the canonical radiative efficiency of 0.1.}.  
As these outflow rates are lower limits, the actual mass loading factors 
($\eta \equiv \dot{M}^{\rm tot}_{\rm out}/{\rm SFR}$) may be even larger. 
Observations of star-forming galaxies find $\eta$ is typically $\sim 1$--3 
over a wide range of SFR and redshift \citep[e.g.,][]{2013Natur.499..450B,2014A&A...562A..21C,2015A&A...580A..35G,2018MNRAS.473.1909G,2019MNRAS.483.4586F,2020A&A...633A..90G}, 
while AGNs can reach $\eta > 5$ \citep[e.g.,][]{2014A&A...562A..21C,2017A&A...601A.143F,2019MNRAS.483.4586F}. 
Therefore, our estimate of $\eta$ is consistent with {\it quasar-driven} outflows. 

This conclusion also holds if we adopt ${\rm SFR}_{\rm TIR}$ instead of 
${\rm SFR}_{\rm [CII]}$ (Table \ref{tbl2}), as our $\dot{M}_{\rm out}$ is merely a lower limit. 
For instance, if we consider a moderate density PDR with $n = 10^3$ cm$^{-3}$, 
we can already achieve $\eta = \dot{M}^{\rm tot}_{\rm out}/{\rm SFR}_{\rm TIR} \sim 5$. 
Furthermore, if we focus on the spatially extended decomposed FIR continuum 
(i.e., use the conservative value of ${\rm SFR}^{\rm cons}_{\rm TIR}$), 
we find a high $\eta > 5$ for the high-density limit, 
and $\eta \sim 12$ for a case of $n = 10^3$ cm$^{-3}$, for example. 
It is also rare for starburst-driven neutral outflows to reach velocities 
greater than $\sim 500$--600 km s$^{-1}$ even for ULIRG-class objects 
\citep[e.g.,][]{2005ApJ...621..227M,2005ApJS..160..115R,2018MNRAS.473.1909G,2020A&A...633A..90G}, 
whereas the observed [\ion{C}{2}] profile of J1243$+$0100 extends to $\sim 900$ km s$^{-1}$. 
Thus we conclude that the fast [\ion{C}{2}] outflow of J1243$+$0100 is {\it quasar-driven}. 

As we pointed out above, this is only the second $z > 6$ quasar known 
with a [\ion{C}{2}] outflow out of dozens of objects observed at submm. 
This may suggest that such outflows have a small duty cycle, 
as seen in simulations \citep[$\lesssim 10$ Myr, e.g.,][]{2014MNRAS.444.2355C,2018MNRAS.479.2079C,2018MNRAS.473.3525Z}. 
Very large-scale (extended) outflows such the $\sim 30$ kpc scale flow 
seen in J1148$+$5251 may also be resolved out 
particularly for the case of high-resolution interferometric observations. 
Note that the currently known highest redshift ($z = 7.64$) quasar J0313$-$1806 \citep{2021arXiv210103179W}, 
for example, shows a smaller SFR than J1243$+$0100 while its quasar nucleus is significantly brighter (Table \ref{tbl4}). 
As J0313$-$1806 also hosts nuclear fast winds, it is intriguing to seek for a galaxy-scale feedback 
that might have quenched the star formation of the host at a further earlier epoch. 
In any case, deep and homogeneous observations toward a large number of quasars, 
with holding sensitivity to extended structures, are necessary to faithfully study outflows.

\subsection{Feedback on the host galaxy}\label{sec4.3}
In order to assess the impact of the outflow on the host galaxy itself, in particular, any quenching of star formation, 
we calculate the outflow kinetic power 
\begin{equation}
\dot{E}_{\rm out} = \frac{1}{2} \dot{M}_{\rm out} v^2_{\rm out}
\end{equation}
and the momentum load normalized by the radiative momentum of the quasar as 
\begin{equation}
\dot{P}_{\rm out}/\dot{P}_{\rm AGN} = \frac{\dot{M}_{\rm out} v_{\rm out}}{L_{\rm Bol}/c}, 
\end{equation}
using the numbers calculated in \S~\ref{sec4.2} (Table \ref{tbl5}). 
By solely using the lower limits of the neutral outflow, we find 
$\dot{E}_{\rm out}/L_{\rm Bol} \gtrsim 0.25\%$ and $\dot{P}_{\rm out}/\dot{P}_{\rm AGN} \gtrsim 3.0$. 
Again assuming that the total outflow rate is $\sim 0.5$ dex higher than the atomic-only value, 
as well as that all phase outflows have comparable velocities, 
$\dot{E}^{\rm tot}_{\rm out}/L_{\rm Bol}$ and $\dot{P}^{\rm tot}_{\rm out}/\dot{P}_{\rm AGN}$ 
approach $\sim 1\%$ and $\sim 10$, respectively. 

It is intriguing that J1243$+$0100 hosts fast winds on the scale of the accretion disk, 
as evidenced by significantly blueshifted \ion{C}{4} emission and \ion{Si}{4} and \ion{C}{4} BALs \citep{2019ApJ...872L...2M}. 
One class of AGN feedback models indeed relies on a coupling between 
the nuclear wind and the galaxy-scale ISM \citep[e.g.,][]{2003ApJ...596L..27K,2015ARA&A..53..115K}. 
The existence of both the nuclear winds and the large-scale [\ion{C}{2}] outflow is a good match to this class of model. 
The {\em lower limits} of $\dot{E}^{\rm tot}_{\rm out}/L_{\rm Bol}$ and $\dot{P}^{\rm tot}_{\rm out}/\dot{P}_{\rm AGN}$ 
for J1243$+$0100 are somewhat smaller than, but on the same order as, the values expected 
in the {\it energy-conserving}\footnote{i.e., the shocked wind flow preserves its thermal energy.} 
coupling mode ($\dot{E}^{\rm tot}_{\rm out}/L_{\rm Bol} \sim 5\%$, $\dot{P}^{\rm tot}_{\rm out}/\dot{P}_{\rm AGN} \sim 20$). 
Such a flow is sufficiently energetic to quench star formation inside the host galaxy 
\citep[e.g.,][]{2012ApJ...745L..34Z,2014MNRAS.444.2355C,2015ARA&A..53..115K}. 

Another class of feedback models explains the large scale outflows as winds 
driven by direct AGN radiation pressure onto dusty clouds 
\citep[e.g.,][]{2005ApJ...618..569M,2015MNRAS.451...93I,2016MNRAS.463.1291I,2018MNRAS.479.2079C,2018MNRAS.473.4197C}, 
i.e., without invoking intermediary winds. 
For example, \citet{2018MNRAS.479.2079C} performed hydrodynamic 
simulations of outflows driven by multi-scattered radiation pressure. 
Their simulation results for, e.g., $v_{\rm out}$, $\dot{E}^{\rm tot}_{\rm out}/L_{\rm Bol}$, and $\dot{P}^{\rm tot}_{\rm out}/\dot{P}_{\rm AGN}$ 
are quite consistent with the values we estimated for J1243$+$0100. 
The simulation also predicts that (i) this mechanism is efficient when the quasar nucleus is obscured 
as radiation pressure requires a dense ISM on which to act, 
and (ii) radiation pressure-driven wind is short-lived ($\sim 10$ Myr) 
as that process loses efficiency once the ISM becomes extended and diffuse. 
In accord with these predictions, the relatively compact size and the short time-scale of the outflow of J1243$+$0100 (Table \ref{tbl5}) 
suggest that this quasar feedback has just begun. 
In a later phase of the evolution of the outflow, the value of $\dot{E}^{\rm tot}_{\rm out}/L_{\rm Bol}$ will drop,
as is seen in the extended [\ion{C}{2}] outflow of J1148$+$5251 \citep{2015A&A...574A..14C}. 
The \citet{2018MNRAS.479.2079C} simulations predict that outflows could be launched  
only in quasars with $L_{\rm Bol} > 10^{47}$ erg s$^{-1}$, an order of magnitude more luminous than J1243$+$0100. 
However, as the outflow can clear out the circumnuclear gas that is the fuel for SMBH accretion, 
the $L_{\rm Bol}$ of J1243$+$0100 may have been much higher at the time the outflow started \citep[e.g.,][]{2010ApJ...717..708C,2010MNRAS.407.1529H}. 

We thus conclude that the outflow properties of J1243$+$0100 
are reasonably consistent with both the energy-conserving wind models 
and the radiation pressure-driven dusty wind models. 
Further observational constraints including outflow geometry, 
observations of other phases of the outflow, and the stellar and gas mass distributions, 
may be required for a better comparison with the models. 
However, we observe a short flow time (Table \ref{tbl5}) 
and on-going active star formation (\S~\ref{sec4.1}), suggesting that no matter what
the underlying model, the outflow of J1243$+$0100 has not yet considerably impacted 
the star formation of the host galaxy, even though it may 
 already have affected the small scale gas accretion leading to a relatively small Eddington ratio. 
Given the high mass loading factor, this outflow should quench at least 
the central kpc-scale starburst in the near future.

\subsection{Gas dynamical modeling}\label{sec4.4}
We saw in \S~\ref{sec3.2.3} that the [\ion{C}{2}] emission shows what seems to be ordered rotation. 
In this section, we model the velocity field in detail to extract a rotation curve and velocity dispersion profile. 
To this end, we fitted six concentric rings with $0\arcsec.1$ width to the [\ion{C}{2}] data cube using the $^{\rm 3D}$Barolo code \citep{2015MNRAS.451.3021D}, 
which has been applied to galaxies at both low and high redshift \citep[e.g.,][]{2020ApJ...898...75I,2020ApJ...900....1F}. 
The parameters we fit to each ring are $V_{\rm rot}$, $\sigma_{\rm disp}$, and the radial velocity in the disk plane ($V_{\rm rad}$). 
We fixed the dynamical center to the quasar position and $V_{\rm sys}$ to 0 km s$^{-1}$, 
and constrained the inclination and position angle of the rings to all be the same, 
with best-fit values of  $i = 25\arcdeg$ and ${\rm PA} = 87\arcdeg$, respectively. 

\begin{figure}
\begin{center}
\includegraphics[width=\linewidth]{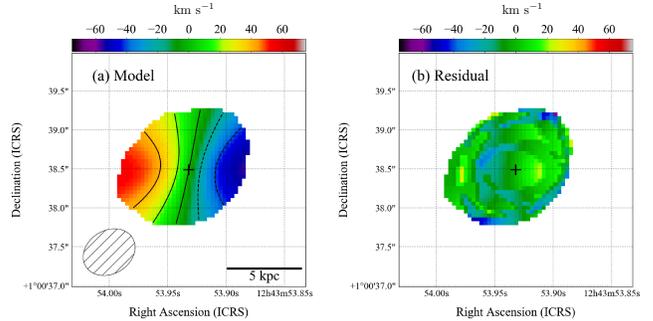}
\caption{
(a) Modeled mean velocity field (moment-1) of J1243$+$0100. 
The contours indicate velocities relative to the systemic, in steps of $\pm 20$ km s$^{-1}$. 
(b) Residual velocity component after subtracting the model from the observed moment-1 map (Figure \ref{fig4}). 
The residual amplitude is mostly $< 20$ km s$^{-1}$, indicating that our fit is good. 
}
\label{fig10}
\end{center}
\end{figure}

A conservative $5\sigma$ clipping was applied to avoid noise contamination, 
hence our model is {\it not sensitive to the faint outflow}. 
We set initial guesses of $V_{\rm rot} = 120$ km s$^{-1}$, $\sigma_{\rm disp} = 40$ km s$^{-1}$, 
and $V_{\rm rad} = 0$ km s$^{-1}$, respectively for all rings. 
The fitting was evaluated by minimizing the residual amplitude, i.e., $|$model$-$observed data$|$. 
Figure \ref{fig10} shows the modeled mean velocity field and the residual map 
after subtracting the model component from the observed one (Figure \ref{fig4}). 
The residuals are mostly small, $< 20$ km s$^{-1}$ 
over the modeled regions, indicating the goodness of our fit. 
We also found that $V_{\rm rad}$ is no larger than $\pm 20$ km s$^{-1}$, 
hence we do not discuss it in further detail hereafter. 

\begin{figure}
\begin{center}
\includegraphics[width=\linewidth]{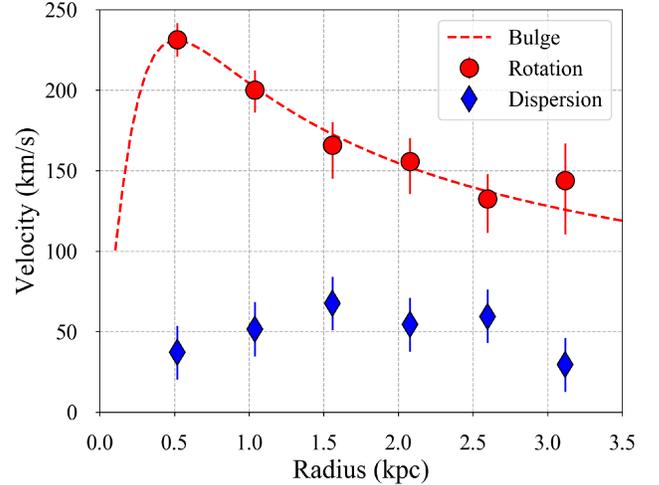}
\caption{
Radial profiles of rotation velocity ($V_{\rm rot}$; red circle) 
and velocity dispersion ($\sigma_{\rm disp}$; blue diamonds) of J1243$+$0100. 
The velocities have been corrected for projection due to the inclination of the galaxy. 
The  best fit Plummer potential (red dashed line) gives a bulge mass of $(3.3 \pm 0.2) \times 10^{10}~M_\odot$ 
with a scale length of $0.36 \pm 0.03$ kpc. 
}
\label{fig11}
\end{center}
\end{figure}

Figure \ref{fig11} shows the derived  $V_{\rm rot}$ and $\sigma_{\rm disp}$ as a function of radius. 
The radial velocity is 3--5 times larger than the velocity dispersion in all bins, clearly indicating 
that the gas dynamics of J1243$+$0100 is {\it rotation-dominated}. 
For comparison, the ratio of rotation velocity to velocity dispersion is considerably lower, 
of order unity, in the well-resolved $z = 6.6$ optically luminous quasar J0305$-$3150 \citep{2019ApJ...874L..30V}. 
Models for high-redshift starburst galaxies (without a central AGN) predict 
small $V_{\rm rot}/\sigma_{\rm disp}$ due to turbulence associated 
with galaxy mergers, inflows, and stellar feedback (for example, \cite{2019MNRAS.490.3196P} 
predict $V_{\rm rot}/\sigma_{\rm disp} \lesssim 2$ at $z \sim 5$). 
Even though $V_{\rm rot}/\sigma_{\rm disp}$ is relatively high in J1243$+$0100, 
it is not quiescent, given its high star formation rate  (\S~\ref{sec3}). 
Indeed, recent ALMA observations have found rotation-dominated but modestly gravitationally unstable galaxies 
at $z > 4$ \citep{2018Natur.560..613T,2020Natur.584..201R}. 
High resolution molecular gas observations would allow us to properly assess the gravitational stability of J1243$+$0100. 

It is noteworthy that $V_{\rm rot}$ is highest near the center of the host, and drops off in the outer regions. 
A rotation curve that rises into the central (sub-)kpc regions of galaxies 
is frequently attributed to the existence of a massive galactic {\it bulge} \citep[][for a review]{2016PASJ...68....2S}. 
Indeed, hydrodynamic simulations of $z \gtrsim 7$ quasars found that their host galaxies 
are typically bulge-dominant massive systems \citep{2019MNRAS.483.1388T,2020MNRAS.499.3819M}. 
We fit the observed velocity profile $V_{\rm rot}(r)$ with 
a simple spherical Plummer potential \citep{1911MNRAS..71..460P}, 
\begin{equation}
\Phi (r) = - \frac{GM_{\rm bulge}}{(r^2 + a^2)^{1/2}}
\end{equation}
where $G$ is the gravitational constant and $a$ is the characteristic Plummer radius, which sets the scale length of the core. 
This simple model fits the observed rotation curve well. 
The best-fit parameters are $M_{\rm bulge} = (3.3 \pm 0.2) \times 10^{10}~M_\odot$ and $a = 0.36 \pm 0.03$ kpc. 
Given the goodness of the fit, any contributions to the dynamics from 
other components should be minor at $R \lesssim 3$ kpc. 
Note that this $M_{\rm bulge}$ is less than half of the $M_{\rm dyn}$ derived in \S~\ref{sec3.2.5}. 
A main cause of this discrepancy may be the crude estimate 
of the disk circular velocity for calculating $M_{\rm dyn}$. 
Indeed, the estimated $v_{\rm circ}$ in \S~\ref{sec3.2.5} is $\sim 500$ km s$^{-1}$, 
which is much higher than what we see here in Figure \ref{fig11}. 
Because of our more detailed modeling in this section, 
we would think the dynamically-modeled $M_{\rm bulge}$ is more robust. 
Our current data therefore suggests that a massive bulge has already formed in this system at $z \sim 7$. 
However, our resolution is limited, thus this estimate of $M_{\rm bulge}$ is {\it tentative}. 
Further higher resolution and higher sensitivity observations of gas dynamics with ALMA, 
as well as direct measurement of the stellar light distribution by the James Webb Space Telescope 
will conclusively determine the structure of the host of J1243$+$0100. 

As another aspect, if we crudely assume that the $V_{\rm rot}$ at the scale length ($\sim 220$ km s$^{-1}$) 
is equivalent to the halo circular velocity, we can roughly estimate the halo mass ($M_{\rm h}$) by 
using the Equation (25) of \citet{2001PhR...349..125B}. 
This resulted in $M_{\rm h} \sim 3 \times 10^{11}~M_\odot$, 
which is also consistent with the hydrodynamic simulation result 
of \citet{2020MNRAS.499.3819M} for this class of quasars.

\subsection{Early co-evolution at $z \sim 7$}\label{sec4.5}
We first treat our measured $M_{\rm dyn}$ (\S~\ref{sec3.2.5}) as a surrogate for the bulge-scale stellar mass 
as usually assumed in $z \gtrsim 6$ quasar studies \citep[e.g.,][]{2013ApJ...773...44W,2016ApJ...816...37V,2019PASJ...71..111I}, 
allowing us to investigate the black hole-bulge mass relation at this early epoch of the universe. 
We adopt this simple treatment as detailed rotation curves (and resultant $M_{\rm bulge}$) 
are not currently available in most of the $z > 6$ quasars. 
Hence to make a fair comparison with the other $z > 6$ objects, we need to use the crudely estimated $M_{\rm dyn}$. 

\begin{figure*}
\begin{center}
\includegraphics[width=\linewidth]{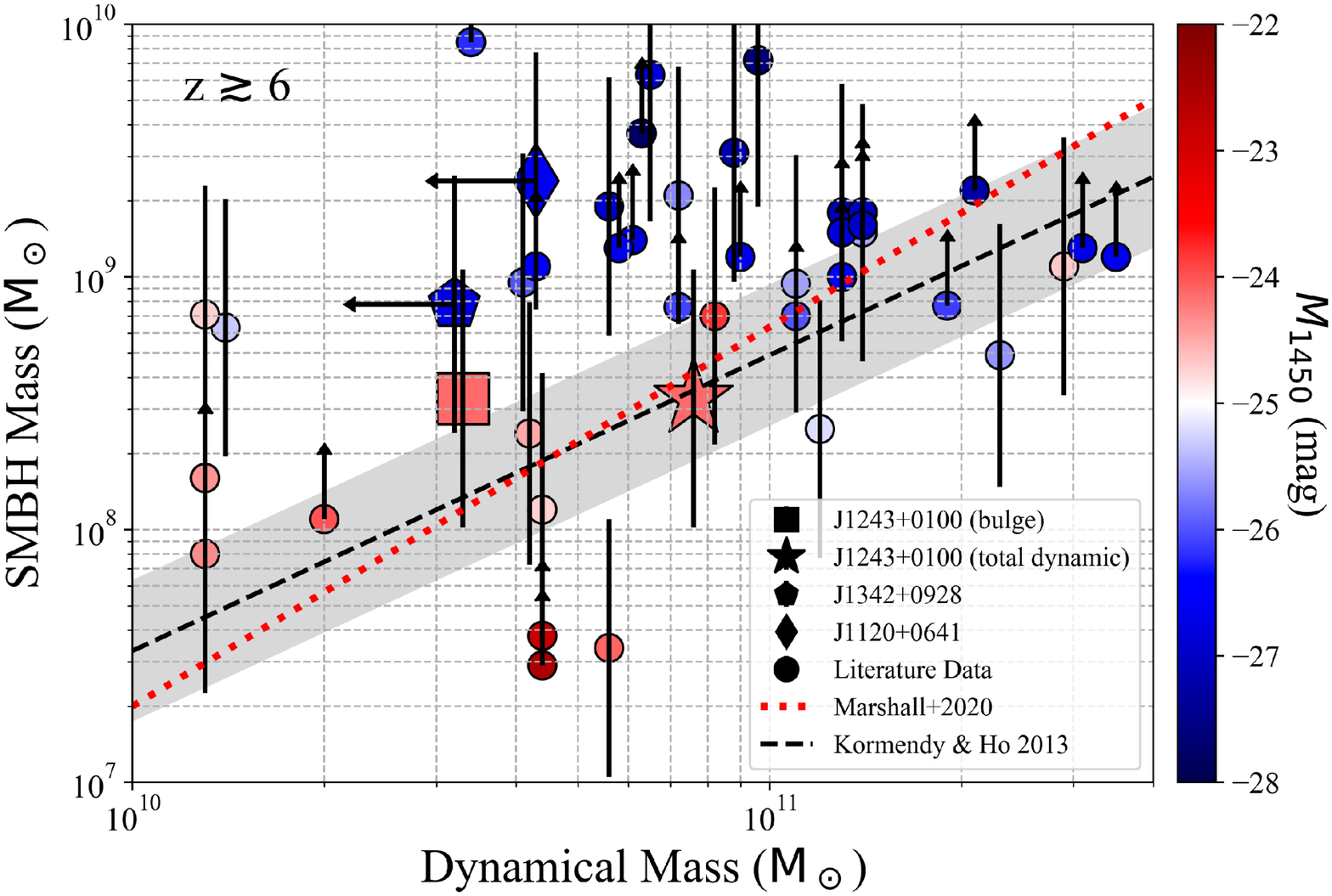}
\caption{
Black hole mass ($M_{\rm BH}$) vs host galaxy dynamical mass ($M_{\rm dyn}$) 
for $z \gtrsim 6$ quasars, using data compiled from \citet{2019PASJ...71..111I}. 
The quasars are color-coded by their $M_{\rm 1450}$. 
Two other $z > 7$ quasars with measured [\ion{C}{2}]-based $M_{\rm dyn}$ are highlighted; 
note upper limits for their dynamical masses are available. 
The black diagonal dashed line indicates the local $M_{\rm BH}$--$M_{\rm bulge}$ relation \citep{2013ARA&A..51..511K}, 
whereas the red dotted line indicates the simulated relation for $z = 7$ quasars \citep{2020MNRAS.499.3819M}.  
}
\label{fig12}
\end{center}
\end{figure*}

Figure \ref{fig12} shows this relation for $z \gtrsim 6$ quasars using data compiled in \citet{2019PASJ...71..111I}. 
We computed their $M_{\rm dyn}$ as we did for J1243$+$0100 in \S~\ref{sec3.2.5}.  
The two $z > 7$ quasars (J1342$+$0928 and J1120$+$0641) are dispersion-dominated, and for them we applied the virial theorem 
\citep{2017ApJ...837..146V,2017ApJ...851L...8V,2019ApJ...881L..23B}. 
The $M_{\rm BH}$ of all quasars were calculated with the common \citet{2009ApJ...699..800V} 
calibration for the \ion{Mg}{2}-based single epoch method. 
Some quasars do not have $M_{\rm BH}$ measurements; for them we assumed 
Eddington-limited accretion to give the lower mass limits.  
The low-luminosity objects ($M_{\rm 1450} \gtrsim -25$ mag) in this plot 
were drawn from the HSC sample \citep{2018PASJ...70...36I,2019PASJ...71..111I} 
and the CFHQS \citep{2013ApJ...770...13W,2015ApJ...801..123W,2017ApJ...850..108W}. 

It is intriguing that J1243$+$0100 shows a $M_{\rm BH}/M_{\rm dyn}$ ratio 
in excellent accord with the local value \citep{2013ARA&A..51..511K}. 
It is also in agreement with the cosmological hydrodynamic simulation by \citet{2020MNRAS.499.3819M}, 
who predict a $M_{\rm BH}$--$M_{\rm bulge}$ relation at $z \sim 7$ that is slightly steeper than, 
but is in a reasonable agreement with, the local relation (see dotted line in Figure \ref{fig12}). 
Note that, however, it is possible that gas contributes significantly to $M_{\rm dyn}$, 
causing to be an over-estimate of $M_{\rm bulge}$. \citet{2019ApJ...881...63N} 
estimated a gas-to-dust mass ratio of the $z = 7.54$ quasar J1342$+$0928 of $< 100$. 
If we apply this number to the value of $M_{\rm dust}$ of J1243$+$0100 that we found in \S~\ref{sec3.1}, we obtain $M_{\rm gas} < 2.5 \times 10^{10}~M_\odot$. 
Thus, a true (total) stellar mass of this quasar host galaxy would be $\sim (5-7) \times 10^{10}~M_\odot$, still in a good agreement with the expectation from the local relation. 
Alternatively, if instead we use the direct $M_{\rm bulge}$ estimate of $(3.3 \pm 0.2) \times 10^{10}~M_\odot$ from  \S~\ref{sec4.4}, 
we still find that within the uncertainties, $M_{\rm BH}/M_{\rm bulge}$ is consistent with the local value. 
Our findings therefore suggest that for J1243$+$0100 at least, the $M_{\rm BH}$--$M_{\rm bulge}$ relation was in place already at $z \sim 7$. 

In contrast, the two other $z > 7$ quasars with [\ion{C}{2}]-based $M_{\rm dyn}$ estimates (J1342$+$0928 and J1120$+$0641) 
have {\it over-massive} black holes relative to these relations, by up to factor $\sim 10$. 
This is also seen in the other luminous ($M_{\rm 1450} \lesssim -26$ mag) quasars at $z\gtrsim6$, 
which may reflect a selection bias to more massive black holes \citep{2007ApJ...670..249L,2014MNRAS.438.3422S}. 
Relatively lower mass SMBHs ($M_{\rm BH} \lesssim 10^{8.5}~M_\odot$), most of which are low-luminosity quasars, 
would not strongly suffer this bias \citep{2007ApJ...670..249L,2020MNRAS.499.3819M} 
and indeed show comparable mass ratios to the local relation (Figure \ref{fig12}). 
Further observations of galaxies hosting less massive SMBHs 
are necessary to  confirm this picture statistically. 
As the co-evolution relation is the end-product of the complex growth of galaxies and SMBH, 
a statistical measurement of this relation at the early universe \citep[see also e.g.,][]{2020ApJ...889...32S,2021arXiv210111273S} 
would help constrain the relative cosmological importance 
of various feeding and feedback processes \citep{2020arXiv200610094H}. 
We also need high resolution observations sensitive to detailed dynamics, 
as well as wide-area deep observations sensitive to the surrounding environments, 
to reveal the driving mechanism of the rapid growth of galaxies and SMBHs in the early universe.

\section{Summary}\label{sec5}
In this paper we present ALMA observations of [\ion{C}{2}] line and underlying rest-frame FIR continuum 
emission toward J1243$+$0100 at $z = 7.07$. 
This object is currently the only low-luminosity quasar known at $z > 7$. 
We clearly detected both the line and continuum, 
from which we determined the following characteristics of this remarkable quasar and its host. 
\begin{itemize}
\item[1.] The FIR continuum is bright, 1.5 mJy, resulting in a total 
$L_{\rm FIR} = (3.5 \pm 0.1) \times 10^{12}~L_\odot$, 
assuming a dust temperature of $T_{\rm dust} = 47$ K 
and an emissivity index of $\beta = 1.6$. 
The inferred area-integrated ${\rm SFR}_{\rm TIR}$ is $742 \pm 16~M_\odot$ yr$^{-1}$ 
if the heating source is entirely attributed to star formation. 
We also estimate the dust mass as $(2.5 \pm 0.1) \times 10^8~M_\odot$. 
This inferred star formation rate is as high at that of optically luminous $z > 6$ quasars, 
and is $>3-10\times$ higher than that of the low-luminosity HSC quasars observed with ALMA. 
\item[2.] However, we also decomposed this FIR continuum emission to a point source and an extended Gaussian component. 
If we regard the former as emission from the quasar nucleus itself, 
our conservative estimate on ${\rm SFR}_{\rm TIR}$ should be that of the extended component, 
i.e., ${\rm SFR^{cons}_{TIR}} = 307 \pm 20~M_\odot$ yr$^{-1}$. 
\item[3.] The [\ion{C}{2}] emission is spatially resolved and is very bright. 
We found a broad wing component ($L_{\rm [CII]} = (1.2 \pm 0.3) \times 10^9~L_\odot$, ${\rm FWHM} = 997 \pm 227$ km s$^{-1}$) 
in addition to a bright core component ($L_{\rm [CII]} = (1.9 \pm 0.2) \times 10^9~L_\odot$, 
${\rm FWHM} = 235 \pm 17$ km s$^{-1}$) in the area-integrated spectrum. 
The inferred ${\rm SFR}_{\rm [CII]}$ from this spectral core component is $165 \pm 17~M_\odot$ yr$^{-1}$. 
\item[4.] We measured the spatial extents of the [\ion{C}{2}] spectral core and wing emission respectively 
by directly modeling the visibilities, and found that $\sim 4.2$ kpc for the core and $< 2.7$ kpc ($3\sigma$ limit) for the wing. 
Thus the broad wing originates from a relatively compact region inside this galaxy. 
\item[5.] The global gas dynamics are governed by rotation. 
We estimate its dynamical mass as $M_{\rm dyn} = (7.6 \pm 0.9) \times 10^{10}~M_\odot$. 
This is $\gtrsim 2-3\times$ larger than the other two $z > 7$ quasars 
(J1343$+$0928 and J1120$+$0641) with $M_{\rm dyn}$ measurements. 
\item[6.] We did not find any significant ($>5\sigma$) companion continuum emitter within our field-of-view. 
This non-detection is, however, consistent with recent 1.2 mm number counts in the field. 
\item[7.] Using the $M_{\rm dyn}$ as a proxy for the host galaxy stellar mass ($M_\star$), 
we found that J1243$+$0100 is located on or even above the star-forming main sequence at $z \sim 6$. 
Considering its low quasar luminosity and low Eddington ratio, 
it is plausible that J1243$+$0100 is in a transition phase: 
feedback may be in the process of turning off the central quasar activity, 
but that has not yet shut down the star formation in the host. 
\item[8.] Various considerations have led us to conclude that the broad [\ion{C}{2}] wing 
is due to a fast neutral outflow, with a rate $\dot{M}_{\rm out} > 447 \pm 137$ $M_\odot$ yr$^{-1}$). 
Including a molecular component would make this value higher, 
leading to a high mass loading factor (e.g., $\gtrsim 9$ relative to ${\rm SFR_{[CII]}}$). 
This high value suggests that this outflow is quasar-driven. 
The outflow kinetic power and momentum load are reasonably 
consistent with the predictions of both the energy-conserving and the radiation pressure-driven wind models. 
The high mass loading factor indicates that this outflow will quench the starburst of this galaxy in the near future. 
\item[9.] By modeling the observed velocity field, we found that the host galaxy dynamics 
is dominated by rotation, with $V_{\rm rot}/\sigma_{\rm disp} \sim 3-5$. 
The rotation curve is highest within 1 kpc, which we model as 
due to a compact (radius $\sim 0.36$ kpc) stellar bulge with a mass of $(3.3 \pm 0.2) \times 10^{10}~M_\odot$. 
While this result is limited by our resolution, we imply that 
massive bulge formation has already occurred at $z \sim 7$, 
in accord with the recent model prediction. 
\item[10.] Using either the (total) dynamical mass or inferred bulge mass 
from our rotation curve modeling, we find a bulge to black hole mass ratio consistent with the local value. 
Our result therefore suggests that the co-evolution relation was already in place at $z \sim 7$. 
\end{itemize}

We have suggested in this paper that a fraction of optically low-luminosity but FIR luminous quasars 
are in a key transition phase, ceasing their nuclear activity due to feedback from a powerful outflow. 
The nuclear fast winds seen in J1243$+$0100 indicate that this quasar provides an outstanding laboratory to study 
quasar-driven feedback processes on scales from the accretion disk to the host galaxy. 
Future ALMA observations will allow us to continue the search for and study of galaxy-scale feedback in these early universe systems.

\acknowledgments 
We appreciate the anonymous reviewer's very constructive comments to improve this manuscript. 
This paper makes use of the following ALMA data: 
ADS/JAO.ALMA\#2019.1.00074.S. 
ALMA is a partnership of ESO (representing its member states), 
NSF (USA) and NINS (Japan), together with NRC (Canada), 
MOST and ASIAA (Taiwan), and KASI (Republic of Korea), 
in cooperation with the Republic of Chile. 
The Joint ALMA Observatory is operated by ESO, AUI/NRAO and NAOJ. 

The Hyper Suprime-Cam (HSC) collaboration includes the astronomical communities 
of Japan and Taiwan, and Princeton University. 
The HSC instrumentation and software were developed by the National Astronomical Observatory of Japan (NAOJ), 
the Kavli Institute for the Physics and Mathematics of the Universe (Kavli IPMU), 
the University of Tokyo, the High Energy Accelerator Research Organization (KEK), 
the Academia Sinica Institute for Astronomy and Astrophysics in Taiwan (ASIAA), and Princeton University. 
Funding was contributed by the FIRST program from the Japanese Cabinet Office, 
the Ministry of Education, Culture, Sports, Science and Technology (MEXT), 
the Japan Society for the Promotion of Science (JSPS), 
the Japan Science and Technology Agency (JST), the Toray Science Foundation, 
NAOJ, Kavli IPMU, KEK, ASIAA, and Princeton University. 

T.H. was supported by the Leading Initiative for Excellent Young Researchers, MEXT, Japan (HJH02007). 
T.I., K.K., A.I., and S.B. were supported by JSPS KAKENHI Grant Number JP20K14531, JP17H06130, JP17H01114, and JP19J00892, respectively. 
K.K. and A.I. were also supported by the NAOJ ALMA Scientific Research Grant Number 2017-06B and 2020-16B, respectively. 
K.I. acknowledges support by the Spanish MICINN under grant PID2019-105510GB-C33 
and ``Unit of Excellence Mar\'ia de Maeztu 2020-2023" awarded to ICCUB (CEX2019-000918-M). 
This work is partially supported by the National Science Foundation of China 
(11721303, 11991052, 11950410493, 12073003) and the National Key R\&D Program of China (2016YFA0400702). 
S.F. acknowledges support from the European Research Council (ERC) Consolidator 
Grant funding scheme (project ConTExt, grant No. 648179) and Independent Research Fund Denmark grant DFF--7014-00017. 
The Cosmic Dawn Center is funded by the Danish National Research Foundation under grant No. 140. 
T.I. is supported by ALMA Japan Research Grant of the NAOJ ALMA Project, NAOJ-ALMA-253.  

\bibliography{Izumi2021b_ref}

\begin{thebibliography}{}
\expandafter\ifx\csname natexlab\endcsname\relax\def\natexlab#1{#1}\fi
\providecommand{\url}[1]{\href{#1}{#1}}
\providecommand{\dodoi}[1]{doi:~\href{http://doi.org/#1}{\nolinkurl{#1}}}
\providecommand{\doeprint}[1]{\href{http://ascl.net/#1}{\nolinkurl{http://ascl.net/#1}}}
\providecommand{\doarXiv}[1]{\href{https://arxiv.org/abs/#1}{\nolinkurl{https://arxiv.org/abs/#1}}}

\bibitem[{{Aalto} {et~al.}(2012){Aalto}, {Garcia-Burillo}, {Muller}, {Winters},
  {van der Werf}, {Henkel}, {Costagliola}, \& {Neri}}]{2012A&A...537A..44A}
{Aalto}, S., {Garcia-Burillo}, S., {Muller}, S., {et~al.} 2012, \aap, 537, A44,
  \dodoi{10.1051/0004-6361/201117919}

\bibitem[{{Aihara} {et~al.}(2018){Aihara}, {Armstrong}, {Bickerton}, {Bosch},
  {Coupon}, {Furusawa}, {Hayashi}, {Ikeda}, {Kamata}, {Karoji}, {Kawanomoto},
  {Koike}, {Komiyama}, {Lang}, {Lupton}, {Mineo}, {Miyatake}, {Miyazaki},
  {Morokuma}, {Obuchi}, {Oishi}, {Okura}, {Price}, {Takata}, {Tanaka},
  {Tanaka}, {Tanaka}, {Uchida}, {Uraguchi}, {Utsumi}, {Wang}, {Yamada},
  {Yamanoi}, {Yasuda}, {Arimoto}, {Chiba}, {Finet}, {Fujimori}, {Fujimoto},
  {Furusawa}, {Goto}, {Goulding}, {Gunn}, {Harikane}, {Hattori}, {Hayashi},
  {He{\l}miniak}, {Higuchi}, {Hikage}, {Ho}, {Hsieh}, {Huang}, {Huang},
  {Imanishi}, {Iwata}, {Jaelani}, {Jian}, {Kashikawa}, {Katayama}, {Kojima},
  {Konno}, {Koshida}, {Kusakabe}, {Leauthaud}, {Lee}, {Lin}, {Lin},
  {Mandelbaum}, {Matsuoka}, {Medezinski}, {Miyama}, {Momose}, {More}, {More},
  {Mukae}, {Murata}, {Murayama}, {Nagao}, {Nakata}, {Niida}, {Niikura},
  {Nishizawa}, {Oguri}, {Okabe}, {Ono}, {Onodera}, {Onoue}, {Ouchi}, {Pyo},
  {Shibuya}, {Shimasaku}, {Simet}, {Speagle}, {Spergel}, {Strauss}, {Sugahara},
  {Sugiyama}, {Suto}, {Suzuki}, {Tait}, {Takada}, {Terai}, {Toba}, {Turner},
  {Uchiyama}, {Umetsu}, {Urata}, {Usuda}, {Yeh}, \&
  {Yuma}}]{2018PASJ...70S...8A}
{Aihara}, H., {Armstrong}, R., {Bickerton}, S., {et~al.} 2018, \pasj, 70, S8,
  \dodoi{10.1093/pasj/psx081}

\bibitem[{{Ba{\~n}ados} {et~al.}(2016){Ba{\~n}ados}, {Venemans}, {Decarli},
  {Farina}, {Mazzucchelli}, {Walter}, {Fan}, {Stern}, {Schlafly}, {Chambers},
  {Rix}, {Jiang}, {McGreer}, {Simcoe}, {Wang}, {Yang}, {Morganson}, {De Rosa},
  {Greiner}, {Balokovi{\'c}}, {Burgett}, {Cooper}, {Draper}, {Flewelling},
  {Hodapp}, {Jun}, {Kaiser}, {Kudritzki}, {Magnier}, {Metcalfe}, {Miller},
  {Schindler}, {Tonry}, {Wainscoat}, {Waters}, \& {Yang}}]{2016ApJS..227...11B}
{Ba{\~n}ados}, E., {Venemans}, B.~P., {Decarli}, R., {et~al.} 2016, \apjs, 227,
  11, \dodoi{10.3847/0067-0049/227/1/11}

\bibitem[{{Ba{\~n}ados} {et~al.}(2018){Ba{\~n}ados}, {Venemans},
  {Mazzucchelli}, {Farina}, {Walter}, {Wang}, {Decarli}, {Stern}, {Fan},
  {Davies}, {Hennawi}, {Simcoe}, {Turner}, {Rix}, {Yang}, {Kelson}, {Rudie}, \&
  {Winters}}]{2018Natur.553..473B}
{Ba{\~n}ados}, E., {Venemans}, B.~P., {Mazzucchelli}, C., {et~al.} 2018, \nat,
  553, 473, \dodoi{10.1038/nature25180}

\bibitem[{{Ba{\~n}ados} {et~al.}(2019){Ba{\~n}ados}, {Novak}, {Neeleman},
  {Walter}, {Decarli}, {Venemans}, {Mazzucchelli}, {Carilli}, {Wang}, {Fan},
  {Farina}, \& {Rix}}]{2019ApJ...881L..23B}
{Ba{\~n}ados}, E., {Novak}, M., {Neeleman}, M., {et~al.} 2019, \apjl, 881, L23,
  \dodoi{10.3847/2041-8213/ab3659}

\bibitem[{{Barkana} \& {Loeb}(2001)}]{2001PhR...349..125B}
{Barkana}, R., \& {Loeb}, A. 2001, \physrep, 349, 125,
  \dodoi{10.1016/S0370-1573(01)00019-9}

\bibitem[{{Beelen} {et~al.}(2006){Beelen}, {Cox}, {Benford}, {Dowell},
  {Kov{\'a}cs}, {Bertoldi}, {Omont}, \& {Carilli}}]{2006ApJ...642..694B}
{Beelen}, A., {Cox}, P., {Benford}, D.~J., {et~al.} 2006, \apj, 642, 694,
  \dodoi{10.1086/500636}

\bibitem[{{B{\'e}thermin} {et~al.}(2020){B{\'e}thermin}, {Fudamoto}, {Ginolfi},
  {Loiacono}, {Khusanova}, {Capak}, {Cassata}, {Faisst}, {Le F{\`e}vre},
  {Schaerer}, {Silverman}, {Yan}, {Amorin}, {Bardelli}, {Boquien}, {Cimatti},
  {Davidzon}, {Dessauges-Zavadsky}, {Fujimoto}, {Gruppioni}, {Hathi}, {Ibar},
  {Jones}, {Koekemoer}, {Lagache}, {Lemaux}, {Moreau}, {Oesch}, {Pozzi},
  {Riechers}, {Talia}, {Toft}, {Vallini}, {Vergani}, {Zamorani}, \&
  {Zucca}}]{2020A&A...643A...2B}
{B{\'e}thermin}, M., {Fudamoto}, Y., {Ginolfi}, M., {et~al.} 2020, \aap, 643,
  A2, \dodoi{10.1051/0004-6361/202037649}

\bibitem[{{Bischetti} {et~al.}(2019){Bischetti}, {Maiolino}, {Carniani},
  {Fiore}, {Piconcelli}, \& {Fluetsch}}]{2019A&A...630A..59B}
{Bischetti}, M., {Maiolino}, R., {Carniani}, S., {et~al.} 2019, \aap, 630, A59,
  \dodoi{10.1051/0004-6361/201833557}

\bibitem[{{Bolatto} {et~al.}(2013){Bolatto}, {Warren}, {Leroy}, {Walter},
  {Veilleux}, {Ostriker}, {Ott}, {Zwaan}, {Fisher}, {Weiss}, {Rosolowsky}, \&
  {Hodge}}]{2013Natur.499..450B}
{Bolatto}, A.~D., {Warren}, S.~R., {Leroy}, A.~K., {et~al.} 2013, \nat, 499,
  450, \dodoi{10.1038/nature12351}

\bibitem[{{Bournaud} {et~al.}(2011){Bournaud}, {Chapon}, {Teyssier}, {Powell},
  {Elmegreen}, {Elmegreen}, {Duc}, {Contini}, {Epinat}, \&
  {Shapiro}}]{2011ApJ...730....4B}
{Bournaud}, F., {Chapon}, D., {Teyssier}, R., {et~al.} 2011, \apj, 730, 4,
  \dodoi{10.1088/0004-637X/730/1/4}

\bibitem[{{Carilli} \& {Walter}(2013)}]{2013ARA&A..51..105C}
{Carilli}, C.~L., \& {Walter}, F. 2013, \araa, 51, 105,
  \dodoi{10.1146/annurev-astro-082812-140953}

\bibitem[{{Carniani} {et~al.}(2016){Carniani}, {Marconi}, {Maiolino},
  {Balmaverde}, {Brusa}, {Cano-D{\'\i}az}, {Cicone}, {Comastri}, {Cresci},
  {Fiore}, {Feruglio}, {La Franca}, {Mainieri}, {Mannucci}, {Nagao}, {Netzer},
  {Piconcelli}, {Risaliti}, {Schneider}, \& {Shemmer}}]{2016A&A...591A..28C}
{Carniani}, S., {Marconi}, A., {Maiolino}, R., {et~al.} 2016, \aap, 591, A28,
  \dodoi{10.1051/0004-6361/201528037}

\bibitem[{{Cicone} {et~al.}(2014){Cicone}, {Maiolino}, {Sturm},
  {Graci{\'a}-Carpio}, {Feruglio}, {Neri}, {Aalto}, {Davies}, {Fiore},
  {Fischer}, {Garc{\'\i}a-Burillo}, {Gonz{\'a}lez-Alfonso}, {Hailey-Dunsheath},
  {Piconcelli}, \& {Veilleux}}]{2014A&A...562A..21C}
{Cicone}, C., {Maiolino}, R., {Sturm}, E., {et~al.} 2014, \aap, 562, A21,
  \dodoi{10.1051/0004-6361/201322464}

\bibitem[{{Cicone} {et~al.}(2015){Cicone}, {Maiolino}, {Gallerani}, {Neri},
  {Ferrara}, {Sturm}, {Fiore}, {Piconcelli}, \&
  {Feruglio}}]{2015A&A...574A..14C}
{Cicone}, C., {Maiolino}, R., {Gallerani}, S., {et~al.} 2015, \aap, 574, A14,
  \dodoi{10.1051/0004-6361/201424980}

\bibitem[{{Ciotti} {et~al.}(2010){Ciotti}, {Ostriker}, \&
  {Proga}}]{2010ApJ...717..708C}
{Ciotti}, L., {Ostriker}, J.~P., \& {Proga}, D. 2010, \apj, 717, 708,
  \dodoi{10.1088/0004-637X/717/2/708}

\bibitem[{{Costa} {et~al.}(2018{\natexlab{a}}){Costa}, {Rosdahl}, {Sijacki}, \&
  {Haehnelt}}]{2018MNRAS.479.2079C}
{Costa}, T., {Rosdahl}, J., {Sijacki}, D., \& {Haehnelt}, M.~G.
  2018{\natexlab{a}}, \mnras, 479, 2079, \dodoi{10.1093/mnras/sty1514}

\bibitem[{{Costa} {et~al.}(2018{\natexlab{b}}){Costa}, {Rosdahl}, {Sijacki}, \&
  {Haehnelt}}]{2018MNRAS.473.4197C}
---. 2018{\natexlab{b}}, \mnras, 473, 4197, \dodoi{10.1093/mnras/stx2598}

\bibitem[{{Costa} {et~al.}(2014){Costa}, {Sijacki}, \&
  {Haehnelt}}]{2014MNRAS.444.2355C}
{Costa}, T., {Sijacki}, D., \& {Haehnelt}, M.~G. 2014, \mnras, 444, 2355,
  \dodoi{10.1093/mnras/stu1632}

\bibitem[{{da Cunha} {et~al.}(2013){da Cunha}, {Groves}, {Walter}, {Decarli},
  {Weiss}, {Bertoldi}, {Carilli}, {Daddi}, {Elbaz}, {Ivison}, {Maiolino},
  {Riechers}, {Rix}, {Sargent}, \& {Smail}}]{2013ApJ...766...13D}
{da Cunha}, E., {Groves}, B., {Walter}, F., {et~al.} 2013, \apj, 766, 13,
  \dodoi{10.1088/0004-637X/766/1/13}

\bibitem[{{Daddi} {et~al.}(2007){Daddi}, {Dickinson}, {Morrison}, {Chary},
  {Cimatti}, {Elbaz}, {Frayer}, {Renzini}, {Pope}, {Alexander}, {Bauer},
  {Giavalisco}, {Huynh}, {Kurk}, \& {Mignoli}}]{2007ApJ...670..156D}
{Daddi}, E., {Dickinson}, M., {Morrison}, G., {et~al.} 2007, \apj, 670, 156,
  \dodoi{10.1086/521818}

\bibitem[{{De Looze} {et~al.}(2014){De Looze}, {Cormier}, {Lebouteiller},
  {Madden}, {Baes}, {Bendo}, {Boquien}, {Boselli}, {Clements}, {Cortese},
  {Cooray}, {Galametz}, {Galliano}, {Graci{\'a}-Carpio}, {Isaak}, {Karczewski},
  {Parkin}, {Pellegrini}, {R{\'e}my-Ruyer}, {Spinoglio}, {Smith}, \&
  {Sturm}}]{2014A&A...568A..62D}
{De Looze}, I., {Cormier}, D., {Lebouteiller}, V., {et~al.} 2014, \aap, 568,
  A62, \dodoi{10.1051/0004-6361/201322489}

\bibitem[{{De Rosa} {et~al.}(2014){De Rosa}, {Venemans}, {Decarli}, {Gennaro},
  {Simcoe}, {Dietrich}, {Peterson}, {Walter}, {Frank}, {McMahon}, {Hewett},
  {Mortlock}, \& {Simpson}}]{2014ApJ...790..145D}
{De Rosa}, G., {Venemans}, B.~P., {Decarli}, R., {et~al.} 2014, \apj, 790, 145,
  \dodoi{10.1088/0004-637X/790/2/145}

\bibitem[{{Decarli} {et~al.}(2017){Decarli}, {Walter}, {Venemans},
  {Ba{\~n}ados}, {Bertoldi}, {Carilli}, {Fan}, {Farina}, {Mazzucchelli},
  {Riechers}, {Rix}, {Strauss}, {Wang}, \& {Yang}}]{2017Natur.545..457D}
{Decarli}, R., {Walter}, F., {Venemans}, B.~P., {et~al.} 2017, \nat, 545, 457,
  \dodoi{10.1038/nature22358}

\bibitem[{{Decarli} {et~al.}(2018){Decarli}, {Walter}, {Venemans},
  {Ba{\~n}ados}, {Bertoldi}, {Carilli}, {Fan}, {Farina}, {Mazzucchelli},
  {Riechers}, {Rix}, {Strauss}, {Wang}, \& {Yang}}]{2018ApJ...854...97D}
---. 2018, \apj, 854, 97, \dodoi{10.3847/1538-4357/aaa5aa}

\bibitem[{{Di Matteo} {et~al.}(2005){Di Matteo}, {Springel}, \&
  {Hernquist}}]{2005Natur.433..604D}
{Di Matteo}, T., {Springel}, V., \& {Hernquist}, L. 2005, \nat, 433, 604,
  \dodoi{10.1038/nature03335}

\bibitem[{{Di Teodoro} \& {Fraternali}(2015)}]{2015MNRAS.451.3021D}
{Di Teodoro}, E.~M., \& {Fraternali}, F. 2015, \mnras, 451, 3021,
  \dodoi{10.1093/mnras/stv1213}

\bibitem[{{D{\'\i}az-Santos} {et~al.}(2013){D{\'\i}az-Santos}, {Armus},
  {Charmandaris}, {Stierwalt}, {Murphy}, {Haan}, {Inami}, {Malhotra},
  {Meijerink}, {Stacey}, {Petric}, {Evans}, {Veilleux}, {van der Werf}, {Lord},
  {Lu}, {Howell}, {Appleton}, {Mazzarella}, {Surace}, {Xu}, {Schulz},
  {Sanders}, {Bridge}, {Chan}, {Frayer}, {Iwasawa}, {Melbourne}, \&
  {Sturm}}]{2013ApJ...774...68D}
{D{\'\i}az-Santos}, T., {Armus}, L., {Charmandaris}, V., {et~al.} 2013, \apj,
  774, 68, \dodoi{10.1088/0004-637X/774/1/68}

\bibitem[{{D{\'\i}az-Santos} {et~al.}(2016){D{\'\i}az-Santos}, {Assef},
  {Blain}, {Tsai}, {Aravena}, {Eisenhardt}, {Wu}, {Stern}, \&
  {Bridge}}]{2016ApJ...816L...6D}
{D{\'\i}az-Santos}, T., {Assef}, R.~J., {Blain}, A.~W., {et~al.} 2016, \apjl,
  816, L6, \dodoi{10.3847/2041-8205/816/1/L6}

\bibitem[{{D{\'\i}az-Santos} {et~al.}(2018){D{\'\i}az-Santos}, {Assef},
  {Blain}, {Aravena}, {Stern}, {Tsai}, {Eisenhardt}, {Wu}, {Jun}, {Dibert},
  {Inami}, {Lansbury}, \& {Leclercq}}]{2018Sci...362.1034D}
---. 2018, Science, 362, 1034, \dodoi{10.1126/science.aap7605}

\bibitem[{{Dudzevi{\v{c}}i{\={u}}t{\.{e}}}
  {et~al.}(2020){Dudzevi{\v{c}}i{\={u}}t{\.{e}}}, {Smail}, {Swinbank}, {Stach},
  {Almaini}, {da Cunha}, {An}, {Arumugam}, {Birkin}, {Blain}, {Chapman},
  {Chen}, {Conselice}, {Coppin}, {Dunlop}, {Farrah}, {Geach}, {Gullberg},
  {Hartley}, {Hodge}, {Ivison}, {Maltby}, {Scott}, {Simpson}, {Simpson},
  {Thomson}, {Walter}, {Wardlow}, {Weiss}, \& {van der
  Werf}}]{2020MNRAS.494.3828D}
{Dudzevi{\v{c}}i{\={u}}t{\.{e}}}, U., {Smail}, I., {Swinbank}, A.~M., {et~al.}
  2020, \mnras, 494, 3828, \dodoi{10.1093/mnras/staa769}

\bibitem[{{Dunne} {et~al.}(2000){Dunne}, {Eales}, {Edmunds}, {Ivison},
  {Alexander}, \& {Clements}}]{2000MNRAS.315..115D}
{Dunne}, L., {Eales}, S., {Edmunds}, M., {et~al.} 2000, \mnras, 315, 115,
  \dodoi{10.1046/j.1365-8711.2000.03386.x}

\bibitem[{{Ellison} {et~al.}(2011){Ellison}, {Patton}, {Mendel}, \&
  {Scudder}}]{2011MNRAS.418.2043E}
{Ellison}, S.~L., {Patton}, D.~R., {Mendel}, J.~T., \& {Scudder}, J.~M. 2011,
  \mnras, 418, 2043, \dodoi{10.1111/j.1365-2966.2011.19624.x}

\bibitem[{{Estrada-Carpenter} {et~al.}(2020){Estrada-Carpenter}, {Papovich},
  {Momcheva}, {Brammer}, {Simons}, {Bridge}, {Cleri}, {Ferguson},
  {Finkelstein}, {Giavalisco}, {Jung}, {Matharu}, {Trump}, \&
  {Weiner}}]{2020ApJ...898..171E}
{Estrada-Carpenter}, V., {Papovich}, C., {Momcheva}, I., {et~al.} 2020, \apj,
  898, 171, \dodoi{10.3847/1538-4357/aba004}

\bibitem[{{Fan} {et~al.}(2001){Fan}, {Narayanan}, {Lupton}, {Strauss}, {Knapp},
  {Becker}, {White}, {Pentericci}, {Leggett}, {Haiman}, {Gunn}, {Ivezi{\'c}},
  {Schneider}, {Anderson}, {Brinkmann}, {Bahcall}, {Connolly}, {Csabai}, {Doi},
  {Fukugita}, {Geballe}, {Grebel}, {Harbeck}, {Hennessy}, {Lamb}, {Miknaitis},
  {Munn}, {Nichol}, {Okamura}, {Pier}, {Prada}, {Richards}, {Szalay}, \&
  {York}}]{2001AJ....122.2833F}
{Fan}, X., {Narayanan}, V.~K., {Lupton}, R.~H., {et~al.} 2001, \aj, 122, 2833,
  \dodoi{10.1086/324111}

\bibitem[{{Fan} {et~al.}(2003){Fan}, {Strauss}, {Schneider}, {Becker}, {White},
  {Haiman}, {Gregg}, {Pentericci}, {Grebel}, {Narayanan}, {Loh}, {Richards},
  {Gunn}, {Lupton}, {Knapp}, {Ivezi{\'c}}, {Brandt}, {Collinge}, {Hao},
  {Harbeck}, {Prada}, {Schaye}, {Strateva}, {Zakamska}, {Anderson},
  {Brinkmann}, {Bahcall}, {Lamb}, {Okamura}, {Szalay}, \&
  {York}}]{2003AJ....125.1649F}
{Fan}, X., {Strauss}, M.~A., {Schneider}, D.~P., {et~al.} 2003, \aj, 125, 1649,
  \dodoi{10.1086/368246}

\bibitem[{{Ferrarese} \& {Merritt}(2000)}]{2000ApJ...539L...9F}
{Ferrarese}, L., \& {Merritt}, D. 2000, \apjl, 539, L9, \dodoi{10.1086/312838}

\bibitem[{{Fiore} {et~al.}(2017){Fiore}, {Feruglio}, {Shankar}, {Bischetti},
  {Bongiorno}, {Brusa}, {Carniani}, {Cicone}, {Duras}, {Lamastra}, {Mainieri},
  {Marconi}, {Menci}, {Maiolino}, {Piconcelli}, {Vietri}, \&
  {Zappacosta}}]{2017A&A...601A.143F}
{Fiore}, F., {Feruglio}, C., {Shankar}, F., {et~al.} 2017, \aap, 601, A143,
  \dodoi{10.1051/0004-6361/201629478}

\bibitem[{{Fluetsch} {et~al.}(2019){Fluetsch}, {Maiolino}, {Carniani},
  {Marconi}, {Cicone}, {Bourne}, {Costa}, {Fabian}, {Ishibashi}, \&
  {Venturi}}]{2019MNRAS.483.4586F}
{Fluetsch}, A., {Maiolino}, R., {Carniani}, S., {et~al.} 2019, \mnras, 483,
  4586, \dodoi{10.1093/mnras/sty3449}

\bibitem[{{Fujimoto} {et~al.}(2016){Fujimoto}, {Ouchi}, {Ono}, {Shibuya},
  {Ishigaki}, {Nagai}, \& {Momose}}]{2016ApJS..222....1F}
{Fujimoto}, S., {Ouchi}, M., {Ono}, Y., {et~al.} 2016, \apjs, 222, 1,
  \dodoi{10.3847/0067-0049/222/1/1}

\bibitem[{{Fujimoto} {et~al.}(2019){Fujimoto}, {Ouchi}, {Ferrara},
  {Pallottini}, {Ivison}, {Behrens}, {Gallerani}, {Arata}, {Yajima}, \&
  {Nagamine}}]{2019ApJ...887..107F}
{Fujimoto}, S., {Ouchi}, M., {Ferrara}, A., {et~al.} 2019, \apj, 887, 107,
  \dodoi{10.3847/1538-4357/ab480f}

\bibitem[{{Fujimoto} {et~al.}(2020){Fujimoto}, {Silverman}, {Bethermin},
  {Ginolfi}, {Jones}, {Le F{\`e}vre}, {Dessauges-Zavadsky}, {Rujopakarn},
  {Faisst}, {Fudamoto}, {Cassata}, {Morselli}, {Maiolino}, {Schaerer}, {Capak},
  {Yan}, {Vallini}, {Toft}, {Loiacono}, {Zamorani}, {Talia}, {Narayanan},
  {Hathi}, {Lemaux}, {Boquien}, {Amorin}, {Ibar}, {Koekemoer},
  {M{\'e}ndez-Hern{\'a}ndez}, {Bardelli}, {Vergani}, {Zucca}, {Romano}, \&
  {Cimatti}}]{2020ApJ...900....1F}
{Fujimoto}, S., {Silverman}, J.~D., {Bethermin}, M., {et~al.} 2020, \apj, 900,
  1, \dodoi{10.3847/1538-4357/ab94b3}

\bibitem[{{Gallerani} {et~al.}(2017){Gallerani}, {Fan}, {Maiolino}, \&
  {Pacucci}}]{2017PASA...34...22G}
{Gallerani}, S., {Fan}, X., {Maiolino}, R., \& {Pacucci}, F. 2017, \pasa, 34,
  e022, \dodoi{10.1017/pasa.2017.14}

\bibitem[{{Gallerani} {et~al.}(2018){Gallerani}, {Pallottini}, {Feruglio},
  {Ferrara}, {Maiolino}, {Vallini}, {Riechers}, \&
  {Pavesi}}]{2018MNRAS.473.1909G}
{Gallerani}, S., {Pallottini}, A., {Feruglio}, C., {et~al.} 2018, \mnras, 473,
  1909, \dodoi{10.1093/mnras/stx2458}

\bibitem[{{Garc{\'\i}a-Burillo} {et~al.}(2015){Garc{\'\i}a-Burillo}, {Combes},
  {Usero}, {Aalto}, {Colina}, {Alonso-Herrero}, {Hunt}, {Arribas},
  {Costagliola}, {Labiano}, {Neri}, {Pereira-Santaella}, {Tacconi}, \& {van der
  Werf}}]{2015A&A...580A..35G}
{Garc{\'\i}a-Burillo}, S., {Combes}, F., {Usero}, A., {et~al.} 2015, \aap, 580,
  A35, \dodoi{10.1051/0004-6361/201526133}

\bibitem[{{Ginolfi} {et~al.}(2020){Ginolfi}, {Jones}, {B{\'e}thermin},
  {Fudamoto}, {Loiacono}, {Fujimoto}, {Le F{\'e}vre}, {Faisst}, {Schaerer},
  {Cassata}, {Silverman}, {Yan}, {Capak}, {Bardelli}, {Boquien}, {Carraro},
  {Dessauges-Zavadsky}, {Giavalisco}, {Gruppioni}, {Ibar}, {Khusanova},
  {Lemaux}, {Maiolino}, {Narayanan}, {Oesch}, {Pozzi}, {Rodighiero}, {Talia},
  {Toft}, {Vallini}, {Vergani}, \& {Zamorani}}]{2020A&A...633A..90G}
{Ginolfi}, M., {Jones}, G.~C., {B{\'e}thermin}, M., {et~al.} 2020, \aap, 633,
  A90, \dodoi{10.1051/0004-6361/201936872}

\bibitem[{{Glazebrook} {et~al.}(2017){Glazebrook}, {Schreiber}, {Labb{\'e}},
  {Nanayakkara}, {Kacprzak}, {Oesch}, {Papovich}, {Spitler}, {Straatman},
  {Tran}, \& {Yuan}}]{2017Natur.544...71G}
{Glazebrook}, K., {Schreiber}, C., {Labb{\'e}}, I., {et~al.} 2017, \nat, 544,
  71, \dodoi{10.1038/nature21680}

\bibitem[{{Goulding} {et~al.}(2018){Goulding}, {Greene}, {Bezanson}, {Greco},
  {Johnson}, {Leauthaud}, {Matsuoka}, {Medezinski}, \&
  {Price-Whelan}}]{2018PASJ...70S..37G}
{Goulding}, A.~D., {Greene}, J.~E., {Bezanson}, R., {et~al.} 2018, \pasj, 70,
  S37, \dodoi{10.1093/pasj/psx135}

\bibitem[{{Greene} {et~al.}(2012){Greene}, {Zakamska}, \&
  {Smith}}]{2012ApJ...746...86G}
{Greene}, J.~E., {Zakamska}, N.~L., \& {Smith}, P.~S. 2012, \apj, 746, 86,
  \dodoi{10.1088/0004-637X/746/1/86}

\bibitem[{{Habouzit} {et~al.}(2021){Habouzit}, {Li}, {Somerville}, {Genel},
  {Pillepich}, {Volonteri}, {Dav{\'e}}, {Rosas-Guevara}, {McAlpine}, {Peirani},
  {Hernquist}, {Angl{\'e}s-Alc{\'a}zar}, {Reines}, {Bower}, {Dubois}, {Nelson},
  {Pichon}, \& {Vogelsberger}}]{2020arXiv200610094H}
{Habouzit}, M., {Li}, Y., {Somerville}, R.~S., {et~al.} 2021, \mnras, 503,
  1940, \dodoi{10.1093/mnras/stab496}

\bibitem[{{Hailey-Dunsheath} {et~al.}(2010){Hailey-Dunsheath}, {Nikola},
  {Stacey}, {Oberst}, {Parshley}, {Benford}, {Staguhn}, \&
  {Tucker}}]{2010ApJ...714L.162H}
{Hailey-Dunsheath}, S., {Nikola}, T., {Stacey}, G.~J., {et~al.} 2010, \apjl,
  714, L162, \dodoi{10.1088/2041-8205/714/1/L162}

\bibitem[{{Hollenbach} \& {Tielens}(1997)}]{1997ARA&A..35..179H}
{Hollenbach}, D.~J., \& {Tielens}, A.~G.~G.~M. 1997, \araa, 35, 179,
  \dodoi{10.1146/annurev.astro.35.1.179}

\bibitem[{{Hopkins} {et~al.}(2006){Hopkins}, {Hernquist}, {Cox}, {Di Matteo},
  {Robertson}, \& {Springel}}]{2006ApJS..163....1H}
{Hopkins}, P.~F., {Hernquist}, L., {Cox}, T.~J., {et~al.} 2006, \apjs, 163, 1,
  \dodoi{10.1086/499298}

\bibitem[{{Hopkins} \& {Quataert}(2010)}]{2010MNRAS.407.1529H}
{Hopkins}, P.~F., \& {Quataert}, E. 2010, \mnras, 407, 1529,
  \dodoi{10.1111/j.1365-2966.2010.17064.x}

\bibitem[{{Hwang} {et~al.}(2010){Hwang}, {Elbaz}, {Magdis}, {Daddi},
  {Symeonidis}, {Altieri}, {Amblard}, {Andreani}, {Arumugam}, {Auld}, {Aussel},
  {Babbedge}, {Berta}, {Blain}, {Bock}, {Bongiovanni}, {Boselli}, {Buat},
  {Burgarella}, {Castro-Rodr{\'\i}guez}, {Cava}, {Cepa}, {Chanial}, {Chapin},
  {Chary}, {Cimatti}, {Clements}, {Conley}, {Conversi}, {Cooray},
  {Dannerbauer}, {Dickinson}, {Dominguez}, {Dowell}, {Dunlop}, {Dwek}, {Eales},
  {Farrah}, {F{\"o}rster Schreiber}, {Fox}, {Franceschini}, {Gear}, {Genzel},
  {Glenn}, {Griffin}, {Gruppioni}, {Halpern}, {Hatziminaoglou}, {Ibar},
  {Isaak}, {Ivison}, {Jeong}, {Lagache}, {Le Borgne}, {Le Floc'h}, {Lee},
  {Lee}, {Lee}, {Levenson}, {Lu}, {Lutz}, {Madden}, {Maffei}, {Magnelli},
  {Mainetti}, {Maiolino}, {Marchetti}, {Mortier}, {Nguyen}, {Nordon},
  {O'Halloran}, {Okumura}, {Oliver}, {Omont}, {Page}, {Panuzzo},
  {Papageorgiou}, {Pearson}, {P{\'e}rez-Fournon}, {Garc{\'\i}a}, {Poglitsch},
  {Pohlen}, {Popesso}, {Pozzi}, {Rawlings}, {Rigopoulou}, {Riguccini}, {Rizzo},
  {Rodighiero}, {Roseboom}, {Rowan-Robinson}, {Saintonge}, {Portal}, {Santini},
  {Sauvage}, {Schulz}, {Scott}, {Seymour}, {Shao}, {Shupe}, {Smith}, {Stevens},
  {Sturm}, {Tacconi}, {Trichas}, {Tugwell}, {Vaccari}, {Valtchanov}, {Vieira},
  {Vigroux}, {Wang}, {Ward}, {Wright}, {Xu}, \& {Zemcov}}]{2010MNRAS.409...75H}
{Hwang}, H.~S., {Elbaz}, D., {Magdis}, G., {et~al.} 2010, \mnras, 409, 75,
  \dodoi{10.1111/j.1365-2966.2010.17645.x}

\bibitem[{{Ikarashi} {et~al.}(2015){Ikarashi}, {Ivison}, {Caputi}, {Aretxaga},
  {Dunlop}, {Hatsukade}, {Hughes}, {Iono}, {Izumi}, {Kawabe}, {Kohno}, {Lagos},
  {Motohara}, {Nakanishi}, {Ohta}, {Tamura}, {Umehata}, {Wilson}, {Yabe}, \&
  {Yun}}]{2015ApJ...810..133I}
{Ikarashi}, S., {Ivison}, R.~J., {Caputi}, K.~I., {et~al.} 2015, \apj, 810,
  133, \dodoi{10.1088/0004-637X/810/2/133}

\bibitem[{{Ikarashi} {et~al.}(2017){Ikarashi}, {Ivison}, {Caputi}, {Nakanishi},
  {Lagos}, {Ashby}, {Aretxaga}, {Dunlop}, {Hatsukade}, {Hughes}, {Iono},
  {Izumi}, {Kawabe}, {Kohno}, {Motohara}, {Ohta}, {Tamura}, {Umehata},
  {Wilson}, {Yabe}, \& {Yun}}]{2017ApJ...835..286I}
---. 2017, \apj, 835, 286, \dodoi{10.3847/1538-4357/835/2/286}

\bibitem[{{Inayoshi} {et~al.}(2020){Inayoshi}, {Visbal}, \&
  {Haiman}}]{2020ARA&A..58...27I}
{Inayoshi}, K., {Visbal}, E., \& {Haiman}, Z. 2020, \araa, 58, 27,
  \dodoi{10.1146/annurev-astro-120419-014455}

\bibitem[{{Ishibashi} \& {Fabian}(2015)}]{2015MNRAS.451...93I}
{Ishibashi}, W., \& {Fabian}, A.~C. 2015, \mnras, 451, 93,
  \dodoi{10.1093/mnras/stv944}

\bibitem[{{Ishibashi} \& {Fabian}(2016)}]{2016MNRAS.463.1291I}
---. 2016, \mnras, 463, 1291, \dodoi{10.1093/mnras/stw2063}

\bibitem[{{Ishiyama} {et~al.}(2015){Ishiyama}, {Enoki}, {Kobayashi}, {Makiya},
  {Nagashima}, \& {Oogi}}]{2015PASJ...67...61I}
{Ishiyama}, T., {Enoki}, M., {Kobayashi}, M. A.~R., {et~al.} 2015, \pasj, 67,
  61, \dodoi{10.1093/pasj/psv021}

\bibitem[{{Izumi} {et~al.}(2018){Izumi}, {Onoue}, {Shirakata}, {Nagao},
  {Kohno}, {Matsuoka}, {Imanishi}, {Strauss}, {Kashikawa}, {Schulze},
  {Silverman}, {Fujimoto}, {Harikane}, {Toba}, {Umehata}, {Nakanishi},
  {Greene}, {Tamura}, {Taniguchi}, {Yamaguchi}, {Goto}, {Hashimoto},
  {Ikarashi}, {Iono}, {Iwasawa}, {Lee}, {Makiya}, {Minezaki}, \&
  {Tang}}]{2018PASJ...70...36I}
{Izumi}, T., {Onoue}, M., {Shirakata}, H., {et~al.} 2018, \pasj, 70, 36,
  \dodoi{10.1093/pasj/psy026}

\bibitem[{{Izumi} {et~al.}(2019){Izumi}, {Onoue}, {Matsuoka}, {Nagao},
  {Strauss}, {Imanishi}, {Kashikawa}, {Fujimoto}, {Kohno}, {Toba}, {Umehata},
  {Goto}, {Ueda}, {Shirakata}, {Silverman}, {Greene}, {Harikane}, {Hashimoto},
  {Ikarashi}, {Iono}, {Iwasawa}, {Lee}, {Minezaki}, {Nakanishi}, {Tamura},
  {Tang}, \& {Taniguchi}}]{2019PASJ...71..111I}
{Izumi}, T., {Onoue}, M., {Matsuoka}, Y., {et~al.} 2019, \pasj, 71, 111,
  \dodoi{10.1093/pasj/psz096}

\bibitem[{{Izumi} {et~al.}(2020){Izumi}, {Nguyen}, {Imanishi}, {Kawamuro},
  {Baba}, {Nakano}, {Kohno}, {Matsushita}, {Meier}, {Turner}, {Michiyama},
  {Harada}, {Mart{\'\i}n}, {Nakanishi}, {Takano}, {Wiklind}, {Nakai}, \&
  {Hsieh}}]{2020ApJ...898...75I}
{Izumi}, T., {Nguyen}, D.~D., {Imanishi}, M., {et~al.} 2020, \apj, 898, 75,
  \dodoi{10.3847/1538-4357/ab9cb1}

\bibitem[{{Izumi} {et~al.}(2021){Izumi}, {Onoue}, {Matsuoka}, {Strauss},
  {Fujimoto}, {Umehata}, {Imanishi}, {Kawamuro}, {Nagao}, {Toba}, {Kohno},
  {Kashikawa}, {Inayoshi}, {Kawaguchi}, {Iwasawa}, {Inoue}, {Goto}, {Baba},
  {Schramm}, {Suh}, {Harikane}, {Ueda}, {Silverman}, {Hashimoto}, {Hashimoto},
  {Ikarashi}, {Iono}, {Lee}, {Lee}, {Minezaki}, {Nakanishi}, {Nakano},
  {Tamura}, \& {Tang}}]{2021arXiv210101199I}
{Izumi}, T., {Onoue}, M., {Matsuoka}, Y., {et~al.} 2021, \apj, 908, 235,
  \dodoi{10.3847/1538-4357/abd7ef}

\bibitem[{{Jiang} {et~al.}(2016){Jiang}, {McGreer}, {Fan}, {Strauss},
  {Ba{\~n}ados}, {Becker}, {Bian}, {Farnsworth}, {Shen}, {Wang}, {Wang},
  {Wang}, {White}, {Wu}, {Wu}, {Yang}, \& {Yang}}]{2016ApJ...833..222J}
{Jiang}, L., {McGreer}, I.~D., {Fan}, X., {et~al.} 2016, \apj, 833, 222,
  \dodoi{10.3847/1538-4357/833/2/222}

\bibitem[{{Kato} {et~al.}(2020){Kato}, {Matsuoka}, {Onoue}, {Koyama}, {Toba},
  {Akiyama}, {Fujimoto}, {Imanishi}, {Iwasawa}, {Izumi}, {Kashikawa},
  {Kawaguchi}, {Lee}, {Minezaki}, {Nagao}, {Noboriguchi}, \&
  {Strauss}}]{2020PASJ...72...84K}
{Kato}, N., {Matsuoka}, Y., {Onoue}, M., {et~al.} 2020, \pasj, 72, 84,
  \dodoi{10.1093/pasj/psaa074}

\bibitem[{{Kennicutt}(1998)}]{1998ARA&A..36..189K}
{Kennicutt}, Robert~C., J. 1998, \araa, 36, 189,
  \dodoi{10.1146/annurev.astro.36.1.189}

\bibitem[{{Kim} \& {Im}(2019)}]{2019ApJ...879..117K}
{Kim}, Y., \& {Im}, M. 2019, \apj, 879, 117, \dodoi{10.3847/1538-4357/ab25ee}

\bibitem[{{King}(2003)}]{2003ApJ...596L..27K}
{King}, A. 2003, \apjl, 596, L27, \dodoi{10.1086/379143}

\bibitem[{{King} \& {Pounds}(2015)}]{2015ARA&A..53..115K}
{King}, A., \& {Pounds}, K. 2015, \araa, 53, 115,
  \dodoi{10.1146/annurev-astro-082214-122316}

\bibitem[{{Kormendy} \& {Ho}(2013)}]{2013ARA&A..51..511K}
{Kormendy}, J., \& {Ho}, L.~C. 2013, \araa, 51, 511,
  \dodoi{10.1146/annurev-astro-082708-101811}

\bibitem[{{Kroupa}(2001)}]{2001MNRAS.322..231K}
{Kroupa}, P. 2001, \mnras, 322, 231, \dodoi{10.1046/j.1365-8711.2001.04022.x}

\bibitem[{{Lauer} {et~al.}(2007){Lauer}, {Tremaine}, {Richstone}, \&
  {Faber}}]{2007ApJ...670..249L}
{Lauer}, T.~R., {Tremaine}, S., {Richstone}, D., \& {Faber}, S.~M. 2007, \apj,
  670, 249, \dodoi{10.1086/522083}

\bibitem[{{Le F{\`e}vre} {et~al.}(2020){Le F{\`e}vre}, {B{\'e}thermin},
  {Faisst}, {Jones}, {Capak}, {Cassata}, {Silverman}, {Schaerer}, {Yan},
  {Amorin}, {Bardelli}, {Boquien}, {Cimatti}, {Dessauges-Zavadsky},
  {Giavalisco}, {Hathi}, {Fudamoto}, {Fujimoto}, {Ginolfi}, {Gruppioni},
  {Hemmati}, {Ibar}, {Koekemoer}, {Khusanova}, {Lagache}, {Lemaux}, {Loiacono},
  {Maiolino}, {Mancini}, {Narayanan}, {Morselli}, {M{\'e}ndez-Hern{\`a}ndez},
  {Oesch}, {Pozzi}, {Romano}, {Riechers}, {Scoville}, {Talia}, {Tasca},
  {Thomas}, {Toft}, {Vallini}, {Vergani}, {Walter}, {Zamorani}, \&
  {Zucca}}]{2020A&A...643A...1L}
{Le F{\`e}vre}, O., {B{\'e}thermin}, M., {Faisst}, A., {et~al.} 2020, \aap,
  643, A1, \dodoi{10.1051/0004-6361/201936965}

\bibitem[{{Leipski} {et~al.}(2013){Leipski}, {Meisenheimer}, {Walter}, {Besel},
  {Dannerbauer}, {Fan}, {Haas}, {Klaas}, {Krause}, \&
  {Rix}}]{2013ApJ...772..103L}
{Leipski}, C., {Meisenheimer}, K., {Walter}, F., {et~al.} 2013, \apj, 772, 103,
  \dodoi{10.1088/0004-637X/772/2/103}

\bibitem[{{Leipski} {et~al.}(2014){Leipski}, {Meisenheimer}, {Walter}, {Klaas},
  {Dannerbauer}, {De Rosa}, {Fan}, {Haas}, {Krause}, \&
  {Rix}}]{2014ApJ...785..154L}
---. 2014, \apj, 785, 154, \dodoi{10.1088/0004-637X/785/2/154}

\bibitem[{{Liang} {et~al.}(2019){Liang}, {Feldmann}, {Kere{\v{s}}}, {Scoville},
  {Hayward}, {Faucher-Gigu{\`e}re}, {Schreiber}, {Ma}, {Hopkins}, \&
  {Quataert}}]{2019MNRAS.489.1397L}
{Liang}, L., {Feldmann}, R., {Kere{\v{s}}}, D., {et~al.} 2019, \mnras, 489,
  1397, \dodoi{10.1093/mnras/stz2134}

\bibitem[{{Liu} {et~al.}(2013){Liu}, {Zakamska}, {Greene}, {Nesvadba}, \&
  {Liu}}]{2013MNRAS.436.2576L}
{Liu}, G., {Zakamska}, N.~L., {Greene}, J.~E., {Nesvadba}, N. P.~H., \& {Liu},
  X. 2013, \mnras, 436, 2576, \dodoi{10.1093/mnras/stt1755}

\bibitem[{{Lupi} {et~al.}(2019){Lupi}, {Volonteri}, {Decarli}, {Bovino},
  {Silk}, \& {Bergeron}}]{2019MNRAS.488.4004L}
{Lupi}, A., {Volonteri}, M., {Decarli}, R., {et~al.} 2019, \mnras, 488, 4004,
  \dodoi{10.1093/mnras/stz1959}

\bibitem[{{Lutz} {et~al.}(2020){Lutz}, {Sturm}, {Janssen}, {Veilleux}, {Aalto},
  {Cicone}, {Contursi}, {Davies}, {Feruglio}, {Fischer}, {Fluetsch},
  {Garcia-Burillo}, {Genzel}, {Gonz{\'a}lez-Alfonso}, {Graci{\'a}-Carpio},
  {Herrera-Camus}, {Maiolino}, {Schruba}, {Shimizu}, {Sternberg}, {Tacconi}, \&
  {Wei{\ss}}}]{2020A&A...633A.134L}
{Lutz}, D., {Sturm}, E., {Janssen}, A., {et~al.} 2020, \aap, 633, A134,
  \dodoi{10.1051/0004-6361/201936803}

\bibitem[{{Madau} \& {Dickinson}(2014)}]{2014ARA&A..52..415M}
{Madau}, P., \& {Dickinson}, M. 2014, \araa, 52, 415,
  \dodoi{10.1146/annurev-astro-081811-125615}

\bibitem[{{Magorrian} {et~al.}(1998){Magorrian}, {Tremaine}, {Richstone},
  {Bender}, {Bower}, {Dressler}, {Faber}, {Gebhardt}, {Green}, {Grillmair},
  {Kormendy}, \& {Lauer}}]{1998AJ....115.2285M}
{Magorrian}, J., {Tremaine}, S., {Richstone}, D., {et~al.} 1998, \aj, 115,
  2285, \dodoi{10.1086/300353}

\bibitem[{{Maiolino} {et~al.}(2012){Maiolino}, {Gallerani}, {Neri}, {Cicone},
  {Ferrara}, {Genzel}, {Lutz}, {Sturm}, {Tacconi}, {Walter}, {Feruglio},
  {Fiore}, \& {Piconcelli}}]{2012MNRAS.425L..66M}
{Maiolino}, R., {Gallerani}, S., {Neri}, R., {et~al.} 2012, \mnras, 425, L66,
  \dodoi{10.1111/j.1745-3933.2012.01303.x}

\bibitem[{{Makiya} {et~al.}(2016){Makiya}, {Enoki}, {Ishiyama}, {Kobayashi},
  {Nagashima}, {Okamoto}, {Okoshi}, {Oogi}, \&
  {Shirakata}}]{2016PASJ...68...25M}
{Makiya}, R., {Enoki}, M., {Ishiyama}, T., {et~al.} 2016, \pasj, 68, 25,
  \dodoi{10.1093/pasj/psw005}

\bibitem[{{Malhotra} {et~al.}(1997){Malhotra}, {Helou}, {Stacey}, {Hollenbach},
  {Lord}, {Beichman}, {Dinerstein}, {Hunter}, {Lo}, {Lu}, {Rubin},
  {Silbermann}, {Thronson}, \& {Werner}}]{1997ApJ...491L..27M}
{Malhotra}, S., {Helou}, G., {Stacey}, G., {et~al.} 1997, \apjl, 491, L27,
  \dodoi{10.1086/311044}

\bibitem[{{Marconi} \& {Hunt}(2003)}]{2003ApJ...589L..21M}
{Marconi}, A., \& {Hunt}, L.~K. 2003, \apjl, 589, L21, \dodoi{10.1086/375804}

\bibitem[{{Marshall} {et~al.}(2020){Marshall}, {Ni}, {Di Matteo}, {Wyithe},
  {Wilkins}, {Croft}, \& {Kuusisto}}]{2020MNRAS.499.3819M}
{Marshall}, M.~A., {Ni}, Y., {Di Matteo}, T., {et~al.} 2020, \mnras, 499, 3819,
  \dodoi{10.1093/mnras/staa2982}

\bibitem[{{Martin}(2005)}]{2005ApJ...621..227M}
{Martin}, C.~L. 2005, \apj, 621, 227, \dodoi{10.1086/427277}

\bibitem[{{Matsuoka} {et~al.}(2016){Matsuoka}, {Onoue}, {Kashikawa}, {Iwasawa},
  {Strauss}, {Nagao}, {Imanishi}, {Niida}, {Toba}, {Akiyama}, {Asami}, {Bosch},
  {Foucaud}, {Furusawa}, {Goto}, {Gunn}, {Harikane}, {Ikeda}, {Kawaguchi},
  {Kikuta}, {Komiyama}, {Lupton}, {Minezaki}, {Miyazaki}, {Morokuma},
  {Murayama}, {Nishizawa}, {Ono}, {Ouchi}, {Price}, {Sameshima}, {Silverman},
  {Sugiyama}, {Tait}, {Takada}, {Takata}, {Tanaka}, {Tang}, \&
  {Utsumi}}]{2016ApJ...828...26M}
{Matsuoka}, Y., {Onoue}, M., {Kashikawa}, N., {et~al.} 2016, \apj, 828, 26,
  \dodoi{10.3847/0004-637X/828/1/26}

\bibitem[{{Matsuoka} {et~al.}(2018{\natexlab{a}}){Matsuoka}, {Onoue},
  {Kashikawa}, {Iwasawa}, {Strauss}, {Nagao}, {Imanishi}, {Lee}, {Akiyama},
  {Asami}, {Bosch}, {Foucaud}, {Furusawa}, {Goto}, {Gunn}, {Harikane}, {Ikeda},
  {Izumi}, {Kawaguchi}, {Kikuta}, {Kohno}, {Komiyama}, {Lupton}, {Minezaki},
  {Miyazaki}, {Morokuma}, {Murayama}, {Niida}, {Nishizawa}, {Oguri}, {Ono},
  {Ouchi}, {Price}, {Sameshima}, {Schulze}, {Shirakata}, {Silverman},
  {Sugiyama}, {Tait}, {Takada}, {Takata}, {Tanaka}, {Tang}, {Toba}, {Utsumi},
  \& {Wang}}]{2018PASJ...70S..35M}
---. 2018{\natexlab{a}}, \pasj, 70, S35, \dodoi{10.1093/pasj/psx046}

\bibitem[{{Matsuoka} {et~al.}(2018{\natexlab{b}}){Matsuoka}, {Iwasawa},
  {Onoue}, {Kashikawa}, {Strauss}, {Lee}, {Imanishi}, {Nagao}, {Akiyama},
  {Asami}, {Bosch}, {Furusawa}, {Goto}, {Gunn}, {Harikane}, {Ikeda}, {Izumi},
  {Kawaguchi}, {Kato}, {Kikuta}, {Kohno}, {Komiyama}, {Lupton}, {Minezaki},
  {Miyazaki}, {Morokuma}, {Murayama}, {Niida}, {Nishizawa}, {Oguri}, {Ono},
  {Ouchi}, {Price}, {Sameshima}, {Schulze}, {Shirakata}, {Silverman},
  {Sugiyama}, {Tait}, {Takada}, {Takata}, {Tanaka}, {Tang}, {Toba}, {Utsumi},
  {Wang}, \& {Yamashita}}]{2018ApJS..237....5M}
{Matsuoka}, Y., {Iwasawa}, K., {Onoue}, M., {et~al.} 2018{\natexlab{b}}, \apjs,
  237, 5, \dodoi{10.3847/1538-4365/aac724}

\bibitem[{{Matsuoka} {et~al.}(2018{\natexlab{c}}){Matsuoka}, {Strauss},
  {Kashikawa}, {Onoue}, {Iwasawa}, {Tang}, {Lee}, {Imanishi}, {Nagao},
  {Akiyama}, {Asami}, {Bosch}, {Furusawa}, {Goto}, {Gunn}, {Harikane}, {Ikeda},
  {Izumi}, {Kawaguchi}, {Kato}, {Kikuta}, {Kohno}, {Komiyama}, {Lupton},
  {Minezaki}, {Miyazaki}, {Murayama}, {Niida}, {Nishizawa}, {Noboriguchi},
  {Oguri}, {Ono}, {Ouchi}, {Price}, {Sameshima}, {Schulze}, {Shirakata},
  {Silverman}, {Sugiyama}, {Tait}, {Takada}, {Takata}, {Tanaka}, {Toba},
  {Utsumi}, {Wang}, \& {Yamashita}}]{2018ApJ...869..150M}
{Matsuoka}, Y., {Strauss}, M.~A., {Kashikawa}, N., {et~al.} 2018{\natexlab{c}},
  \apj, 869, 150, \dodoi{10.3847/1538-4357/aaee7a}

\bibitem[{{Matsuoka} {et~al.}(2019{\natexlab{a}}){Matsuoka}, {Iwasawa},
  {Onoue}, {Kashikawa}, {Strauss}, {Lee}, {Imanishi}, {Nagao}, {Akiyama},
  {Asami}, {Bosch}, {Furusawa}, {Goto}, {Gunn}, {Harikane}, {Ikeda}, {Izumi},
  {Kawaguchi}, {Kato}, {Kikuta}, {Kohno}, {Komiyama}, {Koyama}, {Lupton},
  {Minezaki}, {Miyazaki}, {Murayama}, {Niida}, {Nishizawa}, {Noboriguchi},
  {Oguri}, {Ono}, {Ouchi}, {Price}, {Sameshima}, {Schulze}, {Silverman},
  {Sugiyama}, {Tait}, {Takada}, {Takata}, {Tanaka}, {Tang}, {Toba}, {Utsumi},
  {Wang}, \& {Yamashita}}]{2019ApJ...883..183M}
{Matsuoka}, Y., {Iwasawa}, K., {Onoue}, M., {et~al.} 2019{\natexlab{a}}, \apj,
  883, 183, \dodoi{10.3847/1538-4357/ab3c60}

\bibitem[{{Matsuoka} {et~al.}(2019{\natexlab{b}}){Matsuoka}, {Onoue},
  {Kashikawa}, {Strauss}, {Iwasawa}, {Lee}, {Imanishi}, {Nagao}, {Akiyama},
  {Asami}, {Bosch}, {Furusawa}, {Goto}, {Gunn}, {Harikane}, {Ikeda}, {Izumi},
  {Kawaguchi}, {Kato}, {Kikuta}, {Kohno}, {Komiyama}, {Koyama}, {Lupton},
  {Minezaki}, {Miyazaki}, {Murayama}, {Niida}, {Nishizawa}, {Noboriguchi},
  {Oguri}, {Ono}, {Ouchi}, {Price}, {Sameshima}, {Schulze}, {Shirakata},
  {Silverman}, {Sugiyama}, {Tait}, {Takada}, {Takata}, {Tanaka}, {Tang},
  {Toba}, {Utsumi}, {Wang}, \& {Yamashita}}]{2019ApJ...872L...2M}
{Matsuoka}, Y., {Onoue}, M., {Kashikawa}, N., {et~al.} 2019{\natexlab{b}},
  \apjl, 872, L2, \dodoi{10.3847/2041-8213/ab0216}

\bibitem[{{McMullin} {et~al.}(2007){McMullin}, {Waters}, {Schiebel}, {Young},
  \& {Golap}}]{2007ASPC..376..127M}
{McMullin}, J.~P., {Waters}, B., {Schiebel}, D., {Young}, W., \& {Golap}, K.
  2007, in Astronomical Society of the Pacific Conference Series, Vol. 376,
  Astronomical Data Analysis Software and Systems XVI, ed. R.~A. {Shaw},
  F.~{Hill}, \& D.~J. {Bell}, 127

\bibitem[{{Miyazaki} {et~al.}(2012){Miyazaki}, {Komiyama}, {Nakaya}, {Kamata},
  {Doi}, {Hamana}, {Karoji}, {Furusawa}, {Kawanomoto}, {Morokuma}, {Ishizuka},
  {Nariai}, {Tanaka}, {Uraguchi}, {Utsumi}, {Obuchi}, {Okura}, {Oguri},
  {Takata}, {Tomono}, {Kurakami}, {Namikawa}, {Usuda}, {Yamanoi}, {Terai},
  {Uekiyo}, {Yamada}, {Koike}, {Aihara}, {Fujimori}, {Mineo}, {Miyatake},
  {Yasuda}, {Nishizawa}, {Saito}, {Tanaka}, {Uchida}, {Katayama}, {Wang},
  {Chen}, {Lupton}, {Loomis}, {Bickerton}, {Price}, {Gunn}, {Suzuki},
  {Miyazaki}, {Muramatsu}, {Yamamoto}, {Endo}, {Ezaki}, {Itoh}, {Miwa},
  {Yokota}, {Matsuda}, {Ebinuma}, \& {Takeshi}}]{2012SPIE.8446E..0ZM}
{Miyazaki}, S., {Komiyama}, Y., {Nakaya}, H., {et~al.} 2012, in Society of
  Photo-Optical Instrumentation Engineers (SPIE) Conference Series, Vol. 8446,
  Ground-based and Airborne Instrumentation for Astronomy IV, ed. I.~S.
  {McLean}, S.~K. {Ramsay}, \& H.~{Takami}, 84460Z, \dodoi{10.1117/12.926844}

\bibitem[{{Miyazaki} {et~al.}(2018){Miyazaki}, {Komiyama}, {Kawanomoto}, {Doi},
  {Furusawa}, {Hamana}, {Hayashi}, {Ikeda}, {Kamata}, {Karoji}, {Koike},
  {Kurakami}, {Miyama}, {Morokuma}, {Nakata}, {Namikawa}, {Nakaya}, {Nariai},
  {Obuchi}, {Oishi}, {Okada}, {Okura}, {Tait}, {Takata}, {Tanaka}, {Tanaka},
  {Terai}, {Tomono}, {Uraguchi}, {Usuda}, {Utsumi}, {Yamada}, {Yamanoi},
  {Aihara}, {Fujimori}, {Mineo}, {Miyatake}, {Oguri}, {Uchida}, {Tanaka},
  {Yasuda}, {Takada}, {Murayama}, {Nishizawa}, {Sugiyama}, {Chiba}, {Futamase},
  {Wang}, {Chen}, {Ho}, {Liaw}, {Chiu}, {Ho}, {Lai}, {Lee}, {Jeng}, {Iwamura},
  {Armstrong}, {Bickerton}, {Bosch}, {Gunn}, {Lupton}, {Loomis}, {Price},
  {Smith}, {Strauss}, {Turner}, {Suzuki}, {Miyazaki}, {Muramatsu}, {Yamamoto},
  {Endo}, {Ezaki}, {Ito}, {Kawaguchi}, {Sofuku}, {Taniike}, {Akutsu}, {Dojo},
  {Kasumi}, {Matsuda}, {Imoto}, {Miwa}, {Suzuki}, {Takeshi}, \&
  {Yokota}}]{2018PASJ...70S...1M}
{Miyazaki}, S., {Komiyama}, Y., {Kawanomoto}, S., {et~al.} 2018, \pasj, 70, S1,
  \dodoi{10.1093/pasj/psx063}

\bibitem[{{Mortlock} {et~al.}(2011){Mortlock}, {Warren}, {Venemans}, {Patel},
  {Hewett}, {McMahon}, {Simpson}, {Theuns}, {Gonz{\'a}les-Solares}, {Adamson},
  {Dye}, {Hambly}, {Hirst}, {Irwin}, {Kuiper}, {Lawrence}, \&
  {R{\"o}ttgering}}]{2011Natur.474..616M}
{Mortlock}, D.~J., {Warren}, S.~J., {Venemans}, B.~P., {et~al.} 2011, \nat,
  474, 616, \dodoi{10.1038/nature10159}

\bibitem[{{Murphy} {et~al.}(2011){Murphy}, {Condon}, {Schinnerer}, {Kennicutt},
  {Calzetti}, {Armus}, {Helou}, {Turner}, {Aniano}, {Beir{\~a}o}, {Bolatto},
  {Brandl}, {Croxall}, {Dale}, {Donovan Meyer}, {Draine}, {Engelbracht},
  {Hunt}, {Hao}, {Koda}, {Roussel}, {Skibba}, \& {Smith}}]{2011ApJ...737...67M}
{Murphy}, E.~J., {Condon}, J.~J., {Schinnerer}, E., {et~al.} 2011, \apj, 737,
  67, \dodoi{10.1088/0004-637X/737/2/67}

\bibitem[{{Murray} {et~al.}(2005){Murray}, {Quataert}, \&
  {Thompson}}]{2005ApJ...618..569M}
{Murray}, N., {Quataert}, E., \& {Thompson}, T.~A. 2005, \apj, 618, 569,
  \dodoi{10.1086/426067}

\bibitem[{{Neeleman} {et~al.}(2019){Neeleman}, {Ba{\~n}ados}, {Walter},
  {Decarli}, {Venemans}, {Carilli}, {Fan}, {Farina}, {Mazzucchelli}, {Novak},
  {Riechers}, {Rix}, \& {Wang}}]{2019ApJ...882...10N}
{Neeleman}, M., {Ba{\~n}ados}, E., {Walter}, F., {et~al.} 2019, \apj, 882, 10,
  \dodoi{10.3847/1538-4357/ab2ed3}

\bibitem[{{Nesvadba} {et~al.}(2008){Nesvadba}, {Lehnert}, {De Breuck},
  {Gilbert}, \& {van Breugel}}]{2008A&A...491..407N}
{Nesvadba}, N.~P.~H., {Lehnert}, M.~D., {De Breuck}, C., {Gilbert}, A.~M., \&
  {van Breugel}, W. 2008, \aap, 491, 407, \dodoi{10.1051/0004-6361:200810346}

\bibitem[{{Ni} {et~al.}(2018){Ni}, {Di Matteo}, {Feng}, {Croft}, \&
  {Tenneti}}]{2018MNRAS.481.4877N}
{Ni}, Y., {Di Matteo}, T., {Feng}, Y., {Croft}, R. A.~C., \& {Tenneti}, A.
  2018, \mnras, 481, 4877, \dodoi{10.1093/mnras/sty2616}

\bibitem[{{Noeske} {et~al.}(2007){Noeske}, {Weiner}, {Faber}, {Papovich},
  {Koo}, {Somerville}, {Bundy}, {Conselice}, {Newman}, {Schiminovich}, {Le
  Floc'h}, {Coil}, {Rieke}, {Lotz}, {Primack}, {Barmby}, {Cooper}, {Davis},
  {Ellis}, {Fazio}, {Guhathakurta}, {Huang}, {Kassin}, {Martin}, {Phillips},
  {Rich}, {Small}, {Willmer}, \& {Wilson}}]{2007ApJ...660L..43N}
{Noeske}, K.~G., {Weiner}, B.~J., {Faber}, S.~M., {et~al.} 2007, \apjl, 660,
  L43, \dodoi{10.1086/517926}

\bibitem[{{Novak} {et~al.}(2019){Novak}, {Ba{\~n}ados}, {Decarli}, {Walter},
  {Venemans}, {Neeleman}, {Farina}, {Mazzucchelli}, {Carilli}, {Fan}, {Rix}, \&
  {Wang}}]{2019ApJ...881...63N}
{Novak}, M., {Ba{\~n}ados}, E., {Decarli}, R., {et~al.} 2019, \apj, 881, 63,
  \dodoi{10.3847/1538-4357/ab2beb}

\bibitem[{{Novak} {et~al.}(2020){Novak}, {Venemans}, {Walter}, {Neeleman},
  {Kaasinen}, {Liang}, {Feldmann}, {Ba{\~n}ados}, {Carilli}, {Decarli},
  {Drake}, {Fan}, {Farina}, {Mazzucchelli}, {Rix}, \&
  {Wang}}]{2020ApJ...904..131N}
{Novak}, M., {Venemans}, B.~P., {Walter}, F., {et~al.} 2020, \apj, 904, 131,
  \dodoi{10.3847/1538-4357/abc33f}

\bibitem[{{Onoue} {et~al.}(2019){Onoue}, {Kashikawa}, {Matsuoka}, {Kato},
  {Izumi}, {Nagao}, {Strauss}, {Harikane}, {Imanishi}, {Ito}, {Iwasawa},
  {Kawaguchi}, {Lee}, {Noboriguchi}, {Suh}, {Tanaka}, \&
  {Toba}}]{2019ApJ...880...77O}
{Onoue}, M., {Kashikawa}, N., {Matsuoka}, Y., {et~al.} 2019, \apj, 880, 77,
  \dodoi{10.3847/1538-4357/ab29e9}

\bibitem[{{Onoue} {et~al.}(2020){Onoue}, {Ba{\~n}ados}, {Mazzucchelli},
  {Venemans}, {Schindler}, {Walter}, {Hennawi}, {Andika}, {Davies}, {Decarli},
  {Farina}, {Jahnke}, {Nagao}, {Tominaga}, \& {Wang}}]{2020ApJ...898..105O}
{Onoue}, M., {Ba{\~n}ados}, E., {Mazzucchelli}, C., {et~al.} 2020, \apj, 898,
  105, \dodoi{10.3847/1538-4357/aba193}

\bibitem[{{Pensabene} {et~al.}(2020){Pensabene}, {Carniani}, {Perna}, {Cresci},
  {Decarli}, {Maiolino}, \& {Marconi}}]{2020A&A...637A..84P}
{Pensabene}, A., {Carniani}, S., {Perna}, M., {et~al.} 2020, \aap, 637, A84,
  \dodoi{10.1051/0004-6361/201936634}

\bibitem[{{Pillepich} {et~al.}(2019){Pillepich}, {Nelson}, {Springel},
  {Pakmor}, {Torrey}, {Weinberger}, {Vogelsberger}, {Marinacci}, {Genel}, {van
  der Wel}, \& {Hernquist}}]{2019MNRAS.490.3196P}
{Pillepich}, A., {Nelson}, D., {Springel}, V., {et~al.} 2019, \mnras, 490,
  3196, \dodoi{10.1093/mnras/stz2338}

\bibitem[{{Plummer}(1911)}]{1911MNRAS..71..460P}
{Plummer}, H.~C. 1911, \mnras, 71, 460, \dodoi{10.1093/mnras/71.5.460}

\bibitem[{{Reines} \& {Volonteri}(2015)}]{2015ApJ...813...82R}
{Reines}, A.~E., \& {Volonteri}, M. 2015, \apj, 813, 82,
  \dodoi{10.1088/0004-637X/813/2/82}

\bibitem[{{Rizzo} {et~al.}(2020){Rizzo}, {Vegetti}, {Powell}, {Fraternali},
  {McKean}, {Stacey}, \& {White}}]{2020Natur.584..201R}
{Rizzo}, F., {Vegetti}, S., {Powell}, D., {et~al.} 2020, \nat, 584, 201,
  \dodoi{10.1038/s41586-020-2572-6}

\bibitem[{{Rupke}(2018)}]{2018Galax...6..138R}
{Rupke}, D. 2018, Galaxies, 6, 138, \dodoi{10.3390/galaxies6040138}

\bibitem[{{Rupke} {et~al.}(2005){Rupke}, {Veilleux}, \&
  {Sanders}}]{2005ApJS..160..115R}
{Rupke}, D.~S., {Veilleux}, S., \& {Sanders}, D.~B. 2005, \apjs, 160, 115,
  \dodoi{10.1086/432889}

\bibitem[{{Salmon} {et~al.}(2015){Salmon}, {Papovich}, {Finkelstein}, {Tilvi},
  {Finlator}, {Behroozi}, {Dahlen}, {Dav{\'e}}, {Dekel}, {Dickinson},
  {Ferguson}, {Giavalisco}, {Long}, {Lu}, {Mobasher}, {Reddy}, {Somerville}, \&
  {Wechsler}}]{2015ApJ...799..183S}
{Salmon}, B., {Papovich}, C., {Finkelstein}, S.~L., {et~al.} 2015, \apj, 799,
  183, \dodoi{10.1088/0004-637X/799/2/183}

\bibitem[{{Sanders} {et~al.}(1988){Sanders}, {Soifer}, {Elias}, {Madore},
  {Matthews}, {Neugebauer}, \& {Scoville}}]{1988ApJ...325...74S}
{Sanders}, D.~B., {Soifer}, B.~T., {Elias}, J.~H., {et~al.} 1988, \apj, 325,
  74, \dodoi{10.1086/165983}

\bibitem[{{Sargsyan} {et~al.}(2014){Sargsyan}, {Samsonyan}, {Lebouteiller},
  {Weedman}, {Barry}, {Bernard-Salas}, {Houck}, \&
  {Spoon}}]{2014ApJ...790...15S}
{Sargsyan}, L., {Samsonyan}, A., {Lebouteiller}, V., {et~al.} 2014, \apj, 790,
  15, \dodoi{10.1088/0004-637X/790/1/15}

\bibitem[{{Sault} {et~al.}(1995){Sault}, {Teuben}, \&
  {Wright}}]{1995ASPC...77..433S}
{Sault}, R.~J., {Teuben}, P.~J., \& {Wright}, M.~C.~H. 1995, in Astronomical
  Society of the Pacific Conference Series, Vol.~77, Astronomical Data Analysis
  Software and Systems IV, ed. R.~A. {Shaw}, H.~E. {Payne}, \& J.~J.~E.
  {Hayes}, 433

\bibitem[{{Schaerer} {et~al.}(2020){Schaerer}, {Ginolfi}, {B{\'e}thermin},
  {Fudamoto}, {Oesch}, {Le F{\`e}vre}, {Faisst}, {Capak}, {Cassata},
  {Silverman}, {Yan}, {Jones}, {Amorin}, {Bardelli}, {Boquien}, {Cimatti},
  {Dessauges-Zavadsky}, {Giavalisco}, {Hathi}, {Fujimoto}, {Ibar}, {Koekemoer},
  {Lagache}, {Lemaux}, {Loiacono}, {Maiolino}, {Narayanan}, {Morselli},
  {M{\'e}ndez-Hern{\`a}ndez}, {Pozzi}, {Riechers}, {Talia}, {Toft}, {Vallini},
  {Vergani}, {Zamorani}, \& {Zucca}}]{2020A&A...643A...3S}
{Schaerer}, D., {Ginolfi}, M., {B{\'e}thermin}, M., {et~al.} 2020, \aap, 643,
  A3, \dodoi{10.1051/0004-6361/202037617}

\bibitem[{{Schindler} {et~al.}(2020){Schindler}, {Farina}, {Ba{\~n}ados},
  {Eilers}, {Hennawi}, {Onoue}, {Venemans}, {Walter}, {Wang}, {Davies},
  {Decarli}, {Rosa}, {Drake}, {Fan}, {Mazzucchelli}, {Rix}, {Worseck}, \&
  {Yang}}]{2020ApJ...905...51S}
{Schindler}, J.-T., {Farina}, E.~P., {Ba{\~n}ados}, E., {et~al.} 2020, \apj,
  905, 51, \dodoi{10.3847/1538-4357/abc2d7}

\bibitem[{{Schulze} \& {Wisotzki}(2014)}]{2014MNRAS.438.3422S}
{Schulze}, A., \& {Wisotzki}, L. 2014, \mnras, 438, 3422,
  \dodoi{10.1093/mnras/stt2457}

\bibitem[{{Schweitzer} {et~al.}(2006){Schweitzer}, {Lutz}, {Sturm}, {Contursi},
  {Tacconi}, {Lehnert}, {Dasyra}, {Genzel}, {Veilleux}, {Rupke}, {Kim},
  {Baker}, {Netzer}, {Sternberg}, {Mazzarella}, \&
  {Lord}}]{2006ApJ...649...79S}
{Schweitzer}, M., {Lutz}, D., {Sturm}, E., {et~al.} 2006, \apj, 649, 79,
  \dodoi{10.1086/506510}

\bibitem[{{Setoguchi} {et~al.}(2021){Setoguchi}, {Ueda}, {Toba}, \&
  {Akiyama}}]{2021arXiv210111273S}
{Setoguchi}, K., {Ueda}, Y., {Toba}, Y., \& {Akiyama}, M. 2021, \apj, 909, 188,
  \dodoi{10.3847/1538-4357/abdf55}

\bibitem[{{Shao} {et~al.}(2017){Shao}, {Wang}, {Jones}, {Carilli}, {Walter},
  {Fan}, {Riechers}, {Bertoldi}, {Wagg}, {Strauss}, {Omont}, {Cox}, {Jiang},
  {Narayanan}, \& {Menten}}]{2017ApJ...845..138S}
{Shao}, Y., {Wang}, R., {Jones}, G.~C., {et~al.} 2017, \apj, 845, 138,
  \dodoi{10.3847/1538-4357/aa826c}

\bibitem[{{Shirakata} {et~al.}(2019){Shirakata}, {Okamoto}, {Kawaguchi},
  {Nagashima}, {Ishiyama}, {Makiya}, {Kobayashi}, {Enoki}, {Oogi}, \&
  {Okoshi}}]{2019MNRAS.482.4846S}
{Shirakata}, H., {Okamoto}, T., {Kawaguchi}, T., {et~al.} 2019, \mnras, 482,
  4846, \dodoi{10.1093/mnras/sty2958}

\bibitem[{{Silk} \& {Rees}(1998)}]{1998A&A...331L...1S}
{Silk}, J., \& {Rees}, M.~J. 1998, \aap, 331, L1.
\newblock \doarXiv{astro-ph/9801013}

\bibitem[{{Silverman} {et~al.}(2011){Silverman}, {Kampczyk}, {Jahnke},
  {Andrae}, {Lilly}, {Elvis}, {Civano}, {Mainieri}, {Vignali}, {Zamorani},
  {Nair}, {Le F{\`e}vre}, {de Ravel}, {Bardelli}, {Bongiorno}, {Bolzonella},
  {Cappi}, {Caputi}, {Carollo}, {Contini}, {Coppa}, {Cucciati}, {de la Torre},
  {Franzetti}, {Garilli}, {Halliday}, {Hasinger}, {Iovino}, {Knobel},
  {Koekemoer}, {Kova{\v{c}}}, {Lamareille}, {Le Borgne}, {Le Brun}, {Maier},
  {Mignoli}, {Pello}, {P{\'e}rez-Montero}, {Ricciardelli}, {Peng}, {Scodeggio},
  {Tanaka}, {Tasca}, {Tresse}, {Vergani}, {Zucca}, {Brusa}, {Cappelluti},
  {Comastri}, {Finoguenov}, {Fu}, {Gilli}, {Hao}, {Ho}, \&
  {Salvato}}]{2011ApJ...743....2S}
{Silverman}, J.~D., {Kampczyk}, P., {Jahnke}, K., {et~al.} 2011, \apj, 743, 2,
  \dodoi{10.1088/0004-637X/743/1/2}

\bibitem[{{Sofue}(2016)}]{2016PASJ...68....2S}
{Sofue}, Y. 2016, \pasj, 68, 2, \dodoi{10.1093/pasj/psv103}

\bibitem[{{Solomon} \& {Vanden Bout}(2005)}]{2005ARA&A..43..677S}
{Solomon}, P.~M., \& {Vanden Bout}, P.~A. 2005, \araa, 43, 677,
  \dodoi{10.1146/annurev.astro.43.051804.102221}

\bibitem[{{Speagle} {et~al.}(2014){Speagle}, {Steinhardt}, {Capak}, \&
  {Silverman}}]{2014ApJS..214...15S}
{Speagle}, J.~S., {Steinhardt}, C.~L., {Capak}, P.~L., \& {Silverman}, J.~D.
  2014, \apjs, 214, 15, \dodoi{10.1088/0067-0049/214/2/15}

\bibitem[{{Stacey} {et~al.}(2010){Stacey}, {Hailey-Dunsheath}, {Ferkinhoff},
  {Nikola}, {Parshley}, {Benford}, {Staguhn}, \&
  {Fiolet}}]{2010ApJ...724..957S}
{Stacey}, G.~J., {Hailey-Dunsheath}, S., {Ferkinhoff}, C., {et~al.} 2010, \apj,
  724, 957, \dodoi{10.1088/0004-637X/724/2/957}

\bibitem[{{Steinhardt} {et~al.}(2014){Steinhardt}, {Speagle}, {Capak},
  {Silverman}, {Carollo}, {Dunlop}, {Hashimoto}, {Hsieh}, {Ilbert}, {Le Fevre},
  {Le Floc'h}, {Lee}, {Lin}, {Lin}, {Masters}, {McCracken}, {Nagao}, {Petric},
  {Salvato}, {Sanders}, {Scoville}, {Sheth}, {Strauss}, \&
  {Taniguchi}}]{2014ApJ...791L..25S}
{Steinhardt}, C.~L., {Speagle}, J.~S., {Capak}, P., {et~al.} 2014, \apjl, 791,
  L25, \dodoi{10.1088/2041-8205/791/2/L25}

\bibitem[{{Straatman} {et~al.}(2014){Straatman}, {Labb{\'e}}, {Spitler},
  {Allen}, {Altieri}, {Brammer}, {Dickinson}, {van Dokkum}, {Inami},
  {Glazebrook}, {Kacprzak}, {Kawinwanichakij}, {Kelson}, {McCarthy},
  {Mehrtens}, {Monson}, {Murphy}, {Papovich}, {Persson}, {Quadri}, {Rees},
  {Tomczak}, {Tran}, \& {Tilvi}}]{2014ApJ...783L..14S}
{Straatman}, C. M.~S., {Labb{\'e}}, I., {Spitler}, L.~R., {et~al.} 2014, \apjl,
  783, L14, \dodoi{10.1088/2041-8205/783/1/L14}

\bibitem[{{Suh} {et~al.}(2020){Suh}, {Civano}, {Trakhtenbrot}, {Shankar},
  {Hasinger}, {Sanders}, \& {Allevato}}]{2020ApJ...889...32S}
{Suh}, H., {Civano}, F., {Trakhtenbrot}, B., {et~al.} 2020, \apj, 889, 32,
  \dodoi{10.3847/1538-4357/ab5f5f}

\bibitem[{{Symeonidis} {et~al.}(2016){Symeonidis}, {Giblin}, {Page}, {Pearson},
  {Bendo}, {Seymour}, \& {Oliver}}]{2016MNRAS.459..257S}
{Symeonidis}, M., {Giblin}, B.~M., {Page}, M.~J., {et~al.} 2016, \mnras, 459,
  257, \dodoi{10.1093/mnras/stw667}

\bibitem[{{Tadaki} {et~al.}(2018){Tadaki}, {Iono}, {Yun}, {Aretxaga},
  {Hatsukade}, {Hughes}, {Ikarashi}, {Izumi}, {Kawabe}, {Kohno}, {Lee},
  {Matsuda}, {Nakanishi}, {Saito}, {Tamura}, {Ueda}, {Umehata}, {Wilson},
  {Michiyama}, {Ando}, \& {Kamieneski}}]{2018Natur.560..613T}
{Tadaki}, K., {Iono}, D., {Yun}, M.~S., {et~al.} 2018, \nat, 560, 613,
  \dodoi{10.1038/s41586-018-0443-1}

\bibitem[{{Tenneti} {et~al.}(2019){Tenneti}, {Wilkins}, {Di Matteo}, {Croft},
  \& {Feng}}]{2019MNRAS.483.1388T}
{Tenneti}, A., {Wilkins}, S.~M., {Di Matteo}, T., {Croft}, R. A.~C., \& {Feng},
  Y. 2019, \mnras, 483, 1388, \dodoi{10.1093/mnras/sty3161}

\bibitem[{{Toba} {et~al.}(2017){Toba}, {Bae}, {Nagao}, {Woo}, {Wang}, {Wagner},
  {Sun}, \& {Chang}}]{2017ApJ...850..140T}
{Toba}, Y., {Bae}, H.-J., {Nagao}, T., {et~al.} 2017, \apj, 850, 140,
  \dodoi{10.3847/1538-4357/aa918a}

\bibitem[{{Trakhtenbrot} {et~al.}(2017){Trakhtenbrot}, {Lira}, {Netzer},
  {Cicone}, {Maiolino}, \& {Shemmer}}]{2017ApJ...836....8T}
{Trakhtenbrot}, B., {Lira}, P., {Netzer}, H., {et~al.} 2017, \apj, 836, 8,
  \dodoi{10.3847/1538-4357/836/1/8}

\bibitem[{{Treister} {et~al.}(2020){Treister}, {Messias}, {Privon}, {Nagar},
  {Medling}, {U}, {Bauer}, {Cicone}, {Mu{\~n}oz}, {Evans}, {Muller-Sanchez},
  {Comerford}, {Armus}, {Chang}, {Koss}, {Venturi}, {Schawinski}, {Casey},
  {Urry}, {Sanders}, {Scoville}, \& {Sheth}}]{2020ApJ...890..149T}
{Treister}, E., {Messias}, H., {Privon}, G.~C., {et~al.} 2020, \apj, 890, 149,
  \dodoi{10.3847/1538-4357/ab6b28}

\bibitem[{{Valiante} {et~al.}(2017){Valiante}, {Agarwal}, {Habouzit}, \&
  {Pezzulli}}]{2017PASA...34...31V}
{Valiante}, R., {Agarwal}, B., {Habouzit}, M., \& {Pezzulli}, E. 2017, \pasa,
  34, e031, \dodoi{10.1017/pasa.2017.25}

\bibitem[{{Veilleux} {et~al.}(2005){Veilleux}, {Cecil}, \&
  {Bland-Hawthorn}}]{2005ARA&A..43..769V}
{Veilleux}, S., {Cecil}, G., \& {Bland-Hawthorn}, J. 2005, \araa, 43, 769,
  \dodoi{10.1146/annurev.astro.43.072103.150610}

\bibitem[{{Veilleux} {et~al.}(2020){Veilleux}, {Maiolino}, {Bolatto}, \&
  {Aalto}}]{2020A&ARv..28....2V}
{Veilleux}, S., {Maiolino}, R., {Bolatto}, A.~D., \& {Aalto}, S. 2020, \aapr,
  28, 2, \dodoi{10.1007/s00159-019-0121-9}

\bibitem[{{Venemans} {et~al.}(2019){Venemans}, {Neeleman}, {Walter}, {Novak},
  {Decarli}, {Hennawi}, \& {Rix}}]{2019ApJ...874L..30V}
{Venemans}, B.~P., {Neeleman}, M., {Walter}, F., {et~al.} 2019, \apjl, 874,
  L30, \dodoi{10.3847/2041-8213/ab11cc}

\bibitem[{{Venemans} {et~al.}(2016){Venemans}, {Walter}, {Zschaechner},
  {Decarli}, {De Rosa}, {Findlay}, {McMahon}, \&
  {Sutherland}}]{2016ApJ...816...37V}
{Venemans}, B.~P., {Walter}, F., {Zschaechner}, L., {et~al.} 2016, \apj, 816,
  37, \dodoi{10.3847/0004-637X/816/1/37}

\bibitem[{{Venemans} {et~al.}(2017{\natexlab{a}}){Venemans}, {Walter},
  {Decarli}, {Ferkinhoff}, {Wei{\ss}}, {Findlay}, {McMahon}, {Sutherland}, \&
  {Meijerink}}]{2017ApJ...845..154V}
{Venemans}, B.~P., {Walter}, F., {Decarli}, R., {et~al.} 2017{\natexlab{a}},
  \apj, 845, 154, \dodoi{10.3847/1538-4357/aa81cb}

\bibitem[{{Venemans} {et~al.}(2017{\natexlab{b}}){Venemans}, {Walter},
  {Decarli}, {Ba{\~n}ados}, {Hodge}, {Hewett}, {McMahon}, {Mortlock}, \&
  {Simpson}}]{2017ApJ...837..146V}
---. 2017{\natexlab{b}}, \apj, 837, 146, \dodoi{10.3847/1538-4357/aa62ac}

\bibitem[{{Venemans} {et~al.}(2017{\natexlab{c}}){Venemans}, {Walter},
  {Decarli}, {Ba{\~n}ados}, {Carilli}, {Winters}, {Schuster}, {da Cunha},
  {Fan}, {Farina}, {Mazzucchelli}, {Rix}, \& {Weiss}}]{2017ApJ...851L...8V}
---. 2017{\natexlab{c}}, \apjl, 851, L8, \dodoi{10.3847/2041-8213/aa943a}

\bibitem[{{Venemans} {et~al.}(2018){Venemans}, {Decarli}, {Walter},
  {Ba{\~n}ados}, {Bertoldi}, {Fan}, {Farina}, {Mazzucchelli}, {Riechers},
  {Rix}, {Wang}, \& {Yang}}]{2018ApJ...866..159V}
{Venemans}, B.~P., {Decarli}, R., {Walter}, F., {et~al.} 2018, \apj, 866, 159,
  \dodoi{10.3847/1538-4357/aadf35}

\bibitem[{{Venemans} {et~al.}(2020){Venemans}, {Walter}, {Neeleman}, {Novak},
  {Otter}, {Decarli}, {Ba{\~n}ados}, {Drake}, {Farina}, {Kaasinen},
  {Mazzucchelli}, {Carilli}, {Fan}, {Rix}, \& {Wang}}]{2020ApJ...904..130V}
{Venemans}, B.~P., {Walter}, F., {Neeleman}, M., {et~al.} 2020, \apj, 904, 130,
  \dodoi{10.3847/1538-4357/abc563}

\bibitem[{{Vestergaard} \& {Osmer}(2009)}]{2009ApJ...699..800V}
{Vestergaard}, M., \& {Osmer}, P.~S. 2009, \apj, 699, 800,
  \dodoi{10.1088/0004-637X/699/1/800}

\bibitem[{{Volonteri} \& {Stark}(2011)}]{2011MNRAS.417.2085V}
{Volonteri}, M., \& {Stark}, D.~P. 2011, \mnras, 417, 2085,
  \dodoi{10.1111/j.1365-2966.2011.19391.x}

\bibitem[{{Wang} {et~al.}(2019){Wang}, {Wang}, {Fan}, {Wu}, {Yang}, {Neri}, \&
  {Yue}}]{2019ApJ...880....2W}
{Wang}, F., {Wang}, R., {Fan}, X., {et~al.} 2019, \apj, 880, 2,
  \dodoi{10.3847/1538-4357/ab2717}

\bibitem[{{Wang} {et~al.}(2018){Wang}, {Yang}, {Fan}, {Yue}, {Wu}, {Schindler},
  {Bian}, {Li}, {Farina}, {Ba{\~n}ados}, {Davies}, {Decarli}, {Green}, {Jiang},
  {Hennawi}, {Huang}, {Mazzucchelli}, {McGreer}, {Venemans}, {Walter}, \&
  {Beletsky}}]{2018ApJ...869L...9W}
{Wang}, F., {Yang}, J., {Fan}, X., {et~al.} 2018, \apjl, 869, L9,
  \dodoi{10.3847/2041-8213/aaf1d2}

\bibitem[{{Wang} {et~al.}(2021){Wang}, {Yang}, {Fan}, {Hennawi}, {Barth},
  {Banados}, {Bian}, {Boutsia}, {Connor}, {Davies}, {Decarli}, {Eilers},
  {Farina}, {Green}, {Jiang}, {Li}, {Mazzucchelli}, {Nanni}, {Schindler},
  {Venemans}, {Walter}, {Wu}, \& {Yue}}]{2021arXiv210103179W}
---. 2021, \apjl, 907, L1, \dodoi{10.3847/2041-8213/abd8c6}

\bibitem[{{Wang} {et~al.}(2010){Wang}, {Carilli}, {Neri}, {Riechers}, {Wagg},
  {Walter}, {Bertoldi}, {Menten}, {Omont}, {Cox}, \&
  {Fan}}]{2010ApJ...714..699W}
{Wang}, R., {Carilli}, C.~L., {Neri}, R., {et~al.} 2010, \apj, 714, 699,
  \dodoi{10.1088/0004-637X/714/1/699}

\bibitem[{{Wang} {et~al.}(2013){Wang}, {Wagg}, {Carilli}, {Walter}, {Lentati},
  {Fan}, {Riechers}, {Bertoldi}, {Narayanan}, {Strauss}, {Cox}, {Omont},
  {Menten}, {Knudsen}, {Neri}, \& {Jiang}}]{2013ApJ...773...44W}
{Wang}, R., {Wagg}, J., {Carilli}, C.~L., {et~al.} 2013, \apj, 773, 44,
  \dodoi{10.1088/0004-637X/773/1/44}

\bibitem[{{Willott} {et~al.}(2015){Willott}, {Bergeron}, \&
  {Omont}}]{2015ApJ...801..123W}
{Willott}, C.~J., {Bergeron}, J., \& {Omont}, A. 2015, \apj, 801, 123,
  \dodoi{10.1088/0004-637X/801/2/123}

\bibitem[{{Willott} {et~al.}(2017){Willott}, {Bergeron}, \&
  {Omont}}]{2017ApJ...850..108W}
---. 2017, \apj, 850, 108, \dodoi{10.3847/1538-4357/aa921b}

\bibitem[{{Willott} {et~al.}(2013){Willott}, {Omont}, \&
  {Bergeron}}]{2013ApJ...770...13W}
{Willott}, C.~J., {Omont}, A., \& {Bergeron}, J. 2013, \apj, 770, 13,
  \dodoi{10.1088/0004-637X/770/1/13}

\bibitem[{{Willott} {et~al.}(2007){Willott}, {Delorme}, {Omont}, {Bergeron},
  {Delfosse}, {Forveille}, {Albert}, {Reyl{\'e}}, {Hill}, {Gully-Santiago},
  {Vinten}, {Crampton}, {Hutchings}, {Schade}, {Simard}, {Sawicki}, {Beelen},
  \& {Cox}}]{2007AJ....134.2435W}
{Willott}, C.~J., {Delorme}, P., {Omont}, A., {et~al.} 2007, \aj, 134, 2435,
  \dodoi{10.1086/522962}

\bibitem[{{Willott} {et~al.}(2010){Willott}, {Delorme}, {Reyl{\'e}}, {Albert},
  {Bergeron}, {Crampton}, {Delfosse}, {Forveille}, {Hutchings}, {McLure},
  {Omont}, \& {Schade}}]{2010AJ....139..906W}
{Willott}, C.~J., {Delorme}, P., {Reyl{\'e}}, C., {et~al.} 2010, \aj, 139, 906,
  \dodoi{10.1088/0004-6256/139/3/906}

\bibitem[{{Yang} {et~al.}(2019){Yang}, {Wang}, {Fan}, {Yue}, {Wu}, {Li},
  {Bian}, {Jiang}, {Ba{\~n}ados}, \& {Beletsky}}]{2019AJ....157..236Y}
{Yang}, J., {Wang}, F., {Fan}, X., {et~al.} 2019, \aj, 157, 236,
  \dodoi{10.3847/1538-3881/ab1be1}

\bibitem[{{Yang} {et~al.}(2020){Yang}, {Wang}, {Fan}, {Hennawi}, {Davies},
  {Yue}, {Banados}, {Wu}, {Venemans}, {Barth}, {Bian}, {Boutsia}, {Decarli},
  {Farina}, {Green}, {Jiang}, {Li}, {Mazzucchelli}, \&
  {Walter}}]{2020ApJ...897L..14Y}
---. 2020, \apjl, 897, L14, \dodoi{10.3847/2041-8213/ab9c26}

\bibitem[{{Zubovas}(2018)}]{2018MNRAS.473.3525Z}
{Zubovas}, K. 2018, \mnras, 473, 3525, \dodoi{10.1093/mnras/stx2569}

\bibitem[{{Zubovas} \& {King}(2012)}]{2012ApJ...745L..34Z}
{Zubovas}, K., \& {King}, A. 2012, \apjl, 745, L34,
  \dodoi{10.1088/2041-8205/745/2/L34}

\end{thebibliography}

\end{document}